\documentclass[aps,prf,amsmath,amssymb,amsfonts]{revtex4-2}
\usepackage{epsfig}
\usepackage{graphicx}
\usepackage{dcolumn}
\usepackage{bm}
\usepackage{color}
\usepackage{epstopdf}

\usepackage[titletoc,title]{appendix}
\begin{document}
\title{Intermittency and collisions of fast sedimenting droplets in turbulence}
\author{Itzhak Fouxon$^{1}$}\email{itzhak8@gmail.com}
\author{Seulgi Lee$^{2}$}\email{sg.lee@yonsei.ac.kr}
\author{Changhoon Lee$^{1,2}$}\email{clee@yonsei.ac.kr}
\affiliation{$^1$ School of Mathematics and Computing, Yonsei University, Seoul 03722, South Korea}
\affiliation{$^2$ Department of Mechanical Engineering, Yonsei University, Seoul 03722, South Korea}

\begin{abstract}

We study theoretically and numerically spatial distribution and collision rate of droplets that sediment in homogeneous isotropic Navier-Stokes turbulence. It is assumed that, as it often happens in clouds, typical turbulent accelerations of fluid particles are much smaller than gravity. This was shown to imply that the particles interact weakly with individual vortices and, as a result,  form a smooth flow in most of the space.
In weakly intermittent turbulence with moderate Reynolds number $Re_{\lambda}$, rare regions where the flow breaks down can be neglected in the calculation of space averaged rate of droplet collisions. However, increase of $Re_{\lambda}$ increases probability of rare, large quiescent vortices whose long coherent interaction with the particles destroys the flow. Thus at higher $Re_{\lambda}$, that apparently include those in the clouds, the space averaged collision rate forms in rare regions where the assumption of smooth flow breaks down. This intermittency of collisions implies that rain initiation could be a strongly non-uniform process. We describe the transition between the regimes and provide collision kernel in the case of moderate $Re_{\lambda}$ describable by the flow. The distribution of pairwise distances (radial distribution function or RDF) is shown to obey a separable dependence on the magnitude and the polar angle of the separation vector. Magnitude dependence obeys a power-law with a negative exponent, manifesting multifractality of the droplets' attractor in space. We provide the so far missing numerical confirmation of a relation between this exponent and the Lyapunov exponents and demonstrate that it holds beyond the theoretical range. The angular dependence of the RDF exhibits a maximum at small angles quantifying particles' formation of spatial columns. We provide typical dimensions of the columns, which belong in the inertial range. We derive the droplets' collision kernel using that in the considered limit the gradients of droplets' flow are Gaussian. We demonstrate that as $Re_{\lambda}$ increases the columns' aspect ratio decreases, eventually becoming one when the isotropy is restored. We propose how the theory could be constructed at higher $Re_{\lambda}$ of clouds by using the example of the RDF.

\end{abstract}

\maketitle

\section{Introduction}

Turbulence of air in warm clouds accelerates collisions of water droplets and thus must be included in studies of precipitation \cite{review,kh,pk,nature,kh1,kh2}. This inclusion is of high interest since it could help to resolve the bottleneck problem in rain formation. The bottleneck is caused by the narrowness of the size distribution of droplets created by condensation of vapor. The size proximity implies that the difference in settling velocities of the droplets is so small that their gravitational collisions would take long times, that are incompatible with the observations. Hence, the passage to the later stages of formation of large precipitating drops, where the growth occurs by gravitational collisions of different size droplets, demands something beyond gravity. Turbulence, that creates size dispersion by introducing mechanism for collisions of equal-size droplets, can be the missing factor. However, the relevance of turbulence has not been quantified so far \cite{lars} and assessment of how much turbulence influences the rain formation is an open problem. A main problem is the extremely high Reynolds number that holds in the clouds. Thus numerical studies can be performed only at much smaller Taylor microscale Reynolds numbers $Re_{\lambda}$ than in the clouds. This makes rigorous theoretical studies of droplets' collisions in the Navier-Stokes turbulence (NST), which is the focus of this paper, specially significant. However theoretical studies of droplets' behavior in high-$Re_{\lambda}$ NST are obstructed by having to deal with flow whose statistics is unknown. The problem is aggravated because intermittency produces significant probability of rare events that may locally accelerate the collision rates by a large factor in comparison with estimates using typical events.

The lack of knowledge of statistics of turbulence by itself does not prohibit quantitative predictions for the turbulent transport. In fact, the theory is able to provide accurate quantitative predictions for the NST  \cite{nature}. This is thanks to independence of certain properties of turbulent transport of the details of the statistics (universality). As it is often the case in statistical physics, the universality holds due to the appearance of sum of a large number of independent random variables in the analysis. Here we study the case where the particle sediments so quickly through an individual vortex that the vortex perturbs its motion only weakly. Thus the droplet's velocity is determined by effects of many independent vortices accumulated during the particle velocity relaxation time $\tau$. The issuing universality demands that intermittency is not too strong so that the probability of large quiescent vortices is not too high (intermittency both increases the probability of strong bursts and of large quiescent regions \cite{frisch}).
Otherwise individual vortex is so large that the particle never leaves it during the interaction time $\tau$. The motion would then be determined by interaction with a single vortex destroying the universality. This limits the study to moderate $Re_{\lambda}$ well below those in clouds where the statistics is very intermittent. Still we demonstrate that the results provide a useful reference point and some of them do generalize to high $Re_{\lambda}$.

As far as the case of weak intermittency is concerned, the present work continues the study \cite{fi2015} which provided successful quantitative predictions for correlation dimension and Lyapunov exponents of particles in the Navier-Stokes turbulence. The predictions were confirmed by direct numerical simulations at $Re_{\lambda}=70$ (claimed in \cite{Ireland} contradiction with some data is eliminated below). Here we provide the theory of the collision kernel and perform simulations at the same $Re_{\lambda}=70$, that demonstrate full agreement with the theory. We derive angle-dependent radial distribution function (RDF) that gives probability of finding a pair of particles at a given vector separation. The RDF determines the collision kernel by providing probability of a pair at collision distance. The angle dependence of the RDF allows us to obtain vertical and horizontal dimensions of particles' columns in space whose formation is the signature of the studied fast sedimentation limit, see Fig. \ref{snapshots}. In the remainder of the Introduction we provide
outlook (currently lacking in the literature) at the existing rigorous theories and describe the paper's organization.

\subsection{Theory at weak gravity and small inertia and its breakdown in clouds} \label{wg}

Until \cite{fi2015}, a complete rigorous theory for inertial particles in the NST existed only at negligible gravity and small but finite inertia \cite{nature,itzhak1}. The magnitude of inertia is measured by the dimensionless Stokes number $St$ which is the ratio of $\tau$ and the typical turnover-time of the viscous scale vortices, the Kolmogorov time $\tau_K$, see \cite{frisch}. The inertia is usually considered to be small at $St \ll 1$ where the particle velocity would relax quickly to the local velocity of the flow and the particles would nearly trace the flow \cite{maxey}. However we demonstrate in this subsection that intermittency of turbulence quite certainly
makes the effect of inertia in clouds large, even at $St \ll 1$.

The work \cite{nature} introduced a picture of motion of particles in space that holds at $St\ll 1$ independently of the Reynolds number and, thus, would hold also in the clouds. This picture can be understood by using the concept of the local Stokes number, similar to that of the local Reynolds number \cite{frisch}. This number is necessary in order to describe strong contrasts in the strength of particle-vortex interaction throughout a high Reynolds number flow. The contrasts can be described via the local energy dissipation rate $\epsilon$, defined as the product of the kinematic viscosity $\nu$ and the square of velocity gradients. The rate undergoes strong intermittent fluctuations in clouds, see \cite{review} and references therein. We observe that the Kolmogorov time, defining $St$, is given by $\tau_K\equiv \sqrt{\nu/\epsilon_0}$ where $\epsilon_0$ is the mean energy dissipation.  Correspondingly the local relevance of inertia can be characterised by the local Stokes number $\tau \sqrt{\epsilon/\nu}$. If the local Stokes number is not small, then strong interaction of particles with the local vortices occurs in that region. Due to the intermittency of turbulence, the magnitude of fluctuations of the local Stokes number depends on $Re_{\lambda}$, implying dependence of particles' statistics on both $St$ and $Re_{\lambda}$.

At $St\ll 1$, for the more probable events turbulent vortices are slow, $\tau \sqrt{\epsilon/\nu}\ll 1$.  The particles' motion in these vortices is smooth and ordered in space so that there is no crossing of trajectories of different particles. A flow of particles $\bm v(t, \bm x)$ can be introduced \cite{maxey}, providing the changes of their coordinates $\bm x(t)$ as $\dot{\bm x}=\bm v(t, \bm x(t))$.  However for fast vortices with time-scale $\lesssim \tau$, characterized by non-small local Stokes number, the turbulent driving creates jets of particles that separate from the turbulent flow, as in a sling, and cause the trajectories' crossing \cite{nature}. This ``sling effect" was observed experimentally \cite{gb} and at any $Re_{\lambda}$, including those in clouds, it is confined to well isolated small regions of fast vortices. Increase of $Re_{\lambda}$, at least in the high-$Re_{\lambda}$ limit, would make regions of the sling effect even more rare in space because the regions of quiescent turbulence increase in size with the Reynolds number \cite{frisch}. This creates significant difficulties in experimental and numerical measurements which must have resolution that increases with $Re_{\lambda}$.

The above picture implies that space averages, such as the average rate of droplets' collisions, can be found as sums of contributions of flow and sling regions (termed below "flow contribution" and "sling contribution"). The contributions of these regions into the collision rate were calculated in \cite{nature}. It was demonstrated that, despite that the regions of the sling effect are rare in space, their contribution into the collision kernel can be significant because they create optimal conditions for collisions \cite{nature,fast}.

In contrast with the above qualitative picture, that is valid at any Reynolds number, quantitative predictions of \cite{nature,itzhak1} break down at increasing $Re_{\lambda}$. The predictions for the RDF rely on the assumption that at $St\ll 1$ the RDF is determined by vortices whose turnover time, similarly to the most probable turnover time $\tau_K$, is much smaller than $\tau$. Making that assumption, the theory implies that distribution of inertial particles in the turbulent flow is multifractal. Multifractality manifests itself in the RDF that obeys at small distances a power-law with an exponent $-\alpha$. Here $\alpha$ is positive
- the RDF diverges at zero separation because the particles' concentration is singular. An explicit formula for $\alpha$ via a high order moment of the turbulent velocity gradients is provided in \cite{nature,itzhak1}. Application of the standard phenomenology of turbulence to that formula gives that at large $Re_{\lambda}$ the magnitude of the exponent is proportional to the product of $St^2$ and a positive power of Reynolds number. This implies that in the limit of large $Re_{\lambda}$, however small $St$ is, the exponent, calculated under the described assumptions, is larger than $3$. This however implies that the RDF becomes non-integrable at the origin which contradicts finiteness of the total number of particles (the number's second moment is proportional to the integral of the RDF over distance).

The reason for the described contradiction is that the assumption of the theory that the RDF is determined by vortices with turnover time smaller than $\tau$ becomes inconsistent at large $Re_{\lambda}$. In fact, the theory predicts its own breakdown via the formula for $\alpha$. Intermittency of turbulence implies appreciable presence in the flow of vortices whose turnover time is much smaller than $\tau_K$. More precisely the probability that the turnover time is smaller than $\tau_K$ by a power of the Reynolds number is appreciable \cite{frisch}, cf. the recent \cite{bud}. As a result, in the limit of large Reynolds numbers the gradients that determine $\alpha$ grow with $Re_{\lambda}$ according to a non-trivial power-law \cite{sreeni}. This implies that the time-scale of the vortices that determine $\alpha$ decreases with $Re_{\lambda}$ as $\tau_K Re_{\lambda}^{-q}$ with $q>0$ (according to the standard phenomenology time scales as the inverse gradient). Thus, considering increase of $Re_{\lambda}$ at a fixed small $St$, at large enough $Re_{\lambda}$ the time-scale of the relevant vortices becomes of order $\tau$. We designate the corresponding threshold Reynolds number by $Re_{\lambda}^*$. This number is not sharp and is determined by order of magnitude only via the asymptotic equality $\tau_K \left(Re_{\lambda}^*\right)^{-q}\sim \tau$. Thus at $Re_{\lambda}\gtrsim Re_{\lambda}^*$ the assumption that the relevant vortices have time-scale larger than $\tau$ is inconsistent. We conclude that the RDF is determined at $Re_{\lambda}\gtrsim Re_{\lambda}^*$ by vortices whose characteristic turnover time is of order $\tau$ or less. Anomalously strong, short-lived vortices become relevant to the RDF at higher $Re_{\lambda}$ because intermittency makes their probability appreciable. We also conclude that an approach other than that of \cite{nature,itzhak1} must be devised at $Re_{\lambda}\gtrsim Re_{\lambda}^*$.

The above considerations demonstrate that the actual validity condition of the small inertia theory of \cite{nature,itzhak1} is $St \ll Re^{-q}$  and not $St\ll 1$, cf. \cite{fp,fpsl}. The value of $q$ can hardly be predicted and must be fixed numerically. Thus the theory of \cite{nature,itzhak1} is a so-called asymptotic theory at small $St$ i.e. it holds at $St\to 0$ when $Re_{\lambda}$ is fixed.

The above demonstration of limitations of the theory of \cite{nature} due to intermittency is performed by using the laws holding in the limit of large $Re_{\lambda}$. This does not signify that intermittency becomes relevant only at $Re_{\lambda}\to\infty$. Simulations of \cite{fp} of the particles' motion in the NST demonstrated that for $St\sim 0.1$ the small inertia theory of the RDF of \cite{nature} applies accurately at the moderate $Re_{\lambda}=21$. However, considering higher $Re_{\lambda}$, already at $Re_{\lambda}=47$, that theory fails. We see that smallness of $St$ does not guarantee the smallness of inertial effects already at rather moderate $Re_{\lambda}$. Despite that the criterion of validity of the theory, $St \ll Re^{-q}$, derived at $Re_{\lambda}\to\infty$, cannot be used at $Re_{\lambda}$ of \cite{fp}, we can use it to get very rough idea of magnitude of $q$. We find that $q\sim 0.5$ i.e. it is a number of order one. Using $q=0.5$ we find that for $Re_{\lambda}\sim 10^4$, that can hold in the clouds, the effects of inertia on the RDF are appreciable at $St$ as small as $0.01$ (which covers the whole range of relevant droplet sizes, see Sec. \ref{dar}). More precise considerations demand extensive future numerical work.

The above conclusion transfers to the collision kernel which, under the assumption of small inertia, is given by a sum of two terms \cite{nature}. One term describes the contribution of vortices whose turnover time is much smaller than $\tau$. That term is proportional to the RDF and, as we saw, it becomes determined by vortices with turnover time of order $\tau$ at higher $Re_{\lambda}$ . The other term in the collision kernel describes the contribution of the sling effect which by definition is due to vortices with time-scale $\tau$ or smaller. Thus in the limit of high $Re_{\lambda}$ the collision kernel is due to vortices with turnover time $\tau$ or smaller.

We summarize how increase of $Re_{\lambda}$ results in the breakdown of the theory of \cite{nature}. The theory demonstrates that the clustering rate, holding in the regions of the smooth motion of the droplets, is determined by vortices whose characteristic time-scale decreases with $Re_{\lambda}$ as a power-law. The theory holds at moderate $Re_{\lambda}$ where this time-scale is much larger than $\tau$ as necessary for the self-consistency of the assumption of smooth motion. When $Re_{\lambda}$ increases, faster vortices with decreasing turnover time become relevant, until their time-scale becomes comparable with $\tau$ at some $Re^*_{\lambda}$, cf. \cite{fp}. When this happens the separation of the total rate of collisions into the contributions of the regions of the smooth motion and of the sling effect, that was used in \cite{nature} for the calculations, breaks down. Both contributions are determined by "resonant" vortices with lifetime of order $\tau$. 
The corresponding changes in the theory will be published elsewhere.

\subsection{Fast sedimentation theory}

Applicability of the above theory to clouds is limited by three factors. The first factor is the Stokes number which is not necessarily small for the droplets taking part in the rain formation. The second factor is gravity, which in clouds is strong and not weak. The magnitude of gravity can be measured by the Froude number $Fr$ given by the ratio of the typical (Lagrangian) acceleration of the fluid particles $\epsilon_0^{3/4}/\nu^{1/4}$ and gravitational acceleration $g$, i.e. $Fr\equiv \epsilon_0^{3/4}/(g\nu^{1/4})$. Thus, \cite{nature} applies at $Fr\to \infty$ when gravity is negligible. In contrast, in the clouds even rather strong cloud turbulence with $\epsilon_0=2000 cm^2/s^3$ gives $Fr=0.5$. For typical turbulence with smaller $\epsilon_0$, the Froude number is yet smaller. The last limiting factor is the high $Re_{\lambda}$ of the clouds, as described above. The two former limitations were overcome in \cite{fi2015} who constructed the theory that holds at strong gravity, $Fr\to 0$, and any $St$ including $St\gg 1$. However this theory, similarly to that considered in the previous subsection, is also asymptotic and has the validity condition $Fr\ll Re^{-q'}$ with $q'>0$, as will be discussed in this work. Thus how small $Fr$ must be, for the theory to be valid at a given $Re_{\lambda}$, is not known. It was found in numerical simulations of \cite{fi2015} that at $Re_{\lambda}=70$ the validity of the theory demands $Fr\leq 0.033$ which implies quite weak typical accelerations $\epsilon_0^{3/4}/\nu^{1/4}$ as compared with the gravity. The theory was also shown to apply reasonably well at $Fr=0.05$.

The main limitation of  \cite{fi2015} is again the Reynolds number - the intermittency increases the fluctuations of the gradients and destroys the approximations made. Nevertheless this theory, in comparison with that of \cite{nature}, incorporates the effects of the gravitational sedimentation of the droplets, that are strong in precipitating clouds, giving us qualitative insight and the possibility of consistent interpolation. It seems reasonable, given that clouds are indeed characterized by small $Fr$, to make the small $Fr$ theory a starting point for approaching collisions in clouds. It is to this theory that this paper is devoted. Here we provide more detailed predictions than in \cite{fi2015}, confirm them numerically and describe how the theory breaks down at higher Reynolds number.

Before continuing the development of the approach of \cite{fi2015}, a controversy must be faced. The theoretical formula of \cite{fi2015} for the correlation codimension $\alpha$, studied in detail below, was tested numerically in \cite{Ireland} at $Re_{\lambda}=398$. The comparison was done for $Fr=0.052$ and a range of $St$. They observed that while \cite{fi2015} describe correctly the independence of $\alpha$ of $St$ for $St\geq 1$, quantitatively the predictions are wrong by about $50$ per cent. It is unfortunate that the performed comparison contained two  mistakes. The equation $(4.21)$ of \cite{Ireland} which the authors considered as the prediction of \cite{fi2015} for $\alpha$, is factor of $2$ smaller than the actual prediction made in \cite{fi2015}. If the correct formula is used then the discrepancy is about $30$ and not $50$ per cent. This seems to be as much as one can hope for, because \cite{Ireland} use the predictions outside the domain of validity of the theory, which is $\alpha\ll 1$, cf. \cite{itzhak1}. In fact, the result of \cite{Ireland} completely agrees with the simulations of \cite{fi2015} that demonstrated that at $Fr=0.05$ there are significant deviations from the theory. The inequalities $\alpha\ll 1$ and $Fr\ll 1$ differ much because of the large numerical factor: we have $\alpha\approx 13 Fr$, see below and Sec. \ref{resolution}.

We stress the asymptotic character of the described theories in order to avoid future misunderstandings. Both theories of \cite{nature} and \cite{fi2015} hold rigorously in the limits of $St\to 0$ and $Fr\to 0$, respectively, when the Reynolds number is held fixed. Thus, if it is found that the predictions of \cite{fi2015} are invalid, then it tells that $Fr$ is too large and by decreasing $Fr$ the theory will be made to hold true. Similar situation holds for $St\to 0$ theory of \cite{nature}. Thus the observation of \cite{fp} of breakdown of the theory of \cite{nature} at $Re_{\lambda}\geq 47$ only tells that at this $Re_{\lambda}$ the theory applies at $St$ smaller than those considered in \cite{fp}. These $St$ are determined by the condition that their corresponding $\tau$ are much smaller than the lifetime of the vortices that determine the sum of the Lyapunov exponents in \cite{nature}.
As an example of the use of the asymptotic theory beyond its region of validity we will demonstrate below that the numbers obtained from $Fr\to 0$ theory at $Re_{\lambda}=70$ in \cite{fi2015} can be used for predicting the values observed at $Re_{\lambda}=398$ in \cite{Ireland}.


\subsection{Review of developments before the work \cite{fi2015}}  \label{fb}

We review the developments leading to the small $Fr$ theory of \cite{fi2015}, both to give credit and address concerns of an anonymous referee. This theory was aimed to explain the observations of \cite{parklee} of patterns of particles sedimenting in turbulence. Flow description of the particle motion was employed, which is rigorously valid at $Fr\to 0$ at other parameters, including the Reynolds number, fixed, cf. above. The earliest clue to the possibility of the flow description seemingly was made in \cite{fpsl} who observed that increasing gravity at a fixed, not necessarily small $St$, damps the sling effect. The reason is that faster sedimentation of the particle shortens the interaction time during which an individual vortex swings the particle before the shooting. Thus the sling events, that destroy single-valuedness of the flow \cite{nature}, become more rare at increasing gravity and at a sufficiently strong gravity the flow may become single-valued despite a possibly large $St$.

The value of gravity at which a smooth single-valued flow of droplets exists in the NST was provided in a Master thesis \cite{thesis}. The study was done disregarding the effects of intermittency and predicted that at $Fr\ll 1$ the flow is well-defined for typical vortices, independently of $St$. In application to clouds this implies that all droplets with size within the range from $10$ to $60$ microns, which is the range where turbulence has relevance, move in most of the space according to a size-dependent smooth flow. The condition, that the regions where the smooth flow description breaks down are rare, happens to imply that the flow's compressibility is necessarily weak (this coincidence occurs because the same property of inertia that causes the sling effect also causes the compressibility of the particles' flow). This allows to derive the droplets' distribution from the general solution for tracers' distributions in a weakly compressible random flow \cite{itzhak1,nature}.

Later, the observation that the sling effect is deactivated at large gravity was done in a model two-dimensional flow studied in the regime where the turbulent flow constitutes a small perturbation of the particle's trajectory in \cite{Gustavsson}, see also \cite{bec2014,parklee} and cf. similar expansion in \cite{maxey}. All these works but \cite{bec2014} used for the study of the sling effect the blowup equation introduced in \cite{nature}. The work \cite{bec2014} provided a different outlook by observing numerically that at $Fr\to 0$ average velocity difference of nearby particles scales linearly with the distance between these particles. Thus \cite{bec2014} concluded that in this limit the particles become tracers in an effective flow. Care is needed though. Indeed, if there is a flow, then the linear scaling holds for each realization and hence also statistically. However linear scaling of the average velocity difference would also hold for a velocity field with (effective) discontinuities as in Burgers turbulence \cite{fr}. Thus the observed linear scaling can be used as an indication of the existence of the flow however not as its proof.
Finally \cite{fi2015}, who worked independently of \cite{Gustavsson,bec2014} (the first arxiv version of \cite{fi2015} appeared in the same year with statement of independent work), gave a rigorous theory of the droplets' flow able to provide quantitative predictions for the particles' behavior in the NST. Direct numerical simulations (DNS) of the NST were done for weakly intermittent turbulence with $Re_{\lambda}=70$ and confirmed the theory. The work demonstrated how complete calculations can be performed, despite that there is no explicit formula for the droplets' flow in terms of the underlying fluid flow (the relationship between these flows is non-local both in space and in time).

The limit of large gravity studied in \cite{fi2015} assumes that the droplet's settling velocity is larger than the typical turbulent velocity at the Kolmogorov scale \cite{frisch}. However the settling velocity must still be smaller than the integral scale velocity, to fit the applications in clouds. Thus turbulence is not a small perturbation of the particles' trajectories, as in \cite{Gustavsson} or some qualitative considerations of \cite{bec2014}, because the particle velocity coincides with the local flow velocity in the leading order. The work \cite{fi2015} differed from previous works by aiming at a complete, rigorous $Fr\to 0$ theory for particles in the NST without modelling assumptions. This comes as the next effort to get a rigorous theory for the NST, after the $St\to 0$ theory of \cite{nature}. For the considered $Re_{\lambda}=70$ it was observed numerically that the theory is accurate for $Fr\leq 0.03$, with some theoretical predictions holding up to $Fr=0.1$, cf. above.


Certain aspects of the theory of \cite{fi2015} were observed previously. Due to fast sedimentation the separation of particle pairs is driven effectively by white noise, as it was observed for the NST in \cite{fouxonhorvai}. The authors originally considered the case of zero gravity and large inertia, $St\gg 1$. The large inertia causes the particle to drift fast through the fluid so that the flow "looks to it" as a white noise. However, \cite{fouxonhorvai} observe that, using the gravitational drift instead of inertial one, gives the answer for the case with gravity. The value of the first Lyapunov exponent for the case with gravity was provided. It can also be seen from this work that pair separation is effectively horizontal, the result which was significantly developed further in  \cite{bec2014}. Horizontality of separation implies that the particles spend more time when located one above the other which will be seen as columns in space observed in \cite{bec2014,parklee}.  Independently, the applicability of the white noise model, and preferential alignment of the vector inter-pair distance with the vertical, were observed in the model flow of \cite{Gustavsson}, who also derived numerically the Lyapunov exponents of their model. The observation that gravity enhances preferential concentration at $St\gtrsim 1$ and can result in strong particle clustering was done in \cite{Gustavsson,bec2014,parklee} independently. In contrast, at $St\ll 1$, gravity decreases the clustering where three different asymptotic regimes exist \cite{fi2015}.

\subsection{Organization of the paper}

All studies in this paper are performed for the Navier-Stokes equations of incompressible flow and no model of turbulence is used. Since the text is rather long then for reader's convenience we describe organization of the material. In the next section we introduce the equation of motion, discussing its applicability in the high $Re_{\lambda}$ case of clouds, and the numerical scheme. The main message of this section for future studies is that in high $Re_{\lambda}$ turbulence we need to use a more fundamental description of the particle-vortex interaction than in the usually used equation of particle motion, in order to be certain that we do not miss significant contributions in the collision kernel. This is because
the flow fluctuations with scales smaller than the particle size might be relevant due to intermittency.

In section \ref{lyapunov} we describe how increase of $Re_{\lambda}$ invalidates small $Fr$ theories in the simplest context. We derive the remarkably simple structure of separation of close particles at $Fr\to 0$: the vertical component of the separation is conserved and the horizontal component evolves as separation of two tracers in a white-noise velocity known as the Kraichnan model \cite{reviewt}. This structure is the reason for columns' formation. We derive the spectrum of the Lyapunov exponents extending the results of \cite{fi2015}. We observe that horizontal motions are driven by vortices much larger than the Kolmogorov scale whose characteristic size increases with $Re_{\lambda}$. This leads to breakdown of the theory above certain $Re_{\lambda}$, where the statistics becomes isotropic and particles would not form columns in space.

The next section is devoted to reviewing the developments that preceded the theory of \cite{fi2015} and description of the main predictions of that theory. It is demonstrated, in a significantly more detailed way than in \cite{fi2015}, that the DNS of the NST at $Re_{\lambda}=70$, indicate unequivocally that \cite{fi2015} provides us with a completely valid theory that can be used in the domain of its validity. The contradiction with simulations of \cite{Ireland} at $Re_{\lambda}=398$ is due to the application of the theory outside the domain of validity (this is besides that \cite{Ireland} use a wrong numerical factor in studying the prediction of \cite{fi2015}). Section \ref{clash} is devoted to explanation that the theory of \cite{fi2015} breaks down at increasing $Re_{\lambda}$. The reason is growing intermittency that implies both increasing regions of calm turbulence, allowing long coherent particle-eddy interactions, and increasing relevance of strong bursts of velocity gradients. It must be stressed again that the theory'd breakdown does not mean that there is some critical $Re_{\lambda}$ where it stops to work. Rather, it says that the range of small $Fr$ where \cite{fi2015} holds, shrinks to zero with increasing $Re_{\lambda}$ in a power-law fashion.

The rest of the paper is devoted to the case of $Fr$ so small that the flow description of \cite{fi2015} holds. We provide complete theory of collisions in this limit and its numerical confirmation. Section \ref{properties1} derives the RDF at not too small angles and same size particles. The main difference from \cite{fi2015} is the recognition of the fact that the smoothness scale of the droplets' flow is larger than the Kolmogorov scale by order of magnitude. We provide numerical data for the RDF (which was not done in \cite{fi2015}) and demonstrate that they confirm the theory. We also derive in this section a sum rule. That provides the probability density function (PDF) of the inter-pair distance irrespective of the pair's orientation in space, which equals the angle-averaged RDF. We demonstrate that small angles, that correspond to preferential vertical orientation of the pairs, can be neglected in the PDF of the distances. This imposes a constraint on the magnitude of preferential orientation.

The complete angle dependence of the RDF of equal size particles is derived in section \ref{angl}. Section \ref{ask} extends the calculations to different size particles by providing the bidisperse RDF. Section \ref{RDf} describes reduction of collision kernel of different size particles to the RDF. We apply Yaglom-type relation for rewriting the kernel and calculate the average velocity of approach of colliding droplets. The difference from the classical work \cite{st} is that compressibility demands somewhat different approach in treating the kernel. The approach velocity can be calculated completely due to Gaussianity of gradients of the droplets' flow. The next Section collects the information to complete the calculation of the collision kernel. Section \ref{face} studies the possibilities for interpolation of the results to the Reynolds numbers characteristic of clouds. 

\begin{figure*}
\includegraphics[width=17cm,trim={4mm 4mm 4mm 4mm},clip]{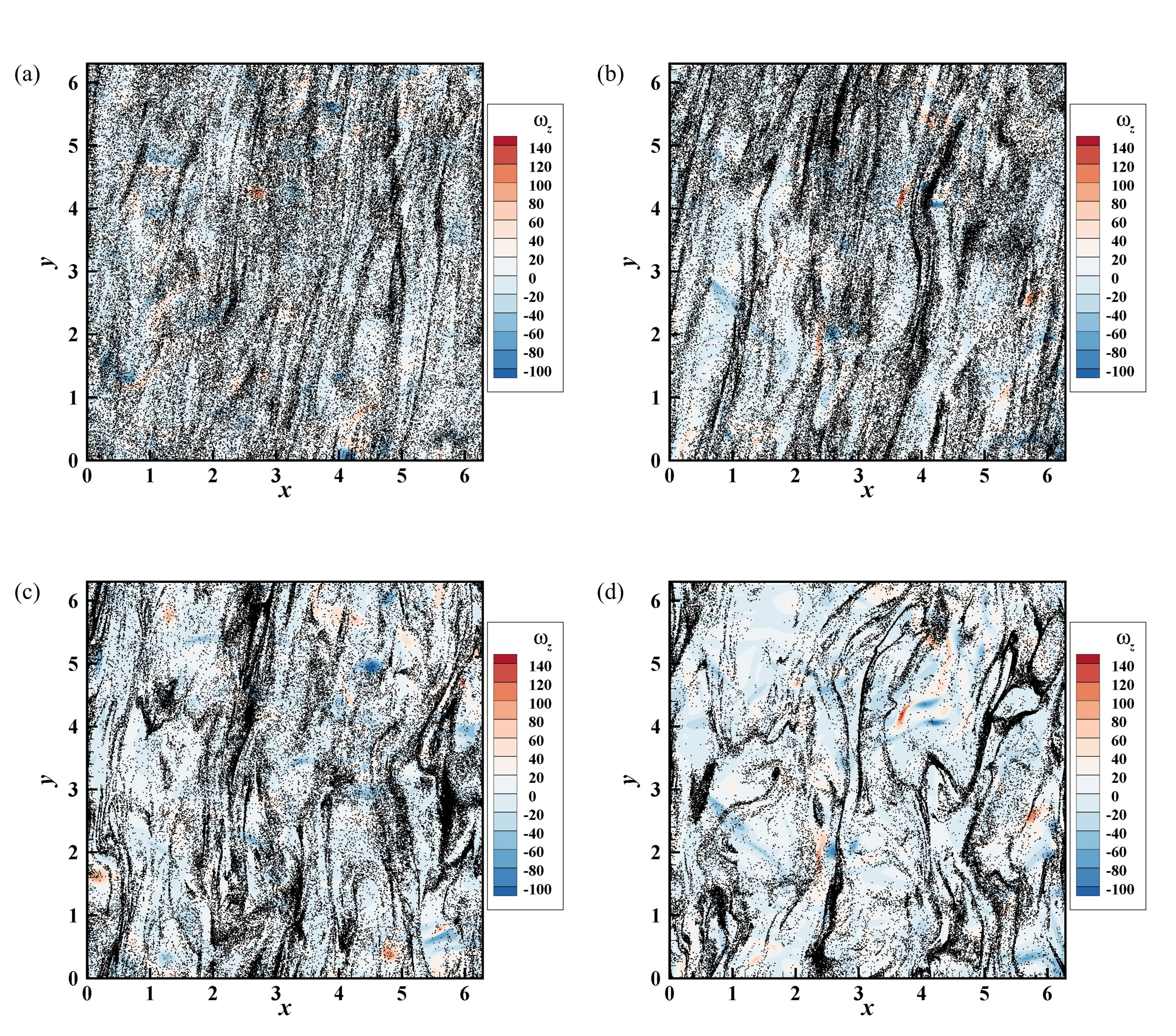}
\caption{DNS snapshots of particles' distributions in space at fixed $St=1$,  $Re_\lambda=70$ and increasing $Fr$ of (a) $Fr=0.0125$,  (b) $Fr=0.0167$, (c) $Fr=0.0333$, and (d) $Fr=0.05$. The distributions are multifractal. The positive difference of $3$ and the fractal dimension\ increases linearly with $Fr$ in the considered range. The fractal structure is anisotropic. Pronounced columns of particles predominantly point in the direction of gravity, that was slightly slanted from the $y$-axis to avoid numerical artifact due to periodicity. We demonstrate in the text that characteristic dimensions of the columns are $g\tau^2$ in the direction of gravity and $l_c\simeq 10 \eta$ in the perpendicular direction. This is confirmed by the Figure. Indeed $St=1$ and Kolmogorov scale $\eta=0.024$, imply $g\tau^2=\eta/Fr$. Thus the theory predicts that characteristic longest dimension of columns is $0.024/Fr$ and the characteristic shortest dimension is $0.24$. This agrees with the snapshots where columns become more convoluted at increasing $Fr$. There is no apparent correlation with the vorticity component in the perpendicular direction, which would hold for the centrifugal mechanism of clustering. }
\label{snapshots}
\end{figure*}

The paper is quite long and for convenience of the reader we provide a rather detailed summary and outlook in Section \ref{outlook}. Appendices provide technical details of calculations. The main results of this work are: demonstration of clash of $Fr\to 0$ and $Re_{\lambda}\to\infty$ limits; theory and numerical confirmation of angle-dependent RDF at $Fr\to 0$; description of dimensions of particles' columns in space (see Fig. \ref{snapshots}) and conjecture on the RDF in the high Reynolds number turbulence in clouds.

\section{Equation of motion and its applicability in clouds} \label{dar}

In this section, we consider the Newton equations of motion of the droplets. The equations are characterized by three dimensionless parameters - the Froude number $Fr=\epsilon_0^{3/4}/[g\nu^{1/4}]$, the Stokes number $St=\tau/\tau_K$ and the Taylor microscale Reynolds number $Re_{\lambda}$, see the Introduction. Our definition of $Fr$ is one of a number of possible definitions. It is chosen here because of its independence of the parameters of the droplets: $Fr$ characterizes turbulence with respect to gravity and is a property of the flow only, cf. \cite{fp} and below. All the three parameters play significant role in defining the nature of the particle trajectories. We describe the parameter ranges that are relevant for the rain formation problem. We stress the possibility that the usually used equation of motion could actually not apply in the clouds because of the strong intermittency.

{\bf Typical values}---We consider spherical droplets with radii $a$ from $10$ up to $60$ microns, which is the size range where turbulence is most relevant in the formation of larger droplets \cite{review}. For a given droplet radius $a$, the Stokes and Froude numbers are not independent. We consider the Stokes number $St_s(a)\equiv \tau_s(a)\sqrt{\epsilon_0/\nu}$ defined with the help of the Stokes time $\tau_s(a)=(2/9) (\rho/\rho_a) (a^2/\nu)$ where $\rho_a$ is the density of air and $\rho$ is the droplet density (this is not $St$ in the rest of the paper, see below). We have $St_s(a)=0.02 (a/15)^2\sqrt{\epsilon_0/10}$ where $a$ is measured in microns, $\epsilon_0$ is measured in $cm^2/s^3$  and we use the numerical values of $\nu=0.15\  cm^2/s$ and water-to-air density ratio of $814$. This formula can be used for studying the size dependence of $St_s$ at different levels of turbulence, as characterized by $\epsilon_0$. For instance, using $a$ of $10$ microns we find $St_s\gtrsim 0.01$ for $\epsilon_0 \gtrsim 10$. We conclude from considerations of subsection \ref{wg} that in the clouds, in the whole range of relevant droplet sizes, the RDF and the collision kernel are determined by vortices with time-scale of order $\tau$.

We also observe that $St_s(a)=0.5 (a/15)^2 Fr^{2/3}$. We find that at $Fr=0.03$ (corresponding to $\epsilon_0=48\  cm^2/s^3$ by $\epsilon_0/10=517 Fr^{4/3}$), the range of $St_s(a)\gtrsim 1$ corresponds to droplets' radii larger than forty microns. When $Fr=0.05$ (which is $\epsilon_0$ about $100 cm^2/s^3$), the range of sizes with $St_s(a)\gtrsim 1$ is somewhat larger than thirty microns. For not weak turbulence with $Fr=0.29$ (with $\epsilon\approx 1000\  cm^2/s^3$), we have $St_s(a)\gtrsim 1$ at size of 15 - 20 microns. Thus allows to see the range of sizes with non-small inertia for different strengths of turbulence.

{\bf Equation}---We use the effective linear drag description within which the droplet's coordinate $\bm x(t)$ and velocity $\bm v(t)$ obey
\begin{eqnarray}&&\!\!\!\!\!\!\!\!\!\!\!\!\!
\frac{d\bm x}{dt}=\bm v, \ \ \frac{d\bm v}{dt}=-\frac{\bm v-\bm u(t, \bm x(t))}{\tau}+\bm g,\label{rd}
\end{eqnarray}
where $\bm u(t, \bm x)$ is the incompressible homogeneous turbulent flow (obeying the Navier-Stokes equation), $\bm g$ is the vector of gravitational acceleration, and $\tau$ is the effective relaxation time, described below. We assume that the droplets form a dilute gas; therefore, the droplet's motion can be considered independently of other droplets. The equation's validity demands that the particles are much denser than the fluid, which is true for liquid droplets in air. The condition of validity of the linear friction force is that the flow changes weakly at the scale of the particle, i.e. $a$ is much smaller than the local viscous scale \cite{frisch}, and the Reynolds number of the flow perturbation due to the particle $Re_p$, given by the product of the drift velocity $|\bm v-\bm u(t, \bm x(t))|$ and $2a/\nu$, is small, $Re_p=2|\bm v-\bm u(t, \bm x(t))|a/\nu\ll 1$. We consider these conditions with the account of intermittency.

{\bf Role of fluctuations of viscous scale}---Due to intermittency, the local viscous scale has strong fluctuations where with appreciable probability its values may be smaller than the Kolmogorov scale $\eta \equiv \nu^{3/4}/\epsilon_0^{1/4}$ by a power of $Re_{\lambda}$, see e.g. \cite{frisch} for theoretical references and \cite{bud} for a recent numerical observation. In clouds, the Kolmogorov scale is one or two orders of magnitudes larger than relevant droplets' sizes $a$, see numerical estimates below. Since the viscous scale associated with extreme events is smaller than the Kolmogorov scale by $Re_{\lambda}^{\delta}$, where $\delta$ is about $1/2$ according to \cite{bud}, then it seems quite certain that Eqs.~(\ref{rd}) do not apply to extreme events in clouds. However, this is not necessarily of concern since these events might be irrelevant for the space-averaged rate of droplets' collisions. What is more relevant is whether the viscous scale associated with those vortices that {\it are} relevant is larger than $a$. The present work indicates that the kernel is determined by events for which the local Froude number $\epsilon^{3/4}/[g\nu^{1/4}]$ is of order one. For these events velocity gradients are larger than the typical value of $\sqrt{\epsilon_0/\nu}$ by a factor of $Fr^{-2/3}$ (with $Fr$ always standing for $\epsilon_0^{3/4}/[g\nu^{1/4}]$). The corresponding local viscous scale is smaller than the Kolmogorov scale by factor of $Fr^{-1/3}$ where we assume that the local viscous scale is of order of $\nu^{3/4}\epsilon^{-1/4}$. Since in practice $Fr^{1/3}\sim 1$ then the assumption that the local viscous scale of cloud vortices relevant for collision kernel is much larger than $a$ appears self-consistent.

Some reservations need to be made. We used in the above estimates the standard phenomenology of turbulence \cite{frisch} which could be invalid \cite{bud}. This might demand a reconsideration. Moreover self-consistency of the assumption that the local viscous scale of relevant vortices is much larger than $a$ may not yet guarantee that vortices with scale smaller than $a$ can be neglected in the collision kernel. For instance, self-consistency argument fails for the sling effect: the calculation of the collision kernel that neglects the sling effect is self-consistent in weakly intermittent turbulence (see the Introduction), yet the sling effect's contribution may be appreciable \cite{nature}. Therefore it might be necessary to perform a separate study of the effect of vortices with spatial scale $\lesssim a$. These vortices could cause a kind of "sling effect" of their own because they could generate large velocity difference of nearby particles. This would demand separate account of these vortices in the collision kernel. It seems that the only way to study this possibility is by performing numerical simulations at high $Re_{\lambda}$. These simulations must use the description of motion that is more fundamental than Eq.~(\ref{rd}) and applies also to vortices with size $\lesssim a$. This is left for future work.

{\bf Remaining condition of $Re_p\ll 1$}---For the study of this condition it is useful to integrate the velocity equation
\begin{eqnarray}&&\!\!\!\!\!\!\!\!
\bm v(t)=\bm v(0)\exp\left(-\frac{t}{\tau}\right) +\int_{0}^t dt' \exp\left[-\frac{t-t'}{\tau}\right]\left(\frac{\bm u[t', \bm x(t')]+\bm g\tau}{\tau}\right).\label{sm}
\end{eqnarray}
We find that, after the particle spent in the flow time larger than $\tau$, we have
\begin{eqnarray}&&\!\!\!\!\!\!\!\!
\bm v-\bm u(t, \bm x(t))=\bm g\tau+\int_{-\infty}^t dt' \exp\left[-\frac{t-t'}{\tau}\right]\left(\frac{\bm u[t', \bm x(t')]-\bm u[t, \bm x(t)]}{\tau}\right), \label{st}
\end{eqnarray}
where we rearranged the terms so that the RHS provides the velocity of the drift with respect to the local flow. The drift has a contribution due to the sedimentation velocity $\bm g\tau$ and the inertial lag behind the flow.
For the study of the RHS we start with instructive case of negligible gravity.

\subsection{Drift velocity at negligible gravity}

Here we consider the self-consistency of the assumption $Re_p\ll 1$ in the case of negligible gravity, $Fr\to \infty$.

{\bf Case of $St\ll 1$}---Provided that the local Stokes number $\tau \sqrt{\epsilon/\nu}$ is much smaller than one, we have the Maxey formula \cite{maxey}
\begin{eqnarray}&&\!\!\!\!\!\!\!\!
\bm v-\bm u(t, \bm x(t))=-\tau \bm a(t, \bm x(t))+O\left(\frac{\tau^2}{\tau_{\nu}^2}\right),\ \ \bm a\equiv \partial_t\bm u+(\bm u\cdot\nabla)\bm u,\  \ \frac{1}{\tau_{\nu}}\equiv \left|\frac{1}{a}\frac{da}{dt}\right|,\label{maxeyf}
\end{eqnarray}
where we introduced the field of Lagrangian accelerations of the fluid particles $\bm a(t, \bm x)$ and local time of variations of viscous scale eddies $\tau_{\nu}$. The above formula is readily confirmed by using Taylor expansion of velocity difference in the integrand in Eq.~(\ref{st}) at $t'=t$. The equation breaks down for fast vortices with $\tau_{\nu}\lesssim \tau$. However we demonstrated in the Introduction that, in the limit of large $Re_{\lambda}$, both the RDF and the collision kernel are determined by vortices with $\tau_{\nu}\sim \tau$. We conclude that for clouds Eq.~(\ref{maxeyf}), despite that it is true in most of the space, see the Introduction, is irrelevant even for $St\ll 1$ (the equation could still be used for calculating the collision kernel of droplets with extremely small $St$, however this would not have practical relevance).

We see from the above that we cannot estimate the drift velocity with the help of Eq.~(\ref{maxeyf}). The velocity is estimated by observing that the relevant vortices with $\tau_{\nu}\sim \tau$ cause the particles to move with respect to the flow at velocity of order $\sqrt{\nu/\tau}$. This is the typical velocity of the (local) viscous scale eddies with turnover time $\tau$ (this assumes the usual phenomenology of turbulence \cite{frisch} which is not obviously true at any $Re_{\lambda}$, cf. above and \cite{bud}). We conclude that the Reynolds number $Re_p=2|\bm v-\bm u(t, \bm x(t))|a/\nu$ of relevant flow perturbations due to the particle is of order $\sqrt{a^2/\nu\tau}$. Using the Stokes formula $\tau_s(a)=2a^2\rho/(9\nu\rho_a)$ and $\tau\sim \tau_s(a)$ we find that the condition $Re_p\ll 1$ gives $\sqrt{\rho_a/\rho}\ll 1$. Here $\rho_a$ is the density of air and $\rho$ is the droplet density. This condition is independent of the particle radius $a$ and it is obeyed by liquid droplets in air. We remark that the obtained condition $\sqrt{a^2/\nu\tau}\ll 1$ coincides with the condition that $a$ is much smaller than the local viscous scale $\sqrt{\nu \tau}$ of vortices with time-scale $\tau$.

We conclude that Eq.~(\ref{rd}) is valid at $St\ll 1$  for all relevant vortices. Moreover we can asymptotically continue this conclusion to $St\sim 1$ since at $St\sim 1$ the statistics is still determined by vortices with $\tau_{\nu}\sim \tau$. Thus at zero gravity and $St\lesssim 1$ the usage of the equation of motion seems valid, up to reservations described above.

{\bf Case of $St\gg 1$}---It remains that we consider the case of $St\gg 1$ and negligible gravity where the time-scale $\tau$ belongs in the inertial range of turbulence (there could also be the case of $\tau$ comparable or larger than the eddy turnover time of the integral scale turbulence. However this case does not seem to have applications in the rain formation problem and will not be considered). If $Re_{\lambda}$ is moderate so that intermittency is negligible, then we can use the dimensional Kolmogorov-type estimate  $\sqrt{\epsilon_0\tau}$ for the drift velocity \cite{fouxonhorvai}. This gives $Re_p=\sqrt{\epsilon_0\tau}(a/\nu) =\sqrt{St}(a/\eta)$ which might not be small for large $St$ particles. At higher  $Re_{\lambda}$, where intermittency is relevant, we can use Landau-type argument \cite{frisch}. Within it, the velocity is estimated by changing $\epsilon_0$ with the local energy dissipation rate $\epsilon$ i.e. is given by $\sqrt{\epsilon\tau}$. The resulting changes depend on which flow fluctuations are relevant in the collision kernel at $St\gg 1$. Theoretical study is yet to be done.

Our conclusion is that $Re_p\ll 1$ is self-consistent for particles with not too large $St$. However for certain $St>1$, whose value depends on the intermittency, we have $Re_p\gtrsim 1$ for relevant fluctuations of turbulence, the flow perturbation due to the particle is non-linear and Eqs.~(\ref{rd}) break down.

\subsection{Reynolds number of flow perturbation $Re_p$ at $Fr\ll 1$ and definition of $St$}

We consider the full Eqs.~(\ref{st}) in the case of non-negligible gravity. We assume $Fr\ll 1$ aiming at asymptotic description of the clouds, as explained in the Introduction. In this case the drift velocity is given in the leading order by the sedimentation velocity in the still air, $\bm v-\bm u[t, \bm x(t)]\approx \bm g\tau$; see \cite{fi2015}. This can be readily demonstrated at $St\ll 1$ where the particles' acceleration $\dot{\bm v}$ is of the order of the acceleration of the fluid particles. This, by definition of $Fr\ll 1$, is much smaller than $\bm g$, so that the LHS of the velocity equation in Eq.~(\ref{rd}) is negligible, giving $\bm v-\bm u[t, \bm x(t)]\approx \bm g\tau$. At $St\gtrsim 1$ we use that Eq.~(\ref{st}) implies $\left|\bm v-\bm u(t, \bm x(t))-\bm g\tau\right|\sim \left|\bm u(t-\tau, \bm x(t-\tau))-\bm u(t, \bm x(t))\right|$ where we estimate the integral as the velocity difference in the particle frame at time lag $\tau$. This difference is either due to turbulence's changes in time or in space. The former are given by the typical velocity $\sqrt{\epsilon_0\tau}$ of eddies with time-scale $\tau$, see above. In turn, the variations in space are
caused by the particle crossing during time $\tau$ of the distance $g\tau^2$ due to sedimentation. At moderate $Re_{\lambda}$ with negligible intermittency, this contributes to $\left|\bm u(t-\tau, \bm x(t-\tau))-\bm u(t, \bm x(t))\right|$
the Kolmogorov velocity difference $(\epsilon_0 g\tau^2)^{1/3}$ between spatial points separated by $g\tau^2$. We find the estimate
\begin{eqnarray}&&\!\!\!\!\!\!\!\!\!\!\!\!\!\!\!\!
\left|\bm v-\bm u(t, \bm x(t))-\bm g\tau\right|\sim max \left[\sqrt{\epsilon_0\tau}, (\epsilon_0 g\tau^2)^{1/3}\right]\sim \sqrt{\epsilon_0\tau} max \left[1, St^{1/6}Fr^{-1/3}\right].
\end{eqnarray}
We conclude from this formula by using $Fr\ll 1$ and $St\gtrsim 1$ that the drift velocity is $\bm g\tau$ by using
\begin{eqnarray}&&\!\!\!\!\!\!\!\!\!\!\!\!\!\!\!\!
\frac{\left|\bm v-\bm u(t, \bm x(t))-\bm g\tau\right|}{g\tau}\sim  max \left[Fr St^{-1/2}, Fr^{2/3} St^{-1/3} \right]\ll 1,
\end{eqnarray}
which proves $\bm v-\bm u(t, \bm x(t))\approx \bm g\tau$. At higher Reynolds numbers where intermittency is relevant, significant changes might be necessary. Their implementation demands the currently missing knowledge of which type of intermittent fluctuations, whose values by themselves cover a wide range of orders of magnitude, are relevant.

We find that the Reynolds number $Re_p$ of the flow perturbation by the particle, $Re_p=2|\bm v-\bm u(t, \bm x(t))|a/\nu$ equals $2a g\tau/\nu$. This number is small at sizes smaller than thirty microns where $\tau$ in Eq.~(\ref{rd}) is the Stokes time $\tau_s(a)=(2/9) (\rho/\rho_a) (a^2/\nu)$, given by the droplet's mass $4\pi \rho a^3/3$ divided by the coefficient of the Stokes force $6\pi \rho_a \nu a$.
In contrast, at radii from $30$ to $60$ microns we have $Re_p\gtrsim 1$. There an effective function $\tau(a)$, different from $\tau_s(a)$ must be used as $\tau$ in Eqs.~(\ref{rd}); see \cite{Wang20081,review}. It is with the help of this function that we define the Stokes number $St(a)=\tau(a)\sqrt{\epsilon_0/\nu}$, which determines the strength of the droplets' inertia, and not $\tau_s(a)$, cf. \cite{Wang20081,fp,fi2015}. The times $\tau(a)$ and $\tau_s(a)$ are of the same order in the size range of interest; the use of $\tau(a)$ in equation of motion, as compared with the use of $\tau_s(a)$, was found to decrease the collision kernel of larger droplets by up to $26$ per cent \cite{Wang20081}.

\subsection{Range of considered parameters}

In the rest of the paper, unless told otherwise, we assume that $St\gtrsim 1$, since there is already a well-developed theory for droplets whose size obeys $St(a)\ll 1$. The traditional small $St$ theory relies on Eq.~(\ref{maxeyf}), see \cite{maxey,nature}. However in some cases Eq.~(\ref{maxeyf}) does not hold at $St\ll 1$ because of the gravity; see the study of all possibilities in \cite{fi2015}. We remark that other sets of dimensionless parameters, different from our $St$, $Fr$ and $Re_{\lambda}$ are in use, cf. e.g. the studies \cite{Wang20081,review} who use the parameter $Sv=g\tau(a)/(\epsilon\nu)^{1/4}=St/Fr$, which is the ratio of the gravitational settling velocity of the particle and the velocity of turbulent eddies at the Kolmogorov scale. This parameter mixes characteristics of particles, gravity, and turbulence and is much larger than one in the range of $Fr\ll 1$ and $St\gtrsim 1$ that we study.

In clouds the velocity at the integral scale of turbulence $L$ is much larger than the sedimentation velocity. Thus we assume that the first term in $\bm v(t)\approx \bm u(t, \bm x(t))+\bm g\tau$ dominates the sum, that
is the typical value of the integral scale velocity $(\epsilon_0 L)^{1/3}$ is much larger than $g\tau$ for relevant $\tau$ (for the limit where
$\bm v(t)$ is dominated by sedimentation and $g\tau\gg (\epsilon_0 L)^{1/3}$ see \cite{maxey} and also \cite{bec2014,Gustavsson}). We must keep $\bm g\tau$ in $\bm v(t)$, despite that it is much smaller than $\bm u(t, \bm x(t))$ since it dominates at small scales where the clustering occurs, exceeding the Kolmogorov scale velocity. The approximation $\bm v(t)\approx \bm u(t, \bm x(t))+\bm g\tau$ is equivalent to stating that the droplets are transported by the flow $\bm u(t, \bm x)+\bm g\tau$. This flow is incompressible and would bring no clustering. Inclusion of further corrections to this approximation is necessary for the description of inhomogeneous spatial distribution of the droplets in the steady state.

\subsection{Direct numerical simulations}

In order to test below the theoretical predictions we performed DNS of the particle-laden isotropic turbulence. The Navier-Stokes equation was numerically solved on $128^3$ grids using a spectral method on the periodic cubic domain to describe the homogeneous isotropic turbulence at $Re_\lambda=70$. The equation of particle motion (Eq. \ref{rd}) was solved by taking into account the linear Stokes drag and gravity together. The initial positions and velocities of the particles are random and local fluid velocities, respectively.  Information of fluid quantities at the particle position was obtained using the fourth-order Hermite interpolation scheme \cite{choi,clee}. The details of the numerics can be found in Refs. \cite{Jung,Ychoi,Ah2012,Ah2013}. Typical snapshots of the particles' distribution in the steady state are shown in Fig. \ref{snapshots}. The angle dependence of the RDFs was computed in a statistically steady condition with five different populations of $N_p=9,971,200$ particles for each Froude number, $Fr=0.0125, 0.0167, 0.025, 0.033, 0.05$. The Stokes number is fixed at $St=1$. Here, $N_p$ was determined by $N_\eta=2.33$, which is the average number of particles within a sphere with radius $\eta$. For the estimation of the correlation codimension $\alpha$ using Eq.~(\ref{eq:alpha}), the Lyapunov exponents were computed by releasing many pairs of particles. The initial distance between the particles is set to $10^{-4}\eta$, and the change in distance between two particles, the area between three particles, and the volume constructed from four particles were measured on the basis of Gram-Schmidt renormalization for a period of $80\tau_\eta$ after the transient period due to the arbitrary initial condition. There are 10,000 sets of pairs released in one flow field, and data are collected over a total of 12 flow fields. The same method was used in our earlier works on droplets and bubbles \cite{fi2015,fsll}. In our simulations the only parameter which we changed was the gravity acceleration that changed $Fr$.

\section{Lyapunov exponents, columns and their disappearance as $Re_{\lambda}$ grows} \label{lyapunov}

In this section we study exponential separation of two particles in the viscous range. This is probably the shortest path to deriving the small-scale columnar structure formed by the particles. We provide a concise formula for the Lyapunov exponent $\lambda_1$ of the droplets via the energy spectrum of turbulence $E(k)$. These results were obtained in \cite{fi2015} where also further references are provided. Here we use a simpler approach, similar to \cite{fouxonhorvai}, cf. \cite{bec2014}. This allows us to address the breakdown of the theory at increasing Reynolds number. The breakdown implies that the columns that were observed in the direct numerical simulation might disappear if $Re_{\lambda}$ is increased at fixed $Fr$ and $St$. We also provide the Lyapunov exponent in a range of non-small $Fr$, that was not derived previously.

It is a direct consequence of the equation of motion (\ref{rd}), that for $r$ deep inside the viscous range, the separation vector of the equal size droplets $\bm r$ obeys
\begin{eqnarray}&&\!\!\!\!\!\!\!\!\!\!\!\!\!
\dot{\bm r}=\bm V,\ \ \dot{\bm V}=-\frac{\bm V-s\bm r}{\tau};\ \  s_{ik}(t)=\nabla_ku_i(t, \bm x(t)),\label{fd1}
\end{eqnarray}
where $s_{ik}$ is the matrix of gradients of turbulent flow in the frame of one of the particles whose trajectory is designated by $\bm x(t)$. We will use this equation for the calculation of $\lambda_1$ which will be seen to be determined by vortices whose size $L_c$ is in the inertial range and much larger than the Kolmogorov scale $\eta$. Therefore the result can be applied to the evolution of $r(t)$ as long as $r(t)\ll L_c$ and not under the more stringent $r\ll \eta$.

We first consider Eq.~(\ref{fd1}) at moderate Reynolds numbers with weak intermittency where rigorous study of $Fr\to 0$ limit is possible. The impact of intermittency on our considerations will be considered later (it must be observed however that $L_c\gg \eta$ is a manifestation of intermittency. Our detailed assumption will be seen below).

\subsection{Derivation at moderate $Re_{\lambda}$}

We study the parameters' range of $Fr\ll 1$ and $St\gtrsim 1$ assuming that $Re_{\lambda}$ is moderate so that several quantities in the study below obey the Kolmogorov-type estimates \cite{frisch}. The particle moves with respect to the local flow at the velocity $\bm g \tau$, as explained above. This implies \cite{nature} that the correlation time of the matrix $s_{ik}(t)$, providing velocity gradients in the frame of the droplet, is the minimum of the Kolmogorov time $\tau_K$ and the time $\tau_g\equiv \eta/(g\tau)$. Here $\tau_g$ is the time during which the droplet moving through the flow at velocity $\bm g\tau$ crosses the correlation length of the turbulent velocity gradients, the Kolmogorov scale $\eta$. Observing that $\tau_g/\tau_K=Fr/St$ is small in the considered range of parameters ($Fr\ll 1$ and $St\gtrsim 1$) we conclude that in our case $s_{ik}(t)$ varies in time at the time-scale $\tau_g$. During this time-scale, which is much less than the turnover time of the viscous scale eddies  $\tau_K$, the field of the velocity gradients does not change in time appreciably if considered in the frame of the {\it fluid} particles. Therefore the temporal correlation function $\left\langle \nabla_ku_i(t, \bm x(t))\nabla_ru_p(t', \bm x(t'))\right\rangle$ equals the spatial correlation function $\left\langle \nabla_ku_i(t, \bm x(t))\nabla_ru_p(t, \bm x(t)+\bm g\tau(t'-t))\right\rangle$. Moreover the product of the typical value of $s_{ik}(t)$, which is $\tau_K^{-1}$, and of the
correlation time $\tau_g$ of $s(t)$ is small. Considering this product as a small parameter, the leading order approximation is the limit of zero correlation time where $s(t)$ in Eq.~(\ref{fd1}) can be replaced by white noise \cite{fouxonhorvai}.

We demonstrate in detail how the white noise description of the effect of $s$ on the evolution of $\bm r$ and $\bm V$ in Eq.~(\ref{fd1}) arises. We observe that Eq.~(\ref{fd1}) gives
\begin{eqnarray}&&\!\!\!\!\!\!\!\!\!\!\!\!\!
\bm V(t)=\bm V(0)\exp\left(-\frac{t-t'}{\tau}\right)+\frac{1}{\tau}\int_0^t\exp\left(-\frac{t-t'}{\tau}\right)s(t')\bm r(t')dt'.\label{sdl}
\end{eqnarray}
We concentrate on the phenomena where the particles stay inside the viscous scale for times much larger than $\tau$. There the above equation holds at $t\gg \tau$ and the first term can be neglected. There are physical phenomena that occur on a time-scale $\tau$ or smaller so that for them $t\gg \tau$ is not the range of interest. An example is the sling effect \cite{nature} where the particles detach from the flow at the time of the beginning of the sling effect, taken to be $t=0$, and then move ballistically toward each other so that the first term in Eq.~(\ref{sdl}) dominates $\bm V(t)$. In this case, despite that $r\ll \eta$ and the Taylor expansion of the velocity difference made in Eq.~(\ref{fd1}) holds, what changes the distance between the particles is not the difference of the turbulent flow velocities at the positions of the particles, which is given by $s\bm r$, but their inertia. The sling effect lasts for times of order $\tau$ and is not describable by the limit of $t\gg \tau$.

Phenomena that happen at times larger than $\tau$ include separation of two infinitesimally close trajectories in the six-dimensional phase space. This defines the first Lyapunov exponent $\lambda_1$ that describes the growth of the separation vector $(\bm r, \bm v)$ between the trajectories via
\begin{eqnarray}&&\!\!\!\!\!\!\!\!\!\!\!\!\!
\lambda_1=\lim_{t\to\infty}\frac{1}{2t}\ln \left(\frac{r^2(t)+\tau^2V^2(t)}{r^2(0)+\tau^2V^2(0)}\right),
\end{eqnarray}
where the dimensional factor $\tau$ is irrelevant and used only for having a dimensionally uniform expression. Here, besides that $r(0)\ll\eta$, also the initial velocity difference is small so that $r(t)$ stays much smaller than $\eta$ during times much larger than $\tau$. The limit above exists and is given by the same constant for almost all trajectories (i.e. with possible exception of initial positions with zero volume in the phase space) \cite{oseledec}. Therefore disregarding the initial period of evolution of duration of order $\tau$ (or simply setting $\bm V(0)=0$ using that the limit is independent of the vector $(\bm r(0), \bm V(0))$ as long as this vector is non-zero) we can use instead of Eq.~(\ref{sdl}) the simplified equation
\begin{eqnarray}&&\!\!\!\!\!\!\!\!\!\!\!\!\!
\bm V(t)=\frac{1}{\tau}\int_{-\infty}^t\exp\left(-\frac{t-t'}{\tau}\right)s(t')\bm r(t') dt'.\label{integr}
\end{eqnarray}
We study the regime determined implicitly by the condition $\lambda_1 \tau_g \ll 1$, whose explicit form will be provided later by obtaining $\lambda_1$. We introduce a separation time $\Delta t$ that obeys $\tau_g\ll \Delta t\ll min[\lambda_1^{-1}, \tau]$. This $\Delta t$ exists because we assumed $\lambda_1 \tau_g \ll 1$ and because $\tau_g/\tau=Fr/St^2$ is small by $Fr\ll 1$ and $St\gtrsim 1$. Then we can write
\begin{eqnarray}&&\!\!\!\!\!\!\!\!\!\!\!\!\!
\bm V(t)\approx \frac{1}{\tau}\int_{-\infty}^t\exp\left(-\frac{t-t'}{\tau}\right)s_{\Delta t}(t')\bm r(t') dt',\ \ s_{\Delta t}(t)\equiv \int_t^{t+\Delta t} s(t')\frac{dt'}{\Delta t},\label{integrc}
\end{eqnarray}
where $s_{\Delta t}$ is $s(t)$ coarse-grained over time-scale $\Delta t$. We used that $\Delta t$ is smaller than both the characteristic time of variations of $\bm r(t)$, which is $\lambda_1^{-1}$ and the characteristic time of variations of the exponent $\tau$.

We observe that $s_{\Delta t}$ is proportional to the integral of the random process $s(t)$ over a time-interval which is much larger than the correlation time of this process, $\tau_g$. This implies by the central limit theorem (CLT) that the statistics of $s_{\Delta t}$ is Gaussian. Indeed we can write the integral in the definition of $s_{\Delta t}$ as sum of contributions of many intervals whose duration is the correlaton time $\tau_g$, i.e. $s_{\Delta t}(t)=\left(\Delta t\right)^{-1}\sum_{i=1}^N \int_{t+(i-1)\tau_g}^{t+i\tau_g}s(t')dt'$ where $N=\Delta t/\tau_g$ is much larger than one (we can use in these considerations $\Delta t$ that is an integer multiple of $\tau_g$). Then, since the contributions of different intervals are independent, we find that $s_{\Delta t}$ is sum of a large number of independent random variables and the CLT applies. A more rigorous proof can be constructed by applying the cumulant expansion theorem \cite{ma} to the characteristic function of $s_{\Delta t}$. We conclude that the distribution of $s_{\Delta t}(t)$ is fully determined by the mean and the dispersion that fix a Gaussian distribution uniquely.

It can be checked that the mean of $s$ is negligible, see \cite{fi2015} and Appendix \ref{spso} (we remark that the usual argument which uses that $\langle s_{ik}\rangle$ is proportional to $\delta_{ik}$ due to isotropy of the small-scale turbulence fails in this case. Gravity makes the statistics anisotropic). Therefore the statistics of $s_{\Delta t}(t)$ coincides with the statistics of $\int_t^{t+\Delta t} s^K(t')dt'/(\Delta t)$ where $s^K(t)$ is a Gaussian matrix process with zero mean and zero correlation time which dispersion is picked so that dispersions of $\int_t^{t+\Delta t} s^K(t')dt'/(\Delta t)$ and $s_{\Delta t}(t)$ agree. The white noise matrix process $s^K(t)$ arises in the Kraichnan model of the turbulent transport as indicated by the superscipt \cite{reviewt}. It is readily seen that our condition on the dispersion of $s^K(t)$ demands that $\int_{-\infty}^{\infty} \left\langle s_{ik}(t)s_{pr}(t')\right\rangle dt'$ equals $\int_{-\infty}^{\infty} \left\langle s^K_{ik}(t)s^K_{pr}(t')\right\rangle dt'$.
Finally using our previous consideration of correlations of $s_{ik}(t)$ we find
\begin{eqnarray}&&\!\!\!\!\!\!\!\!\!\!\!\!\!
\left\langle s^K_{ik}(t)s^K_{pr}(t')\right\rangle=\delta(t-t')\kappa_{ikpr},\ \ \kappa_{ikpr}\equiv \int_{-\infty}^{\infty} \left\langle \nabla_ku_i(0)\nabla_ru_p(\bm g\tau  t)\right\rangle dt, \label{akpa}
\end{eqnarray}
where the correlation function in the integrand is the equal-time correlation function of turbulent velocity gradients. Calculation of $\kappa_{ikpr}$ made in \cite{fi2015} reveals that in the leading order $s^K_{ik}$ is a $2\times 2$ random matrix since all the entries of the matrix that contain index $z$ vanish. This $2\times 2$ matrix obeys the usual statistics of two-dimensional Kraichnan model determined by restriction of $\kappa_{ikpr}$ to indices different from $z$
\begin{eqnarray}&&\!\!\!\!\!\!\!\!\!\!\!\!\!
\kappa_{\alpha \beta \gamma \delta}=D\left(3\delta_{\alpha \gamma}\delta_{\beta \delta}-\delta_{\alpha \beta}\delta_{\gamma \delta}-\delta_{\alpha \delta}\delta_{\gamma\beta}\right), \ \ D\tau=\frac{\pi\int_0^{\infty} E(k)kdk}{8g}\sim Fr,
\end{eqnarray}
where $E(k)$ is the energy spectrum of turbulence, cf. \cite{fouxonhorvai}. Here and below, the Greek indices take values of $1$ or $2$ and summation over repeated indices is implied (no confusion can be caused between $\alpha$ as index and $\alpha$ as exponent). Two-dimensional Kraichan model $\dot{\bm v}_{\perp}=-\left(\bm v_{\perp}-s^K\bm r_{\perp}\right)/\tau$, where the subscript stands for horizontal components of the vectors, is well studied \cite{piterbarg,bh}. It is seen from the formula above that $D\tau\sim Fr\ll 1$. This corresponds to the limit of small inertia where the approximation $\dot {\bm r}_{\perp}\approx s^K\bm r_{\perp}$ holds. Thus the problem reduces to solved evolution of distance between the two fluid particles in the Kraichnan model \cite{reviewt}. This gives immediately the spectrum of the Lyapunov exponents $\lambda_i$ defined by asymptotic growth rates of hypersurfaces composed of particles. Thus $\lambda_1$, $\lambda_1+\lambda_2$, and $\sum_{i=1}^3\lambda_i$ provide the asymptotic logarithmic growth rates of the infinitesimal line, surface, and volume elements at large times, respectively \cite{reviewt}. We have $\lambda_1=-\lambda_3$, that describes behavior of $\bm r_{\perp}$ and $\lambda_2=0$ that describes conservation of the vertical component of $\bm r$. The value of $\lambda_1$ is \cite{reviewt,fi2015}
\begin{eqnarray}&&\!\!\!\!\!\!\!\!\!\!\!\!\!
\lambda_1=D=\frac{\pi\int_0^{\infty} E(k)kdk}{8g \tau}. \label{ps}
\end{eqnarray}
Self-consistency of the derivation demands that the obtained $\lambda_1$ obeys $\lambda_1 \tau_g\ll 1$. This gives the condition $D\tau\ll 1$ which is equivalent to  $Fr^2/St^2\ll 1$. The inequality holds in the studied range of parameters proving the self-consistency. The above prediction was confirmed in \cite{fi2015} numerically by direct numerical simulations of the motion of inertial particles in the Navier-Stokes turbulence with $Re_{\lambda}=70$ . The result was found to be quantitatively accurate for $Fr\leq 0.05$, where linear dependence on $Fr$ holds, and reasonably good for $Fr\leq 0.1$, see Fig. \ref{breakdown}. For higher $Fr$ the equation's validity depends on $St$. The higher the $St$ is, the more accurate the above formula is.

The general validity condition of the effective white-noise description above is $Fr/St\ll 1$. This condition guarantees that correlation time of velocity gradients in the particle's frame is $\tau_g$. It also ensures that both $\lambda_1\tau_g\ll 1$ and $\tau_g/\tau=Fr/St^2\ll 1$ hold, allowing to pass from Eq.~(\ref{integr}) to Eq.~(\ref{integrc}).

We observe that $Fr/St\ll 1$ can hold also for non-small $D\tau\propto Fr$ if $St\gg 1$. Since the Froude number in clouds can be of order one then the case of $St\gg Fr\sim 1$ might have practical applications. Therefore we provide the results in this case for possible future use. We assume non-small Froude number, $Fr\gtrsim 1$, and $St\gg Fr$, so that the white noise description holds. If $D\tau\geq 1$ then $\lambda_1$ of the Kraichan model can be written as $\lambda_1=D^{1/3}\tau^{-2/3}{\tilde \lambda}_1\left[(D\tau)^{-1/3}\right]$. Here ${\tilde \lambda}_1$
is a function that takes values of order one at $D\tau\geq 1$ slowly changing with its argument from $0.5$ at $D\tau=1$ to $2$ at zero argument corresponding to $D\tau=\infty$, see \cite{fouxonhorvai} and references therein. We find then that the self-consistency condition $\lambda_1\tau_g\ll 1$ holds by $\lambda_1 \tau_g\sim  Fr^{4/3}/St^2$. We conclude that
\begin{eqnarray}&&\!\!\!\!\!\!\!\!\!\!\!\!\!
\lambda_1=\frac{1}{\tau}\left(\frac{\pi\int_0^{\infty} E(k)kdk}{8g}\right)^{1/3}{\tilde \lambda}_1\left[\left(\frac{\pi\int_0^{\infty} E(k)kdk}{8g}\right)^{-1/3}\right],\ \ 1\lesssim Fr\ll St. \label{refined}
\end{eqnarray}
Testing this prediction in the DNS is left for future work.

The above results imply that the droplets will form columns in space. Indeed in the leading order the droplets separate horizontally keeping their vertical separation conserved. Once the random trajectories of the particles bring one of the particles on the top of the other, this pair configuration is preserved in time indefinitely (more precisely for long time due to instability). The particles form meta-stable bound states and after some time the space will be divided into columns of particles that would coalesce and form a single column. In reality the next order corrections cause gradual dissolution of the vertical pair configuration so that columns are only the more probable configuration of the particles. This phenomenon is quantitatively described in later sections by studying the angular dependence of the radial distribution function.

\subsection{Implications of dissipation range statistics and breakdown at large $Re_{\lambda}$}  \label{fbreakdown}

Here we assume that $Fr$ is fixed at a value much smaller than $1$ and study the impact of increasing intermittency at growing $Re_{\lambda}$ on the considerations above. The increase of $Re_{\lambda}$ can have two-fold effect on $\lambda_1$. It can influence the estimate $\lambda_1\tau\sim Fr$, implied by Eqs.~(\ref{ps})-(\ref{refined}), and it can also invalidate the assumptions made in the derivation. We start from considering the former effect.

{\bf Comparison with Lyapunov exponent of tracers}---It is useful to make comparison with first Lyapunov exponent of tracer particles in the Navier-Stokes turbulence $\lambda_1^t$. The dimensionless product $\lambda_1^t \tau_K$, decays with $Re_{\lambda}$. This was predicted by using the multifractal model in \cite{cr} and confirmed numerically in \cite{jeremy}, see also \cite{mj}. The reason for the decay is that increasing intermittency of turbulence increases size of regions with quiescent quasilaminar turbulence where the separating pair of particles consequently stays longer \cite{frisch}. This depletes chaos as measured by the dimensionless Lyapunov exponent \cite{cr}. We consider if there is a similar dependence of $\lambda_1$ in Eq.~(\ref{ps}) on $Re_{\lambda}$.

{\bf Influence of non-trivial structure of the dissipation range}---The dimensionless Lyapunov exponent $\lambda_1\tau$ obeys
\begin{eqnarray}&&\!\!\!\!\!\!\!\!\!\!\!\!\!
\lambda_1\tau=\frac{\pi\int_0^{\infty} E(k)kdk}{8g}=c_0 Fr; \ \ \ c_0\equiv \frac{\pi\nu^{1/4}\int_0^{\infty}\!\!E(k)kdk}{8\epsilon_0^{3/4}}, \label{c0}
\end{eqnarray}
where we defined dimensionless constant $c_0$. The integral in $c_0$ is determined by the form of the turbulence energy spectrum $E(k)$ in the dissipative range. Therefore it can be assumed to be independent of the large-scale forcing so that $c_0$ is a dimensionless function of $Re_{\lambda}$. It can be seen from the numerical data of \cite{fi2015} and Fig. \ref{breakdown} that $c_0\approx 1.5$ at $Re_{\lambda}=70$. The existing data on the spectrum can be used to study the dependence of $c_0$ on $Re_{\lambda}$. It was found in \cite{ishihara,gotoh} that the dissipation range spectrum observed in their DNS was well described with
\begin{eqnarray}&&\!\!\!\!\!\!\!\!\!\!\!\!\!
E(k)=C (k\eta)^{{\tilde \alpha}}\exp\left[-\beta k \eta \right],\label{spf}
\end{eqnarray} 
where $C$, ${\tilde \alpha}$, and $\beta$ are functions of the Reynolds number $Re_{\lambda}$.  We find assuming that the contribution of smaller wavenumbers can neglected and using the equation above that
\begin{eqnarray}&&\!\!\!\!\!\!\!\!\!\!\!\!\!
c_0= \frac{\pi\int_0^{\infty}\!\!E(k)kdk}{8\eta\int_0^{\infty}\!\!E(k)k^2dk}=\frac{\pi\int_0^{\infty}\!\!k^{1+{\tilde \alpha}}\exp\left[-\beta k \eta \right]dk}{8\eta\int_0^{\infty}\!\!k^{2+{\tilde \alpha}}\exp\left[-\beta k \eta \right]dk}=\frac{\pi\beta}{8(2+{\tilde \alpha})},
\end{eqnarray}
where we used $\int_0^{\infty}\!\!E(k)k^2dk=\epsilon_0/\nu$ and $\eta=\nu^{3/4}/\epsilon_0^{1/4}$. 
The calculation is self-consistent provided that ${\tilde \alpha}>-2$ since otherwise $\int_0^{\infty}\!\!k^{1+{\tilde \alpha}}\exp\left[-\beta k \eta \right]dk$ diverges at small $k$ where Eq.~(\ref{spf}) does not apply. Moreover $2+{\tilde \alpha}$ must be not too small since otherwise the contribution of small wavenumbers would still be appreciable. The measurement of $\alpha$ provided in \cite{ishihara} and described by the fit $\alpha=7.3 Re_{\lambda}^{-0.47}-2.9$ gives that the condition ${\tilde \alpha}>-2$ is obeyed by $Re_{\lambda}$ smaller than about $85$. For $Re_{\lambda}=70$ we have $2+\alpha=0.09$ which is too small for the calculation to be self-consistent so the fits for $\alpha$ and $\beta$ provided in \cite{ishihara,gotoh} cannot be used for evaluating $c_0$.
However these fits give unequivocal indication that $c_0$ has appreciable growth with $Re_{\lambda}$. For further details on the spectrum see \cite{gotoh} for the DNS aspects and \cite{dis} for the theory.

{\bf Correlation length of gradients of particles' flow is much larger than the Kolmogorov scale}--- Further insight into $\lambda_1$ is reached by rewriting the above prediction for the Lyapunov exponent as %
\begin{eqnarray}&&\!\!\!\!\!\!\!\!\!\!\!\!\!
\lambda_1=D=\frac{ \kappa_{\alpha\beta\alpha\beta}}{8}=\frac{1}{8}\int_{-\infty}^{\infty} \left\langle \nabla_{\beta}u_{\alpha}(0)\nabla_{\beta}u_{\alpha}(\bm g\tau  t)\right\rangle dt
=\frac{1}{4g\tau}\int_0^{\infty} \left\langle \nabla_{\beta}u_{\alpha}(0)\nabla_{\beta}u_{\alpha}(z{\hat z})\right\rangle dz,\label{ls}
\end{eqnarray}
where ${\hat z}$ is directed upwards. The RHS is similar to the dispersion of the finite-time Lyapunov exponent studied in \cite{cr}, which behaves as time integral of the different time Lagrangian correlation function of velocity gradients $\int_{-\infty}^{\infty} \left\langle \nabla_ku_i(0)\nabla_ku_i(t)\right\rangle dt$. The last integral can be written as $2t_c/\tau_K^2$ where $t_c\equiv \int_{0}^{\infty} \left\langle \nabla_ku_i(0)\nabla_ku_i(t)\right\rangle dt/\left\langle (\nabla \bm u)^2\right\rangle$ is the effective correlation time $t_c$. It was found that $t_c/\tau_K$ obeys rather strong increase with the Reynolds number given by $Re_{\lambda}^{\kappa}$ with $\kappa\sim 0.1$ which qualitatively agrees with the numerical observations in \cite{jeremy}. The reason is that at increasing $Re_{\lambda}$, due to intermittency, the regions of moderate velocity gradients of order $\tau_K^{-1}$ become larger both in space and in time \cite{frisch}. This results in power-law increase of $t_c$ with $Re_{\lambda}$. This makes it highly probable that also the spatial correlation length $L_c$ defined by
\begin{eqnarray}&&\!\!\!\!\!\!\!\!\!\!\!\!\!
L_c\equiv \frac{\int_0^{\infty} \left\langle \nabla_{\beta}u_{\alpha}(0)\nabla_{\beta}u_{\alpha}(z{\hat z})\right\rangle dz}{\left\langle \nabla_{\beta}u_{\alpha}\nabla_{\beta}u_{\alpha}\right\rangle}=\frac{5\nu\int_0^{\infty} \left\langle \nabla_{\beta}u_{\alpha}(0)\nabla_{\beta}u_{\alpha}(z{\hat z})\right\rangle dz}{2\epsilon_0}, \label{corl}
\end{eqnarray}
increases with $Re_{\lambda}$ so that $L_c/\eta\propto Re_{\lambda}^{\Delta_{\lambda}}$ with $\Delta_{\lambda}\sim 0.1$ (the power-law dependence is probably valid quantitatively at large $Re_{\lambda}$ and at moderate $Re_{\lambda}$ is valid qualitatively only). Here we used that small-scale isotropy, incompressibility and spatial homogeneity imply that single-point statistics of turbulent velocity gradients obeys
\begin{eqnarray}&&\!\!\!\!\!\!\!\!\!\!\!\!\!
\left\langle \nabla_ku_i \nabla_ru_p \right\rangle=\frac{\epsilon_0}{30\nu}\left(4\delta_{ip}\delta_{kr}-\delta_{ik}\delta_{pr}-\delta_{ir}\delta_{pk}\right), \ \
\left\langle \nabla_{\beta}u_{\alpha}\nabla_{\beta}u_{\alpha}\right\rangle=\frac{2\epsilon_0}{5\nu}.
\end{eqnarray}
(This formula is found by differentiation of the velocity pair correlation function in \cite{ll6}.) 
We find from Eq.~(\ref{ls}) that
\begin{eqnarray}&&\!\!\!\!\!\!\!\!\!\!\!\!\!
\lambda_1\tau=\frac{\epsilon_0L_c }{10g\tau \nu}=Fr \frac{L_c}{10\eta},\label{lau}
\end{eqnarray}
which would give $\lambda_1\tau \sim Fr Re_{\lambda}^{\Delta_{\lambda}}$ on assuming the power-law dependence of $L_c/\eta$ on $Re_{\lambda}$. Using $\lambda_1\tau=1.5 Fr$, which was observed in the simulations of  \cite{fi2015} at $Fr\leq 0.033$, see Fig. \ref{breakdown}, we find that $L_c=15\eta$ at $Re_{\lambda}=70$. The large numerical factor demonstrates failure of dimensional estimates. The reason for this factor is the non-trivial structure of the energy spectrum in the dissipation range. Indeed, $L_c$ is proportional to $c_0$ considered previously.


{\bf Theory breakdown at large $Re_{\lambda}$}---The observation that the scale of relevant flow configurations $L_c$ is much larger than $\eta$ and grows with $Re_{\lambda}$ as $L_c/\eta\propto Re_{\lambda}^{\Delta_{\lambda}}$ implies breakdown of the assumptions of the theory at large $Re_{\lambda}$. The derivation of $\lambda_1$ assumed that the correlation time of velocity gradients in the frame of the particle is the time during which the droplet crosses the spatial correlation length $\eta$. However the correlation length of relevant gradients is $L_c$ and not $\eta$, which is much larger than $\eta$ already at moderate $Re_{\lambda}=70$. Using in the previous considerations $L_c/(g\tau)$ as the correlation time of $s(t)$, we find that the condition that $s_{\Delta t}$ is a sum of large number of independent random variables demands that $L_c/(g\tau)\ll \tau$ and $L_c/(g\tau)\ll \lambda_1^{-1}$. We observe that the former condition is stronger because $\lambda_1\tau$ is never large, $\lambda_1\tau\lesssim 1$, at Stokes numbers of order one considered here. Indeed, inertia causes the Lyapunov exponent of the particles, that tend to move ballistically, to be smaller than the Lyapunov exponent of tracer particles $\lambda_1^t$. Therefore $\lambda_1^t\tau\sim St$ implies $\lambda_1\tau\lesssim 1$. We conclude that the condition of the theory applicability boils down to $L_c/(g\tau)\ll \tau$. We find assuming the power-law dependence of $L_c/\eta$ on $Re$ that our derivation of $\lambda_1$ holds provided that
\begin{eqnarray}&&\!\!\!\!\!\!\!\!\!\!\!\!\!
L_c\ll g\tau^2,\ \ \ \ \
15 \left(\frac{Re_{\lambda}}{70}\right)^{\Delta_{\lambda}}\ll \frac{St^2}{Fr}. \label{lv}
\end{eqnarray}
We observe that there is large numerical factor in Eq.~(\ref{lv}). This limits the theory's applicability to rather small $Fr$: at $Re_{\lambda}=70$ deviations from the prediction for $\lambda_1$ are observed at $Fr$ as small as $0.05$, see \cite{fi2015}. The increase of $Re_{\lambda}$ will further decrease $Fr$ at which the theory applies.

We saw that increase of $Re_{\lambda}$ makes our calculation of $\lambda_1$ inconsistent starting from $Re_{\lambda}$ for which $L_c\sim g\tau^2$. However Eqs.~(\ref{ps})-(\ref{refined}) could also become invalid because of breakdown of the assumed Gaussianity of $s_{\Delta t}(t)$ in Eq.~(\ref{integrc}), despite $\Delta t$ being much larger than the correlation time of $s(t)$. This could happen because cumulants of order higher than two, which the Gaussian approximation neglects \cite{ma}, involve higher order moments of the velocity gradients. These moments due to intermittency would contain higher powers of $Re_{\lambda}$. This would invalidate their discarding at large enough $Re_{\lambda}$. The resulting criterion is similar to Eq.~(\ref{lv}) and brings the same conclusion that the derivation breaks down at increasing  $Re_{\lambda}$.

We describe how the predictions of the effective white-noise description are used in practice assuming that the validity conditions hold and $St\sim 1$. The first step is to derive $D$ from the spectrum of turbulence by using Eq.~(\ref{ps}) and check if $D\tau\ll 1$. If yes, then $D$ is the predicted value of the Lyapunov exponent. If not, then the white-noise description fails and other treatment is necessary. Thus if we consider $\lambda_1\tau$ as a function of $Re_{\lambda}$ at fixed $Fr\ll 1$ and $St\sim 1$, then $\lambda_1\tau$ increases with $Re_{\lambda}$ until $\lambda_1\tau$ becomes of order one. It is seen that this happens when the correlation
time $L_c/(g\tau)\sim \lambda_1\tau_K^2$ of $s(t)$ is of order $\tau_K$. It seems that at further increase of $Re_{\lambda}$ the correlation time, which cannot be larger than $\tau_K$, gets fixed at the Lagrangian correlation time $\tau_K$ and the statistics of $s(t)$ becomes similar to the isotropic statistics of turbulent velocity gradients. This would lift anisotropy of particles' distribution in space and make columnar structures disappear.

We conclude that the above considerations indicate that increase of $Re_{\lambda}$ would smear the columnar structure and make it to disappear altogether at the high $Re_{\lambda}$ holding in clouds. In fact we demonstrate below that $L_c$ and $g\tau^2$ provide horizontal and vertical dimensions of the particles' columns respectively. Here $L_c$ must be considered as the correlation length of the velocity gradients of the particles' flow which need not and is not the same as its counterpart for turbulence. Starting from $Re_{\lambda}$ for which $L_c\sim g\tau^2$, which is where the calculation of $\lambda_1$ becomes invalid, vertical and horizontal dimensions of the columns become similar, the particles' structures are isotropic and columns are no longer preferential. Numerical studies of dependence of $L_c$ on $Re_{\lambda}$ would provide further insight into the discussion of this section and are left for future work.

\section{Predictions for $Fr\to 0$ limit and confirmation} \label{resolution}

In this section we revisit predictions of \cite{fi2015} preparing the ground for the study of how intermittency affects validity of \cite{fi2015} performed in the next Section. The main observation of \cite{fi2015} is that smallness of Lagrangian acceleration of the fluid particles in comparison with $g$ results in the smooth spatial motion of the droplets. The particles' velocities after transients are uniquely determined by their spatial positions on which they depend in a differentiable manner. This conclusion is reached by estimating the accelerations with the typical value $\epsilon_0^{3/4}/\nu^{1/4}$. The smoothness holds irrespective of the Stokes number, which is significant since droplets in the clouds often have $St\gtrsim 1$ where without gravity particles' motion would not be spatially smooth.

The limitations of the above observation are in the usage of typical accelerations which might not be relevant in view of intermittency of small-scale turbulence. Thus at $Re_{\lambda}=690$, which is much smaller than $Re_{\lambda}$ of the clouds, the accelerations' flatness of $55$ was observed in \cite{bod}. It is then not obvious at all if the estimation of the role of gravity by using typical turbulent accelerations is adequate in the clouds.

The above issue can be considered similarly to the case of negligible gravity and $St\ll 1$. We can decompose the flow domain into the major part, where turbulent accelerations are much smaller than gravity, and rare regions of vigorous turbulence where turbulent accelerations are larger or comparable with $g$. For qualitative considerations we can estimate local turbulent acceleration of the fluid particles as $\epsilon^{3/4}/\nu^{1/4}$, cf. \cite{sreme}. The local Froude number $\epsilon^{3/4}/(g\nu^{1/4})$ is small in most of the space. In this calm major part of the space the droplets form a smooth spatial flow as explained in \cite{fi2015}. In contrast, in regions with local Froude number of order one, sling effect with particles' jets holds. The rate of collisions is then given by the sum of contributions of the large calm region of turbulence, where smooth flow holds, and widely spaced regions of jets. This decomposition is identical to that at weak gravity and small inertia, see the Introduction and \cite{nature}.

For small enough $Fr$, however large $Re_{\lambda}$ is, the collision kernel is due to the major volume fraction with smooth flow and the theory of \cite{fi2015} applies. In contrast, if we fix $Fr$ and increase $Re_{\lambda}$ then the characteristic acceleration of relevant vortices of the smooth portion of the flow increases until it becomes of order $g$ (the sling effect would become non-negligible already at smaller $Re_{\lambda}$, cf. \cite{nature}). At higher $Re_{\lambda}$ the collisions due to both smooth flow and sling effect, occur predominantly in the rare regions of "resonant" vortices with acceleration of order $g$ and the theory of \cite{fi2015} does not apply. We provide details on this non-commutativity of $Fr\to 0$ and $Re_{\lambda}\to\infty$ limits below.

\subsection{Theory of \cite{fi2015}: predictions, confirmation, contradiction}

\begin{figure*}
\includegraphics[width=17cm,trim={1mm 2mm 1mm 2mm},clip]{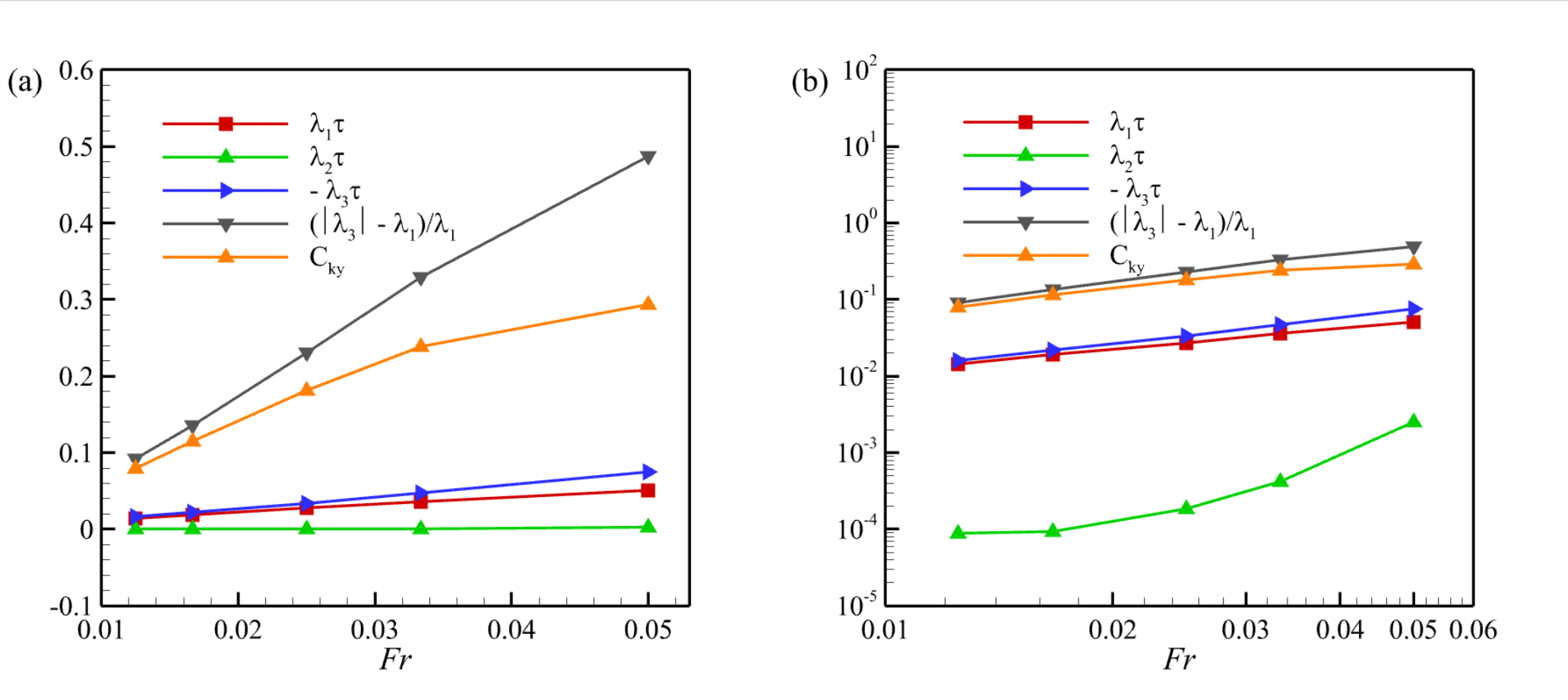}
\caption{Lyapunov exponents: theory vs. measurement, (a) linear and (b) log-log plots. The small $Fr$ theory of \cite{fi2015} is valid at smallest $Fr$, however breaks down at increasing $Fr$. The linear growth $|\lambda_3|\tau=1.5 Fr$ predicted in \cite{fi2015} is observed. However $\lambda_1$, predicted to be equal to $-|\lambda_3|$, obeys a different  behavior, describable by the fit $\lambda_1\tau=1.5 Fr^{0.9}$ (we used in the text that $\lambda_1\tau=1.5 Fr$ is valid at $Fr\leq 0.033$). The ratio $(|\lambda_3|-\lambda_1)/\lambda_1$, that vanishes in the leading order of the theory, grows roughly linearly with $Fr$ remaining reasonably small at $Fr=0.033$ however its value $\simeq 0.5$ at $Fr=0.05$ indicates the theory breakdown. Similarly the compressibility ratio $C_{KY}\equiv |\sum_{i=1}^3\lambda_i/\lambda_3|$ deviates from the predicted linear dependence on $Fr$ at $Fr=0.05$, remaining, as predicted, quite small. The exponent $\lambda_2$ corresponds to vertical separation of the particles and it remains small at all $Fr$ with $max[\lambda_2/\lambda_1]=0.05$ and $max[\lambda_2/|\sum_{i=1}^3\lambda_i|]=0.11$, both attained at $Fr=0.05$. Thus $\lambda_2$ can be neglected in $\sum_{i=1}^3\lambda_i$. Compressibility of the flow grows with $Fr$ because the horizontal flow, described by $\lambda_1$ and $\lambda_3$, becomes more compressible. }
\label{breakdown}
\end{figure*}

\begin{figure*}
\includegraphics[width=8.5cm,trim={2mm 2mm 2mm 2mm},clip]{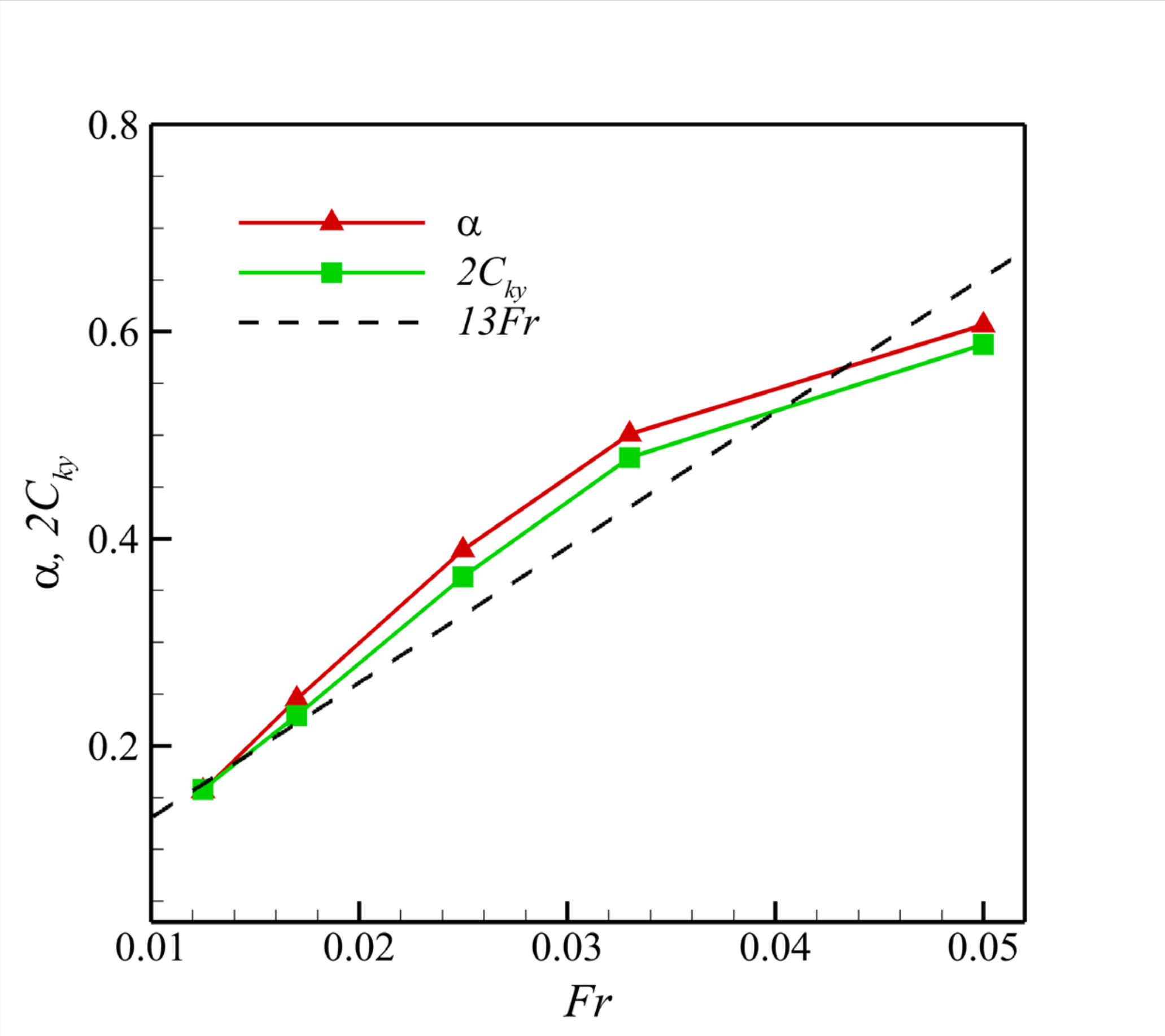}
\caption{Shown are the DNS values of the correlation codimension $\alpha$, obtained from numerical observations of the RDF, twice the Kaplan-Yorke (Lyapunov) codimension $C_{KY}$, fixed from the numerical observations of the Lyapunov exponents, and the low-$Fr$ theory prediction $3\pi\int_0^{\infty} E(k)kdk/(2g)$ that gives $13 Fr$.  All figures in the paper use $St=1$ and $Re_{\lambda}=70$. }
\label{ouralpha}
\end{figure*}

We describe the results for the spectrum of the Lyapunov exponents $\lambda_i$ of spatial motion of the particles in the light of more detailed numerical simulations performed for this work, see Fig. \ref{breakdown}.
It is immediate consequence of the white noise description introduced in the previous section that the spectrum has time-reversal symmetry \cite{reviewt}; that is $\lambda_1=-\lambda_3$ and $\lambda_2=0$.
Here the vanishing of $\lambda_2$ describes conservation of the vertical component of the separation in the leading order, see above. This approximation provides the linear order term of the dependence of $\lambda_i$ in $Fr$.
The first Lyapunov exponent vanishes at $Fr=0$ where the particles' trajectories in space stop to be chaotic. This is because they become effectively the trajectories of particles sedimenting in still air. This is not completely trivial because the particles' velocity in the leading order is still the local turbulent flow $\bm u(\bm x(t), t)$.

To linear order in $Fr$ the sum of the Lyapunov exponents $\sum_{i=1}^3\lambda_i$, which is a main measure of clustering, is zero and there is no clustering. The leading order term in $\sum_{i=1}^3\lambda_i$ is quadratic in $Fr$ and it was derived in \cite{fi2015} via the energy spectrum of the turbulent flow $E(k)$. The theoretical predictions are confirmed in Fig. \ref{breakdown}. The Figure also shows that $\left|\sum_{i=1}^3\lambda_i\right|$ grows with $Fr$ not because of the growth of $\lambda_2$ but rather because of growing compressibility of the horizontal flow, $max(\lambda_2/|\sum_{i=1}^3\lambda_i|)\ll 1$.

The main application of the Lyapunov exponents to the distribution of the particles is to the calculation of the Kaplan-Yorke codimension $C_{KY}$, i.e. the difference of the space dimension and the Kaplan-Yorke dimension. It can be seen by using $\lambda_1+\lambda_2>0$ and $\sum_{i=1}^3 \lambda_i<0$ in the dimension's definition in \cite{ky} that in our case $C_{KY}=\left|\left(\sum_{i=1}^3 \lambda_i\right)/\lambda_3\right|$. The codimension in this case has interpretation of compressibility ratio and it is small because of the flow's weak compressibility. The results of \cite{itzhak1} for attractors of weakly compressible flows give that the steady state fluctuations of the particles' concentration are lognormal and the rest of fractal dimensions can be obtained from the Kaplan-Yorke dimension. For instance the so-called correlation codimension $\alpha$, that is defined as the scaling exponent in the power-law $r^{-\alpha}$ for the RDF (probability to find a pair of droplets at distance $r$; see the Introduction) obeys
\begin{eqnarray}&&\!\!\!\!\!\!\!\!\!\!\!\!\!
\alpha=2C_{KY}=\frac{2\left|\sum_{i=1}^3 \lambda_i\right|}{\left|\lambda_3\right|}, \label{alpf}
\end{eqnarray}
cf. \cite{nature}. It is significant for the sequel that this is a universal relation that  holds for any weakly compressible flow \cite{itzhak1}. This explains the central role of the effective spatial flow of the droplets in the theory.

We determined the RDF numerically, which was not done in \cite{fi2015}, in order to test Eq.~(\ref{alpf}), see below and Fig. \ref{ouralpha}. The usage in the above formula of the (corroborated) formulas of \cite{fi2015} that provide $\lambda_i$ via $E(k)$ gives,
\begin{eqnarray}&&\!\!\!\!\!\!\!\!\!\!\!\!\!
\alpha=\frac{3\pi\int_0^{\infty} E(k)kdk}{2g},\ \ Fr\to 0. \label{sp}
\end{eqnarray}
We find by comparison with Eqs.~(\ref{ps}) and (\ref{lau}) that
\begin{eqnarray}&&\!\!\!\!\!\!\!\!\!\!\!\!\!
\alpha=12\lambda_1\tau=Fr \frac{6L_c}{5\eta}.
\end{eqnarray}
These are remarkably simple formulas that must however be taken with a grain of salt: it is an asymptotic result and at high $Re_{\lambda}$ its validity may demand unpractically small $Fr$ as in the discussion of $\lambda_1$ above, see also below.

We observe that $\alpha$ predicted by Eq.~(\ref{sp}) is independent of the Stokes number: it is a property of turbulence and not of particles. The reason for the particle size independence is that the stretching rate and compression rate of infinitesimal volumes of particles, which ratio determines the fractal dimensions, have identical dependence on $St$. Thus statistics of attractors of particles with different sizes is identical. This does not tell that these attractors coincide in space. Spatial locations of these attractors differ and only their average properties agree.

{\bf Comparison with other works}---The above size-independence explains the observations of \cite{bec2014} in the relevant range; see also \cite{Ireland}. It was found in \cite{bec2014} that at $St \gtrsim 1$ the correlation codimension is $St-$independent, which was explained by further developing the formalism of \cite{fouxonhorvai}. The authors managed to demonstrate that $\alpha=C Fr$ where $C$ is a constant prefactor which is independent of $St$. The calculation of $C$ was not provided and it could be plausibly thought that $C\sim 1$. However \cite{fi2015} demonstrated that Eq.~(\ref{sp}) gives $C\simeq 13$ at $Re_{\lambda}=70$. This is a large value that would be difficult to obtain without calculation. This large numerical factor implies that clustering, whose strength is measured by $\alpha$, can be strong at very small $Fr$.
%

{\bf Highest $Fr$ describable by \cite{fi2015}}---The predictions of the $Fr\to 0$ theory allow detailed comparison with the DNS data. It was observed in \cite{fi2015} that, as already mentioned, at $Re_{\lambda}=70$, the predictions for the Lyapunov exponents hold at $Fr\leq 0.033$. However some of them break down already at $Fr=0.05$, see Fig. \ref{breakdown} where data, more detailed than in \cite{fi2015}, are presented. At the same time,  for some quantities, the theory was found to apply at larger $Fr$ also. This is because deviations of different quantities may compensate each other as happens to be the case of $C_{KY}/(\lambda_1\tau)$ predicted in \cite{fi2015} to be equal to six, independently of the details of turbulence, of $Fr$ and of $St$. This quantity is accidentally close to the theoretical value at all $Fr$ within fifteen per cent discrepancy. The prediction $\alpha=2C_{KY}\propto Fr$ holds within four per cent discrepancy up to $Fr=0.05$, see Fig. \ref{ouralpha}. Moreover, the data of \cite{fi2015} demonstrate that the theoretical predictions for $C_{KY}$ and $\lambda_1$ hold at $St=1$ or $St=2$ up to $Fr\leq 0.1$ and break down only at $Fr=0.2$. In contrast, at $St=0.5$ the prediction for $\lambda_1$ fails at $Fr=0.1$. These observations are in agreement with \cite{bec2014} who observed different behavior at $St\geq 1$ and $St<1$ for $Fr=0.05$.

{\bf Examination of the claim that \cite{fi2015} is invalid}---The success of the theory of \cite{fi2015} is evident from the above. However \cite{Ireland} observed that Eq.~(\ref{sp}) does not work at $St=1$ and $Fr=0.052$ and concluded that the equation is wrong. This is despite that the equation was confirmed in \cite{fi2015} (strictly speaking \cite{fi2015} confirmed $C_{KY}=3\pi\int_0^{\infty} E(k)kdk/(4g)$ which together with $\alpha=2C_{KY}$, confirmed in the present work, validates Eq.~(\ref{sp}) at $Re_{\lambda}=70$). The discrepancy however is the indication of incorrect use of Eq.~(\ref{sp}), rather than its invalidity. The calculations of \cite{fi2015} are rigorous asymptotic calculations at $Fr\to 0$ and can be checked. Their application depends on the Reynolds number, since the theory holds non-uniformly in $Re_{\lambda}$ as we stress in this work. In the case of \cite{Ireland} the discrepancy is both due to factor of $2$ mistake in the formula and testing of the theory at $Fr$ which was already claimed to be beyond the theory in \cite{fi2015}. The simple rule of the thumb is that the theoretical prediction for $\alpha$ applies provided that $\alpha\ll 1$, cf. below. Since $\alpha$ studied in \cite{Ireland} does not obey the inequality then the observed deviation is more than reasonable, cf. above.

\section{Theory's limitation: clash of $Fr\to 0$ and $Re_{\lambda}\to\infty$ limits} \label{clash}


In this section we provide detailed exposition of the theory of \cite{fi2015} in order to demonstrate that increase of the Reynolds number at fixed $Fr$ invalidates this theory. The main observation of \cite{fi2015} is that in the $Fr\to 0$ limit, one can define the field $\bm v_a(t, \bm x)$, which provides the velocity of the droplet of radius $a$ located at time $t$ at point $\bm x(t)$:
\begin{eqnarray}&&\!\!\!\!\!\!\!\!\!\!\!\!\!
\dot{\bm x}=\bm v_a(t, \bm x(t)),\ \ \partial_t \bm v_a+(\bm v_a\cdot\nabla)\bm v_a=-\frac{\bm v_a-\bm u}{\tau(a)}+\bm g, \label{flo}
\end{eqnarray}
where the partial differential equation (PDE) on $\bm v_a(t, \bm x)$ is implied by the equation of motion on differentiating $\dot{\bm x}=\bm v_a(t, \bm x(t))$ over time \cite{nature}.  Below we omit the subscript of $\bm v_a(t, \bm x)$ unless the radius to which the flow pertains needs to be referred. The flow is introduced implicitly as the solution of the above PDE. An explicit formula for $\bm v(t, \bm x)$ via $\bm u(t, \bm x)$, similar to the small Stokes numbers' Eq.~(\ref{maxeyf}), is unavailable. The solution can only become a well-defined single-valued field after transients, lasting for times of order $\tau(a)$, during which the initial condition is forgotten. Indeed, one can devise initial conditions for Eq.~(\ref{flo}) producing multi-valued flow after a short time.

It is readily seen that at finite $Fr$ the assumption of single-valued field is inconsistent due to the blowup (sling) events at which $\nabla\bm v$ explodes. Indeed, spatial differentiation of Eq.~(\ref{flo}) , and the passage in the resulting equation to the frame moving with the particle, gives a closed matrix ODE for the gradients $\sigma_{ik}(t)\equiv \nabla_k v_i(t, \bm x(t))$ \cite{nature},
\begin{eqnarray}&&\!\!\!\!\!\!\!\!\!\!\!\!\!
\dot{\sigma} \!+\!\sigma^2\!=\!-\frac{\sigma\!-\!s}{\tau},\label{gradients11}
\end{eqnarray}
where $s_{ik}(t)\equiv \nabla_k u_i(t, \bm x(t))$. If the quadratic term in the LHS is not small then it would cause finite-time explosion of the gradients. That explosion would signal the breakdown of the flow description due to the flow becoming multi-valued \cite{nature,fp}. In contrast, if $\sigma^2$ in Eq.~(\ref{gradients11}) is small then, in the leading order, the gradients of the droplets' flow in the particle frame are given by the finite expression,
\begin{eqnarray}&&\!\!\!\!\!\!\!\!\!\!\!\!\!
\sigma\approx \sigma_l, \ \
\sigma_l\equiv \int_{-\infty}^t \exp\left(-\frac{t-t'}{\tau}\right)s(t')\frac{dt'}{\tau}, \label{lin}
\end{eqnarray}
where the subscript stands for linearization of Eq.~(\ref{gradients11}).

We first disregard intermittency as in our study of the Lyapunov exponent. As in that study, the correlation time of $s_{ik}(t)$ is given by the smallest of the Kolmogorov time $\tau_{K}$ and the sedimentation time $\tau_g=\eta/g\tau$. We find as previously that the time is $\tau_g$ due to $\tau_g/\tau_{\eta}=Fr/St\ll 1$. Moreover, we have $\tau_g/\tau=Fr/St^2\ll 1$; therefore, the effective integration interval $\tau$ in Eq.~(\ref{lin}) is much larger than the correlation time of $s(t)$. Thus $\sigma$ is effectively a sum of large number of independent identically distributed random variables and is Gaussian, cf. the study of $\lambda_1$.

The Gaussianity implies that the condition that the probability of sling events is small is tantamount to the condition that the dispersion $\langle \sigma_l^2\rangle$ is much smaller than $1/\tau^2$ since the bulk of the probability is determined by the dispersion (in writing matrix as a scalar, as $\sigma_l^2$, the characteristic value is implied). Here we disregard the non-zero average of $s_{ik}$ that exists due to the combination of preferential concentration and anisotropy. This average can be excluded by considering $\sigma_l-\langle\sigma_l\rangle$ instead of $\sigma_l$ and observing that smallness of $\langle\sigma_l\rangle$ makes it irrelevant in the estimates here and below, see Appendix \ref{spso}. Averaging the square of Eq.~(\ref{lin}), it is found that $\langle \sigma_l^2\rangle\tau^2\sim Fr$, see \cite{fi2015} and Eq.~(\ref{d01}) below. Therefore, in the leading order in $Fr\ll 1$, the nonlinear term in Eq.~(\ref{gradients11}) can be self-consistently neglected. The flow is single-valued and well-defined in most of the space.
This does not guarantee though that the rare regions of slings cannot provide appreciable contribution to some quantities, see the Introduction and \cite{nature}.

We observe that the effective domain of integration in Eq.~(\ref{lin}) is $(t-\tau, t)$ so that $\sigma_l(t)\tau \!\sim \int_{t-\tau}^t\! s(t') dt'$. Then the considerations of the last paragraph imply that $\sigma_l$ is statistically similar to instantaneous turbulent flow gradient $\nabla\bm u$ averaged over spatial interval of order $g\tau^2$ passed by sedimenting particle in time $\tau$. This object resembles a similar average of local energy dissipation rate, which is a standardly measured quantity \cite{srea}. The measurements of \cite{se} indicate that this quantity must undergo strong fluctuations in clouds. The fluctuations, that are weak at $Re_{\lambda}=70$ considered in \cite{fi2015}, destroy the particles' flow at larger Reynolds numbers.

\subsection{Flow breakdown at increasing $Re_{\lambda}$}  \label{flowbreakdown}

The Kolmogorov-type estimates used after Eq.~(\ref{lin}) are changed profoundly by intermittency. Below we perform order of magnitude calculations only, not writing the matrix indices.

{\bf Condition for well-defined flow of the particles}---It is immediate consequence of our study in Sec. \ref{lyapunov} that the relevant spatial scale in the calculation of $\langle \sigma_l^2\rangle$ is $L_c$ and not $\eta$. We have assuming that $L_c/(g\tau)\lesssim \tau$, cf. Eq.~(\ref{lv})
\begin{eqnarray}&&\!\!\!\!\!\!\!\!\!\!\!\!\!
\langle \sigma_l^2\rangle\tau^2\sim \int_{t-\tau}^t dt_1dt_2 \langle s(t_1)s(t_2)\rangle\sim \frac{\tau}{\tau_K^2}\frac{L_c}{g\tau}=Fr\frac{L_c}{\eta}\sim 15 Fr  \left(\frac{Re_{\lambda}}{70}\right)^{\Delta_{\lambda}}, \label{sl}
\end{eqnarray}
where we observed that $\int \langle s(0)s(t)\rangle dt$ is estimated as the product of characteristic value $\tau_K^{-2}$ of $s^2$ and the correlation time $L_c/(g\tau)$ and we used notations of Sec. \ref{lyapunov}. We observe that for $St\sim 1$ the condition $\langle \sigma_l^2\rangle\tau^2\ll 1$ coincides with the validity condition of the calculation of $\lambda_1$ given by Eq.~(\ref{lv}). Since the last condition is $L_c/(g\tau)\ll \tau$ then our assumption $L_c/(g\tau)\lesssim \tau$ made in the calculation for Eq.~(\ref{sl}) is self-consistent. If $\langle \sigma_l^2\rangle\tau^2\ll 1$ holds then, by Chebyshev's inequality, the probability of the sling effect, $|\sigma_l|\tau\sim 1$, is small and motion in most of the space is smooth, cf. the discussion after Eq.~(\ref{lv}). In contrast, at high $Re_{\lambda}$ obeying $Fr Re_{\lambda}^{\Delta_{\lambda}}\gtrsim 1$, the sling effect is typical. Due to $St\sim 1$ the particle strongly interacts with vortex correlated over the correlation length $L_c$ and the flow description breaks down. We consider it highly plausible that this transition happens at $Re$ lower than those in the clouds.

\subsection{Strong anisotropy of the particles' flow at small scales} \label{asgr}

We have introduced in subsection \ref{fbreakdown} the scale $L_c$ which is the spatial scale of turbulent flow gradients that determine the Lyapunov exponent. We indicated that this scale must be understood as the correlation length of the particles' flow gradients. Here we refine this statement by demonstrating that this is the correlation length only for generically oriented distances between the points where correlations are studied. For vertical distances the correlations are of longer range. As a result gradients of the particles' flow are correlated over spatial region similar to a vertically oriented column.  This anisotropy leaves direct mark on the multifractal distribution of particles, considered later, where the multifractal structures extend over similar columns.

The flow of the particles integrates non-trivially the turbulent flow via Eq.~(\ref{flo}). This implies that spatio-temporal properties of $\bm v(t, \bm x)$ and $\bm u(t, \bm x)$ are quite different. We consider pair correlations of $\nabla_iv_k(\bm x)$ at $t=0$. We have according to Eq.~(\ref{lin}) that
\begin{eqnarray}&&\!\!\!\!\!\!\!\!\!\!\!\!\!
\nabla_iv_k(\bm x)\approx \int_{-\infty}^0 \exp\left(\frac{t}{\tau}\right)s_{ik}(t)\frac{dt}{\tau},\ \ s_{ik}(t)\equiv\nabla_k u_i(t, \bm q(t, \bm x)),
\end{eqnarray}
where we introduced the Lagrangian trajectories of the particles' flow $\bm q(t, \bm x)$. These are particles' trajectories that pass $\bm x$ at $t=0$. The flow's existence implies that these trajectories are unique and solve
\begin{eqnarray}&&\!\!\!\!\!\!\!\!\!\!\!\!\!
\partial_t\bm q(t, \bm x)=\bm v(t, \bm q(t, \bm x)),\ \ \bm q(t=0, \bm x)=\bm x. \label{Lagra}
\end{eqnarray}
We have
\begin{eqnarray}&&\!\!\!\!\!\!\!\!\!\!\!\!\!
\left\langle \nabla_iv_k(0) \nabla_lv_m(\bm x)\right\rangle \approx \int_{-\infty}^0 \exp\left(\frac{t_1+t_2}{\tau}\right)\left\langle \nabla_k u_i(t_1, \bm q(t_1, 0))\nabla_l u_m(t_2, \bm q(t_2, \bm x))\right\rangle\frac{dt_1dt_2}{\tau^2}.
\end{eqnarray}
The exponential decay factor in above integral effectively cuts the integration domain off at times $|t_i|$ smaller or of order $\tau$. However in the studied regime the particles' separation during times of order $\tau$ is negligible due to $\lambda_1\tau\sim Fr\ll 1 $, see Section \ref{lyapunov}. Thus we have $\bm q(t, \bm x)-\bm q(t, 0)\approx x$ for $|t|\lesssim \tau$. The trajectories' positions with respect to each other are determined by the gravitational sweeping, similarly to the previous sections, and we may write
\begin{eqnarray}&&\!\!\!\!\!\!\!\!\!\!\!\!\!
\left\langle \nabla_iv_k(0) \nabla_lv_m(\bm x)\right\rangle \approx \int_{-\infty}^0 \exp\left(\frac{t_1+t_2}{\tau}\right)\left\langle \nabla_k u_i(\bm g \tau t_1)\nabla_l u_m(\bm x+\bm g \tau t_2)\right\rangle\frac{dt_1dt_2}{\tau^2}, \label{real}
\end{eqnarray}
where the angular brackets here stand for usual spatial averaging. We see that, provided that $|\bm x|$ is much smaller the correlation length $L_c$ of typical turbulent velocity gradients that determine the above average, we can omit $\bm x$ in the above equation. We find that $\left\langle \nabla_iv_k(0) \nabla_lv_m(\bm x)\right\rangle\approx \left\langle \nabla_iv_k(0) \nabla_lv_m(0)\right\rangle$ for $x\ll L_c$. In contrast, if $x\gg L_c$ then, for a generic direction of $\bm x$, the arguments of $\nabla\bm u$ in Eq.~(\ref{real}) are separated at all relevant $t_i$ by distance much larger than $L_c$ and the correlation function $\left\langle \nabla_iv_k(0) \nabla_lv_m(\bm x)\right\rangle$ is much smaller than $\left\langle \nabla_iv_k(0) \nabla_lv_m(0)\right\rangle$. There is however an exception. If the "initial" separation $\bm x$ is vertical and much smaller than $g\tau^2$ then within time much smaller than $\tau$ the argument $\bm x+\bm g \tau t_2$ becomes similar to $\bm g \tau t_2$ and we still have $\left\langle \nabla_iv_k(0) \nabla_lv_m(\bm x)\right\rangle \approx \left\langle \nabla_iv_k(0) \nabla_lv_m(0)\right\rangle$. It is readily seen that these considerations extend to those $\bm x$ that, besides the vertical component smaller than $g\tau^2$, have a non-zero horizontal component that is smaller than $L_c$.  We conclude that velocity gradients are correlated over cylinders (columns) with height of order $g\tau^2$ and radius of the base of order $L_c$ (of course the shape is not sharply defined). The cylinder is elongated, with height larger than diameter by a large factor studied in detail later.

The significant difference in spatial correlations of flows of the fluid and of the particles indicates that most probable gradients of droplets' flow are not created by the most probable gradients of turbulence, because the connection between the gradients is not local in time. Rather gradients of the particles' flow are created by eddies with local viscous scale of order $L_c$.

\subsection{Breakdown of Gaussianity due to intermittency}

The other effect of the intermittency is that it destroys Gaussianity of $\sigma_l$. This can occur when the probability of the sling effect, as measured by $\langle \sigma_l^2\rangle\tau^2$, is still small, $\langle \sigma_l^2\rangle\tau^2\ll 1$. In that case the flow description applies however the flow gradients are non-Gaussian. We demonstrate breakdown of Gaussianity  by considering the fourth moment. We have using Eq.~(\ref{lin})
\begin{eqnarray}&&\!\!\!\!\!\!\!\!\!\!\!\!\!
\langle \sigma_l^4\rangle\tau^4\sim 3\int_{t-\tau}^t\! \langle s(t_1)s(t_2)\rangle \langle s(t_3)s(t_4)\rangle dt_1dt_2dt_3dt_4+\int_{t-\tau}^t \!\langle s(t_1)s(t_2)s(t_3)s(t_4)\rangle_c dt_1dt_2dt_3dt_4, \label{decompo}
\end{eqnarray}
where we decomposed the fourth-order correlation function of $s(t)$ into the sum of the reducible part, given by the first term in the RHS, and the irreducible part or the cumulant \cite{ma}, designated by the subscript $c$ and given by the last term. We have using properties of the different-time correlations of $s(t)$, described previously, the estimate
\begin{eqnarray}&&\!\!\!\!\!\!\!\!\!\!\!\!\!
\int_{t-\tau}^t \langle s(t_1)s(t_2)\rangle \langle s(t_3)s(t_4)\rangle dt_1dt_2dt_3dt_4\sim  Fr^2 Re_{\lambda}^{2\Delta_{\lambda}}.
\end{eqnarray}
If intermittency is negligible (so that $Re_{\lambda}^{\Delta_{\lambda}}\sim 1$ and the last term in Eq.~(\ref{decompo}) obeys $Re_{\lambda}-$independent estimate $\tau (\eta/g\tau)^3/\tau_K^4\sim Fr^3/St^2$) then the last term in Eq.~(\ref{decompo}) would be smaller than the term coming from the reducible part of the correlation function by the factor of $\eta/(g\tau^2)\sim Fr/St^2$. Indeed, the irreducible correlation function $\langle s(t_1)s(t_2)s(t_3)s(t_4)\rangle_c$ decays quickly when the separation of any pair of $t_i$ becomes larger than the correlation time $\eta/(g\tau)$. Thus the effective domain of integration over $t_i$ is smaller than for the reducible contribution by $Fr/St^2$, the fact that lies at the origin of the Gaussianity considered previously.

The intermittency causes the probability of $s$ much larger than the rms value $\tau_{K}^{-1}$ to be non-negligible. Thus equal-time fourth order moment of $s$ is proportional to $\tau_{K}^{-4}$ times a positive power of $Re_{\lambda}$. As a result, at large $Re_{\lambda}$, the fourth order-cumulant is larger than the reducible contribution $3\langle s^2\rangle^2\sim \tau_{K}^{-4}$ by a power of the Reynolds number. Since correlation time also depends on $Re_{\lambda}$ then it is not completely obvious, however very plausible, that the last term of the LHS of Eq.~(\ref{decompo}) contains a higher power of $Re_{\lambda}$ than $Re_{\lambda}^{2\Delta_{\lambda}}$ of the reducible contribution. We obtain
\begin{eqnarray}&&\!\!\!\!\!\!\!\!\!\!\!\!\!
\frac{\int_{t-\tau}^t \langle s(t_1)s(t_2)s(t_3)s(t_4)\rangle_c dt_1dt_2dt_3dt_4}{\int_{t-\tau}^t \langle s(t_1)s(t_2)\rangle \langle s(t_3)s(t_4)\rangle dt_1dt_2dt_3dt_4}\sim \frac{Fr Re_{\lambda}^{\Delta}}{St^2}, \label{fsa}
\end{eqnarray}
where the exponent $\Delta$ is positive and absorbs all dependencies on $Re_{\lambda}$.
Then the increase of $Re_{\lambda}$ at fixed $Fr$ and $St$ causes the irreducible contribution in $\langle \sigma_l^4\rangle$ to become dominant. This happens at $Re_{\lambda}^{\Delta}\sim St^2/Fr$ which at $St\sim 1$ occurs when the flow is still well-defined provided that $\Delta>\Delta_{\lambda}$, cf. the comment after Eq.~(\ref{sl}). Thus in this case we will have a flow with non-Gaussian statistics of the gradients similarly to the turbulence itself. In contrast, if $\Delta<\Delta_{\lambda}$ then, as long as the flow is well-defined at all, the gradients' statistics is Gaussian. Similar observations hold for higher order moments $\langle \sigma_l^{2n}\rangle$ with $n>2$. The converse of the above statements is that decrease of $Fr$, at fixed $St$ and however large $Re_{\lambda}$, causes the statistics of $\sigma_l$ to become Gaussian and makes the flow description valid.

\section{Radial distribution function at $Fr\to 0$: spatial correlations and sum rule} \label{properties1}

We have considered above the general properties of the flow of the particles and how increase in the intermittency due to growth of the Reynolds number can change its properties. The rest of the paper is devoted to the case of $Fr$ so small that the flow description is valid. This range depends on $St$ and $Re_{\lambda}$ as explained above.

In this section we consider the radial distribution function (RDF). Since the flow is weakly compressible then, generally, the RDF obeys scaling below the correlation length of the gradients of the droplets' flow \cite{itzhak1}. We saw
in subsection \ref{asgr} that the gradients are correlated over columns whose height is of order $g\tau^2$ and diameter of order of $L_c$. Here we demonstrate numerically that, indeed, the scaling of the RDF holds over these columns, all of whose dimensions are much larger than the Kolmogorov scale $\eta$. This effect is solely due to gravity: multifractality at negligible gravity and small inertia holds below the same correlation length $\eta$ as the turbulent velocity gradients have \cite{nature}. We also derive here a sum rule demonstrating that despite that the RDF, due to strong anisotropy, has a non-trivial angular dependence, still the angle-averaged RDF is rather simple.

\subsection{Small-scale flow is two-dimensional}

Main properties of the droplets' flow at $Fr\to 0$ are weak compressibility and small scale two-dimensionality. The weak compressibility can be seen by taking the trace of Eq.~(\ref{lin}) and using $tr s=0$. This gives that in the leading order $tr\sigma=0$ and the droplets' flow is incompressible. Clustering occurs due to small compressible component of the flow. The leading order term for $tr \sigma$ can be found by taking the trace of Eq.~(\ref{gradients11}). The solution of the obtained equation gives (this equation in \cite{fi2015} contains a typo with a wrong sign that is irrelevant for the calculations there)
\begin{eqnarray}&&\!\!\!\!\!\!\!\!\!\!\!\!\!
tr\sigma(t)=-\int_{-\infty}^t dt'\exp\left(-\frac{t-t'}{\tau}\right)tr \sigma^2(t') dt's\approx -\int_{-\infty}^t dt'\exp\left(-\frac{t-t'}{\tau}\right)tr \sigma_l^2(t') dt',\label{trace}
\end{eqnarray}
where the last term is estimated as $\tau \sigma_l^2$. We find that compressibility ratio $tr \sigma/\sigma_l$ is of order $\tau \sigma_l\ll 1$.

Droplets' flow at small scales is described by the Taylor expansion of the velocity difference below the smoothness scale $\eta_p(\theta)$ of the particles' flow ($\theta$ is the polar angle of $\bm r$, with $z-$axis directed upwards)
\begin{eqnarray}&&\!\!\!\!\!\!\!\!\!\!\!\!\!
\bm v(\bm x_2)-\bm v(\bm x_1)\approx (\bm r\cdot\nabla)\bm v(x_1),\ \ r\ll \eta_p(\theta),\ \ \bm r\equiv \bm x_2\bm x_1. \label{smoothnes}
\end{eqnarray}
Here we introduced the angle-dependent correlation length $\eta_p(\theta)$. The study that we performed in subsection \ref{asgr} implies that $\eta_p(\theta)\sim L_c$ for $\theta\gg L_c/(g\tau^2)$ and $\eta_p(\theta)\sim g\tau^2$ for $\theta\lesssim L_c/(g\tau^2)$. Here $L_c/(g\tau^2)\ll 1$, see below, and $0\leq \theta\leq \pi/2$ (the range of $\pi/2\leq \theta\leq \pi$ is obtained by interchange of $\bm x_i$ in Eq.~(\ref{smoothnes})).

The matrix $\nabla_iv_k$ is a Gaussian $2\times 2$ matrix in the leading order in $Fr$. This observation was made in \cite{fi2015} for the Lagrangian statistics in the particle frame, see also Sec. \ref{lyapunov}. We demonstrate in Appendix \ref{gradients} that the same proof applies for the Eulerian statistics at a fixed spatial point. One finds that $\nabla \bm v$ does not have vertical components or derivatives of horizontal components in the vertical direction in the leading order in $Fr$. The statistics of planar gradients is Gaussian so that they are described fully by zero mean and the dispersion,
\begin{eqnarray}&&\!\!\!\!\!\!\!\!\!\!\!\!\!
\langle \nabla_k v_i \nabla_r v_p\rangle
=\frac{c_0Fr\left(3\delta_{ip}\delta_{kr}-\delta_{ik}\delta_{pr}-\delta_{ir}\delta_{kp}\right)}{2\tau^2},
\label{d01}
\end{eqnarray}
where $c_0$ is defined in Eq.~(\ref{c0}). This formula confirms the order of magnitude estimate $\langle (\nabla\bm v)^2\rangle\tau^2\sim Fr$ made above.
Taking its trace over $x$ and $y$ gives $\langle \left(\nabla_xv_x+\nabla_yv_y\right)^2\rangle=0$ so that the planar flow is incompressible. There are only three random variables that describe the gradients given by the components of the zero trace $2\times 2$ matrix. We stress that these properties describe small scales only: the absolute motion of the droplet is three-dimensional, resembling that of the tracer, by our assumption that the droplet velocity is dominated by large-scale turbulence.

\subsection{RDF}

The RDF, designated by $g_{12}(\bm r)$, determines the probability density function (PDF) $P_{12}(\bm r)$ of finding a droplet of radius $a_2$ at a distance $\bm r$ from a droplet with radius $a_1$.
The PDF is proportional to the mean concentration $\langle n_2\rangle$ so that we can write $P_{12}(\bm r)=\langle n_2\rangle g_{12}(\bm r)$ where $g_{12}(\bm r)$ is a property of the two-particle motion.
Here and below $n_i(\bm x, t)$ is the concentration of droplets with radius $a_i$. The RDF also provides the pair-correlation function of concentrations, $\langle n_1(0)n_2(\bm r)\rangle=\langle n_1\rangle\langle n_2\rangle g_{12}(\bm r)$, see Appendix \ref{kernel}.

Pair correlations of concentration $\langle n_1(0)n_2(\bm r)\rangle$ are determined quite similarly to the pair correlation function of passive scalar in incompressible turbulence in the presence of the scalar source \cite{reviewt}. The reason is that the continuity equation, obeyed by the droplets' concentration $n$, can be written as,
\begin{eqnarray}&&\!\!\!\!\!\!\!\!\!\!\!\!\!
\partial_t \ln n+(\bm v\cdot\nabla)\ln n=-\nabla\cdot\bm v,
\end{eqnarray}
We see that the flow's divergence, whose correlation length is $\eta_p$, can be considered as the source of concentration fluctuations. In the case of passive scalar the pair correlation is determined by the time that the particles spent below the correlation length of the source before the observation, when the correlations accumulate \cite{reviewt}. Similarly here $\langle n_1(0)n_2(\bm r)\rangle$ is determined by the mean duration $t^*_{12}(\bm r)$ of the time interval that the particles spent below the correlation length of the source $\eta_p$ before they approach by distance $r$ where $r\ll \eta_p$ (here $\eta_p$ is much smaller than the vessel size which is of order of integral scale of turbulence or larger).

The relevance of $t^*_{12}(\bm r)$ is seen by considering motion of only two droplets
of radii $a_1$ and $a_2$ in a finite-size vessel over long time. The droplets' separation is most of the time of order of the size of the vessel and is much larger than $\eta_p$. At these scales the flow is incompressible so that there is no build-up of correlations - the particles approach or separate with the same probability without creating large-scale inhomogeneities of the concentration. However occasionally the particles approach by a small distance $\bm r$. This happens after they approach by distance $\eta_p$. The mean time between the last two events is $t^*_{12}(\bm r)$ and during this time-interval the particles have effective attraction  because they experience the same local gradient of the flow whose divergence, on average, is negative, which leads to correlations' growth \cite{arxiv}.

The explicit representation of $g_{12}(\bm r)$ via $t^*_{12}(\bm r)$ is found by starting from straightforward generalization of the formula of \cite{itzhak1} that uses cumulant expansion theorem and weakness of compressibility
\begin{eqnarray}&&\!\!\!\!\!\!\!\!\!\!\!\!\!
\ln g_{12}(\bm r)=\int_{-\infty}^0 dt_1dt_2
\langle \nabla\cdot\bm v_1(t_1, \bm q_1(t_1, 0))\nabla\cdot\bm v_2(t_2, \bm q_2(t_2, \bm r))\rangle, \label{reps}
\end{eqnarray}
where $\bm v_i(t, \bm x)$ is short-hand notation for $\bm v_{a_i}(t, \bm x)$, see Eq.~(\ref{flo}), and $q_i(t, \bm x)$ are the Lagrangian trajectories of the flow $\bm v_i(t, \bm x)$, see Eq.~(\ref{Lagra}).
Here in the leading order in the small compressibility we can use for $\bm q_i$ the Lagrangian trajectories of the solenoidal component of $\bm v_i$ which guarantees that $\left\langle \nabla\cdot\bm v_i(t, \bm q_i(t, \bm r))\right\rangle=0$ (otherwise dispersion must be used in the above equation, see \cite{itzhak1}).
We consider first the case of $a_1=a_2$. This is the only case where $g_{12}(0)$ is infinite because the trajectories $q_i(t, 0)$ coincide causing divergence of the integral in Eq.~(\ref{reps}) at $r=0$.
 We find from Eq.~(\ref{reps}) by using the formula of \cite{ff} for the sum of Lyapunov exponents $\sum\lambda_i$ at weak compressibility that \cite{itzhak1}
\begin{eqnarray}&&\!\!\!\!\!\!\!\!\!\!\!\!\!
\ln g(\bm r)=2 \left|\sum\lambda_i\right| t^*(\bm r), \ \
2\left|\sum\lambda_i\right|=\int_{-\infty}^{\infty} \langle \nabla\cdot\bm v(t, \bm q(0, 0))\nabla\cdot\bm v(t, \bm q(t, 0))\rangle dt,
\label{time}
\end{eqnarray}
where $g(\bm r)=g_{11}(\bm r)$ and $\bm q(t, \bm r)\equiv \bm q_1(t, \bm r)$. Here the time $t^*(\bm r)\equiv t^*_{11}(\bm r)$ can be defined as the time of backward in time divergence of trajectories starting at distance $r$ to a distance $\eta_p$, which is equivalent to the definition above. Beyond $t^*(\bm r)$ velocity divergence factors in Eq.~(\ref{reps}) decorrelate and the integral converges rapidly.
We use that the PDF of $t^*(\bm r)$ is strongly peaked at the average value so the fluctuations of $t^*(\bm r)$ can be neglected, see \cite{itzhak1} and below.  We also observe that since the product of concentration and infinitesimal volume is the conserved mass then $|\sum\lambda_i|$ is also asymptotic growth rate of the concentration. This can be used to show that $g(\bm r)$ equals squared density accumulated by compression of initially unit density from volume of size $\eta_p$ during time $t^*(\bm r)$, see \cite{nature,fi2015}.

\begin{figure}
 \includegraphics[width=17cm,trim={1mm 2mm 1mm 2mm},clip]{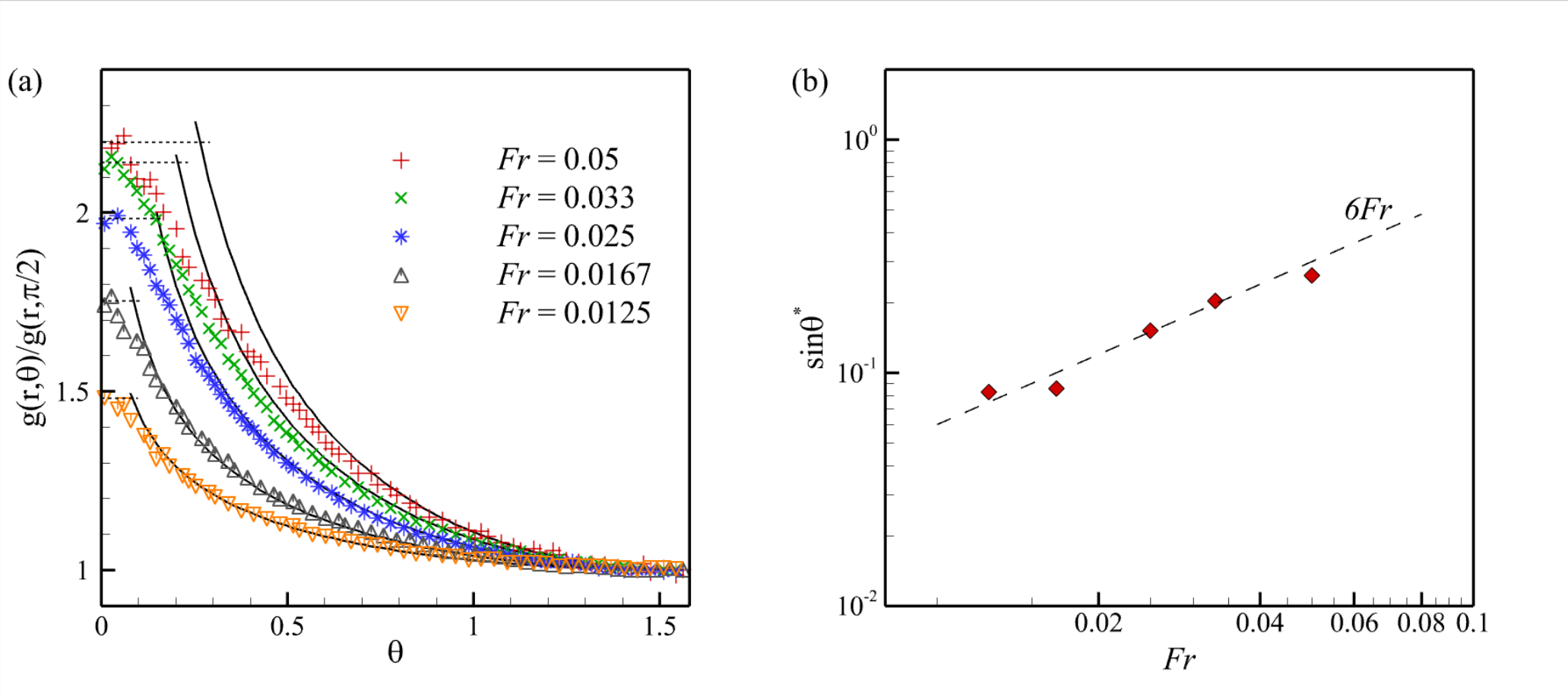}

\caption{ Angle dependence of the RDF at fixed distance $r$ between the particles deep in the smoothness range, (a) normalized RDF and (b) $\sin \theta^* (Fr)$. The smoothness range is about $10\eta$, and the considered range of $r$ was an order of magnitude smaller.
The calculations were performed separately for $r$ in small windows centered around different values of $r$, and the results for different $r$ produced identical angle dependence. Numerical observations for five different $Fr$ at fixed $St=1$ are presented by different colors and compared with the $\sin^{-\alpha}\theta$ law (solid line), that holds in the range of angles near $\pi/2$ as predicted asymptotically in \cite{fi2015}. The indefinite growth of the RDF toward smaller $\theta$ is stopped at angles where separation rates dictated by the vertical and horizontal components are of the same order. This causes flattening of the RDF at the characteristic angle $\theta^*$ determined from $g(r, 0)/g(r, \pi/2)=\sin^{-\alpha}\theta^*$ studied in section \ref{angl}.}
\label{fig:angular}
\end{figure}

Since the divergence of the trajectories backward in time is horizontal, then here $\eta_p$ is an appropriately defined correlation length $l_c$ of $\nabla\cdot \bm v$ in the horizontal direction. We saw that $l_c$ must be of order $L_c$ and we keep it as a free parameter for fitting numerical results. The time $t^*(\bm r)$ diverges logarithmically at small $r$ since the divergence of the trajectories is exponential. Therefore, the divergence of $g(r)$ at small $r$ is a power-law, see Eq.~(\ref{time}). For finding this law we use that the fluctuations of time $t^*(\bm r)$ defined as solution of $|\bm q(t, \bm r)-\bm q(t, 0)|=l_c$ are weak. Indeed, similarly to the limit for the Lyapunov exponent in a forward in time flow, we also have a realization-independent limit for the first Lyapunov exponent of the time-reversed flow  $\lim_{t\to-\infty} \lim_{r\to 0}|t|^{-1}\ln (|\bm q(t, \bm r)-\bm q(t, 0)|/r)$. This logarithmic rate of backward in time separation can be obtained by considering time-reversed evolution of small volumes of particles, and using the compressibility smallness. It is found that the rate equals the same $|\lambda_3|$ that characterizes the rate of particles approach forward in time, see details in \cite{itzhak1,reviewt,fi2015,arxiv,reviewt}. Thus at large finite $|t|$ we have, with probability close to one, $|t|^{-1}\ln (|\bm q(t, \bm r)-\bm q(t, 0)|/r)\approx |\lambda_3|$ (we saw previously that $|\lambda_3|\approx \lambda_1$). Moreover, in the leading order the separation is two-dimensional and is determined by the horizontal component $r\sin\theta$ of the initial separation (the vertical component $r\cos\theta$ is conserved). We find at $t<0$ the separation law $|t|^{-1}\ln (|\bm q(t, \bm r)-\bm q(t, 0)|/r\sin\theta)\approx |\lambda_3|$. This gives $t^*(\bm r)\approx |\lambda_3|^{-1}\ln(l_c/r\sin\theta)$.
We find from Eq.~(\ref{time})
\begin{eqnarray}&&\!\!\!\!\!\!\!\!\!\!\!\!\!
g(\bm r)= \left(\frac{l_c}{r\sin\theta}\right)^{\alpha},\ \ \theta\gg \theta^*, \ \ \alpha=2\left|\frac{\sum_{i=1}^3\lambda_i}{\lambda_3}\right|=\frac{3\pi\int_0^{\infty} E(k)kdk}{2g}=12 c_0 Fr.
\label{eq:alpha}
\end{eqnarray}
Here we introduced a lower threshold for the above considerations,  a small angle $\theta^*$. If the backward in time separation of $\bm q(t, \bm r)$ and $\bm q(t, 0)$ was rigorously two-dimensional, the form of $g(\bm r)$ in Eq.~(\ref{eq:alpha}) would hold at any $\theta$, producing divergence at $\theta=0$. In fact the next order corrections make the separation weakly three-dimensional. Thus particles that are initially separated vertically disperse due to the components of $\sigma$ neglected in Eq.~(\ref{d01}). This regularizes the divergence and introduces the lower limit $\theta^*$ of applicability in Eq.~(\ref{eq:alpha}). The small angle $\theta^*$, studied in section \ref{angl}, is defined by the condition that the separation velocities due do horizontal and vertical components of the separation vector are of the same order when the particle separation forms angle $\theta^*$ with the vertical. Eq.~(\ref{eq:alpha}) was derived in \cite{fi2015} using the smoothness scale of turbulence, the viscous scale $\eta$, instead of the smoothness scale of the droplets' flow $l_c$. The difference is insignificant at $Fr\to 0$ where $\alpha$ is so small that $(l_c/\eta)^{\alpha}\approx 1$ and the formulas coincide. However when larger $Fr$ are considered, the difference becomes relevant, see below. We stress that the formula for $\alpha$ is independent of the particle size and holds for droplets driven by the Navier-Stokes turbulence without approximations \cite{fi2015}. In this work we provide the numerical confirmation of the prediction (which was not done in \cite{fi2015}), see Fig. \ref{fig:angular}. We also observe the large numerical factor of $12$ above.

\subsection{Correlation dimension and angle-averaged RDF: sum rule} \label{sumr}

The angle-averaged (equal-size) RDF can be derived from the result of \cite{fi2015} that the concentration $n_r$ coarse-grained over scale $r$, which is much smaller than $l_c$, obeys
\begin{eqnarray}&&
\frac{\langle n_r^2\rangle}{\langle n\rangle^2}=\left(\frac{l_c}{r}\right)^{\alpha},\ \ r\ll l_c;\ \  n_r(\bm x)=\int_{|\bm x'-\bm x|<r} n(\bm x')\frac{d\bm x'}{4\pi r^3/3}.\label{coars}
\end{eqnarray}
This result holds provided that $\alpha\ll 1$ so that concentration fluctuations at $r=l_c$ are weak \cite{itzhak1}. It derives from the observation that mass $\int_{|\bm x'-\bm x|<r} n(\bm x') d\bm x'$ observed in the ball at $t=0$ is compressed by the flow effectively from an ellipsoid whose initial mass is given by the mean field's estimate $\langle n\rangle$ times the volume and whose largest dimension is $l_c$. The axes of the ellipsoid change during the compression to become equal to $r$ at $t=0$. The concentration of the mass is then due to fluctuations of the two smaller axes of the initial ellipsoid, see details in \cite{fi2015,itzhak1}.

We remark that the correlation length is a quantity defined up to some factor $f$ of order one. This indeterminacy does not prohibit us from writing equality in Eq.~(\ref{coars}) because $\alpha\ll 1$ guarantees that $f^{\alpha}\approx 1$. Thus, as remarked above, \cite{fi2015} used $\eta$ instead of $l_c$ in Eq.~(\ref{coars})  since the distinction between $l_c$ and $\eta$, even though it constitutes an order of magnitude, is irrelevant at $Fr\to 0$ where $\alpha\to 0$. However at finite $Fr$ the difference of $l_c$ and $\eta$ is significant already at $Fr=0.033$ as we demonstrate below using the DNS. The usage of $l_c$ in Eq.~(\ref{coars}) is significant.

Correlation dimension of the multifractal attractor of droplets is defined \cite{hp} with the help of mass in ball of radius $r$. It is readily seen from the definition that Eq.~(\ref{coars}) tells that $\alpha$ is the correlation codimension of the droplets' attractor, that is the difference between the space dimension three and the correlation dimension. Thus Eq.~(\ref{eq:alpha}) tells that correlation codimension is twice the Kaplan-Yorke codimension \cite{nature,itzhak1,fi2015} as claimed in Eq.~(\ref{alpf}).

It is quite obvious that Eq.~(\ref{coars}) implies that angle-averaged RDF ${\tilde g}(r)$ defined by (we use $g(r, \pi-\theta)=g(r, \theta)$)
\begin{eqnarray}&&\!\!\!\!\!\!\!\!\!\!\!\!\!\!
{\tilde g}(r)\equiv \int g(r{\hat r})\frac{d{\hat r}}{4\pi}=\int_0^{\pi/2} g(r, \theta)\sin\theta d\theta, \ \ {\hat r}\equiv \frac{\bm r}{|\bm r|}, \label{sdb}
\end{eqnarray}
obeys the same scaling $r^{-\alpha}$ law. The RDF ${\tilde g}(r)$  describes probability of finding a pair of particles at distance $r$ irrespective of the angle between the pair orientation and the vertical. We perform the calculation of proportionality constant in ${\tilde g}(r)\propto r^{-\alpha}$. We have by squaring the definition of $n_r$ in Eq.~(\ref{coars}) and averaging,
\begin{eqnarray}&&\!\!\!\!\!\!\!\!\!\!\!\!\!\!
\frac{\langle n_r^2\rangle}{\langle n\rangle^2}= \int_{x_1<r,\  x_2<r} g(\bm x_2-\bm x_1)\frac{d \bm x_1d \bm x_2}{[4\pi r^3/3]^2}=\int_{|\bm x_2-\bm x|<r,\  x_2<r} g(\bm x)\frac{d \bm xd \bm x_2}{[4\pi r^3/3]^2}=\int_{x<2r} g(\bm x)d\bm x\frac{\pi(4r+x)(2r-x)^2}{12[4\pi r^3/3]^2}.
\end{eqnarray}
We used the formula for the volume of intersection of two balls of radii $r$ whose centers are separated by $x$. We find
\begin{eqnarray}&&\!\!\!\!\!\!\!\!\!\!\!\!\!\!
\frac{16}{3}\frac{\langle n_r^2\rangle}{\langle n\rangle^2}=\frac{1}{r^6}\int_0^{2r}x^2 (4r+x)(2r-x)^2{\tilde g}(x) dx=\int_0^{2}y^2 (4+y)(2-y)^2{\tilde g}(ry) dy,
\end{eqnarray}
where $y=x/r$. The above connection between the second moment of $n_r$ and the angle-averaged RDF is general. The solution of this equation for the moment given by Eq.~(\ref{coars}) with $\alpha\ll 1$ (that fails for $\alpha\sim 1$) is,
\begin{eqnarray}&&
{\tilde g}(r)=\left(\frac{l_c}{r}\right)^{\alpha},\ \ r\ll l_c. \label{angel}
\end{eqnarray}
We call this relation "sum rule" because it provides global information on the angle-dependence of the RDF. In fact, it demonstrates that contribution of angles $\theta\lesssim \theta*$ into ${\tilde g}(r)$ is negligible since the integral for ${\tilde g}(r)$ in Eq.~(\ref{sdb}) is determined by $\theta\gg \theta^*$. We have from Eq.~(\ref{eq:alpha}) using $\alpha\ll 1$ that
\begin{eqnarray}&&\!\!\!\!\!\!\!\!\!\!\!\!\!\!
\int_{\theta^*}^{\pi/2} g(r, \theta)\sin\theta d\theta=\left(\frac{l_c}{r}\right)^{\alpha}\int_{\theta^*}^{\pi/2}\sin^{1-\alpha}\theta d\theta\approx \left(\frac{l_c}{r}\right)^{\alpha}, \label{RDFa}
\end{eqnarray}
where the integral is independent of $\theta^*$ by $\theta^*\ll 1$. This implies that the probability of finding a pair of particles at distance $r$ is determined by the events of horizontal approach to angles $\theta\gg \theta^*$.

The theoretical considerations above hold for $\alpha\ll 1$ and cannot predict the value of $l_c$. It is anticipated that the power-law will hold also for $\alpha\sim 1$ and we would like to see whether the prediction for $\alpha\ll 1$ can be extended. These issues are dealt with numerically.

\begin{figure*}
\includegraphics[width=8.5cm,trim={2mm 2mm 2mm 2mm},clip]{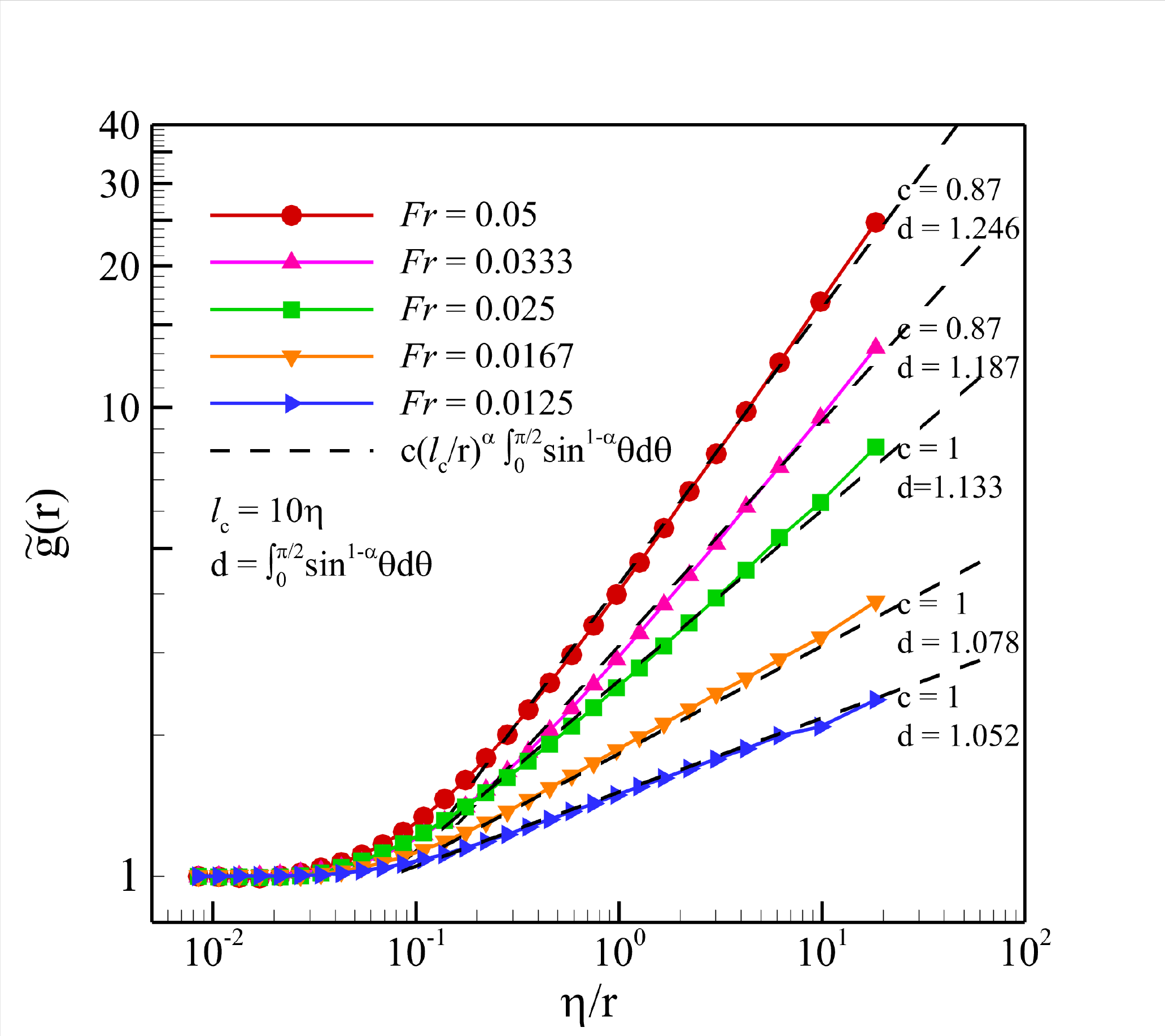}
\caption{The angle-averaged RDF for the five considered values of $Fr$ and $St=1$. We use $\alpha=2|\sum_{i=1}^3 \lambda_i/\lambda_3|$ where the Lyapunov exponents were determined numerically and $l_c=10\eta$.}
\label{angles}
\end{figure*}

\subsection{Comparison with the DNS}

The results of the simulations for the angle-averaged RDF are shown in Fig. \ref{angles}. It is seen that the power-law extends inside the inertial range of turbulence holding up to the scale  $l_c=10\eta$. Here the same cutoff scale $l_c$, which can be considered as the smoothness scale of the particles' flow, describes a range over which Froude numbers vary significantly. The scale is of order $L_c$ independently of $Fr$ because the same turbulent fluctuations with characteristic scale $L_c$ determine the RDF, see above. For $r<10\eta$ the data are described perfectly by the power-law exponent $\alpha=2|\sum_{i=1}^3 \lambda_i/\lambda_3|$. Here $\lambda_i$ are determined numerically, in accord with Eq.~(\ref{eq:alpha}).
For $r>10\eta$ the RDF relaxes to the asymptotic value $g(r=\infty)=1$ which describes lack of pair correlations beyond $l_c$.

The DNS are performed at a finite $Fr$ and this demands refinements of the theory. It was observed previously that order of one uncertainty in the definition of the correlation length does not influence the RDF in $Fr\to 0$ limit due to smallness of $\alpha$. However a correction factor $c$ must be introduced at a finite $Fr$, when $\alpha$ is not so small, changing Eq.~(\ref{angel}) to
\begin{eqnarray}&&
{\tilde g}(r)=c\left(\frac{l_c}{r}\right)^{\alpha},\ \ r\ll l_c. \label{angel}
\end{eqnarray}
The efficiency of this fit of the RDF is known since \cite{lance}. The factor $c$ is a result of fluctuations and incomplete determinacy of the smoothness scale. The theory then tells that $c=1$ in the limit of $Fr\to 0$. Fig. \ref{angles} confirms this.

It was already demonstrated in \cite{fi2015} that the Lyapunov exponents are accurately described by the formulas via $E(k)$ for $Fr\leq 0.033$. This implies, by combining with the above, that the formula for $\alpha$ via $E(k)$ in Eq.~(\ref{eq:alpha}) can be used for $Fr<0.05$. For $Fr=0.05$ the spectral formula is not accurate however $\alpha=2|\sum_{i=1}^3 \lambda_i/\lambda_3|$ is still valid. Finally we stress that usage of $\eta$ instead of $l_c$ in Eq.~(\ref{coars}) would give wrong results. For instance there would be factor of three mistake at $Fr=0.033$ where we have $10^{\alpha}\approx 3$.


\section{Angle dependence of the RDF for the same size particles} \label{angl}

In the previous section we described the RDF at not too small angles and also provided the sum rule which describes the angle dependence globally. Here we complete the study in the case of equal size particles by providing the theory of the angle dependence of the RDF . This is done by calculating $t^*(\bm r)$ in Eq.~(\ref{time}). This time describes separation of trajectories backward in time. However we demonstrated previously that small-scale flow is time-reversible in the leading order. This allows us to find $t^*(\bm r)$ by studying forward in time separation of particles to distance $\eta_p$ starting from initial separation $\bm r$. We designate the positions of the particles by $\bm x(t)$ and $\bm x(t)+\bm r(t)$ so that $\bm r(0)=\bm r$. Then, $r\ll \eta_p$ implies that in the distance equation $\dot {\bm r}=\bm v(t, \bm x+\bm r)-\bm v(t, \bm x)$, we can use the Taylor expansion that gives
\begin{eqnarray}&&
\dot{\bm r}=\sigma \bm r, \ \ \sigma_{ik}(t, \bm x_0)=\nabla_k v_i[t, \bm x(t)]. \label{sep}
\end{eqnarray}
It is useful to separate the evolution into the evolution of the distance $r$ and orientation described by unit vector ${\hat n}$. We have using $\bm r=r{\hat n}$ that
\begin{eqnarray}
&&\frac{d\ln r}{dt}={\hat n}\sigma{\hat n},\ \ \frac{d{\hat n}}{dt}=\sigma{\hat n}-{\hat n}[{\hat n} \sigma{\hat n}];\ \ {\hat n}\equiv\frac{\bm r}{r}.\label{orien}
\end{eqnarray}
In the leading order the horizontal separation $\bm r_{\perp}(t)$ is identically zero for particles that are initially separated only vertically, ${\hat n}(t=0)={\hat z}$, since the small-scale flow is planar. However higher order corrections cause $r_{\perp}(t)$ to increase until it reaches a crossover value at which the leading order horizontal separation takes over. During the transient we have ${\hat n}\approx (n_1, n_2, 1)$, where $|n_{\alpha}|\ll 1$ (here $|{\hat n}|=1$ holds in linear order in $|n_{\alpha}|$). The angle $\theta$ between $\bm r$ and ${\hat z}$ equals the magnitude $\sqrt{n_1^2+n_2^2}$ of $(n_1, n_2)$. In the leading order from Eq.~(\ref{orien}), we have
\begin{eqnarray}&&\!\!\!\!\!\!\!
\dot{n}_{\alpha}\approx \sigma_{\alpha z}+\sigma_{\alpha\beta}n_{\beta},\label{alp}
\end{eqnarray}
where we use $|\sigma_{zz}|\ll |\sigma_{\alpha\beta}|$ (remind that the Greek indices are used for $1$ or $2$). The form of this equation is identical with that for the two-dimensional separation vector $(n_1, n_2)$ of two particles whose motion is the superposition of diffusion and the flow with local gradient $\sigma_{\alpha\beta}$. Here, the diffusion is given by the white noise $\sigma_{\alpha z}$, with the diffusion coefficient $\kappa$ obeying $4\kappa=\int \left\langle \sigma_{\alpha z}(0)\sigma_{\alpha z}(t)\right\rangle dt=\kappa_{\alpha z \alpha z}$ where $\kappa_{ikpr}$ is defined in Eq.~(\ref{akpa}).
This equation is well studied; see e. g. \cite{fb}. The diffusive term dominates at small $n_{\alpha}$ and the gradient term dominates at large $n_{\alpha}$. Therefore, if the growth starts from small separations, then it is initially diffusive $n_{\alpha}(t)-n_{\alpha}(0)\approx \int_0^t\sigma_{\alpha z}(t')dt'$. The crossover to the regime of separation due to the gradients $\sigma_{\alpha\beta}n_{\beta}$ occurs at the ``diffusive scale" $\theta^*=\sqrt{\kappa/\lambda_1}$, where $\lambda_1$ is the Lyapunov exponent of $\sigma_{\alpha\beta}$, which was calculated in \cite{fi2015} and Sec. \ref{lyapunov}. We have
\begin{eqnarray}&&\!\!\!\!\!\!\!
\theta^*\sim\sqrt{\frac{\kappa_{\alpha z \alpha z}}{\kappa_{\alpha \beta \alpha \beta}}}\ll 1,
\end{eqnarray}
where we used $\lambda_1\sim \kappa_{\alpha \beta \alpha \beta}$, which was derived previously, and the smallness of $\kappa_{\alpha z \alpha z}/\kappa_{\alpha \beta \alpha \beta}$.  The time of reaching the crossover scale $\theta^*$ is of order $1/\lambda_1$, independently of $\kappa$. Direct inference of $\theta^*$ from the above formula demands the calculation of $\kappa_{\alpha z \alpha z}$ which is a formidable task. The estimate for $\theta^*$ is obtained below differently.

{\bf Distance evolution}---It is useful to consider the evolution to the crossover directly in terms of $\bm r$. We observe that due to the smallness of the $z-$components of $\sigma_{ik}$, the stretching exponent in the $z-$direction is much smaller than $\lambda_1$.
Therefore, during the time $t_c\sim \lambda_1^{-1}$, we can consider $z$ as a constant. The approximate law of evolution until the crossover is
\begin{eqnarray}&&\!\!\!\!\!\!\!
\dot{r}_{\alpha}=\sigma_{\alpha z}z,\ \ z=const, \ \ \langle r_{\perp}^2(t)\rangle=z^2 \kappa_{\alpha z \alpha z} t ,\ \ t\lesssim \lambda_1^{-1},\label{grd}
\end{eqnarray}
where $r_{\perp}^2(0)$ is assumed to have $\theta$ much less than $\theta^*$. On using ${\hat n}_{\alpha}{\hat n}_{\alpha}\approx r_{\perp}^2(t)/z^2$ the above reproduces diffusive growth of ${\hat n}_{\alpha}$. This evolution continues until the neglected last term in the complete equation $\dot{r}_{\alpha}=\sigma_{\alpha z}z+\sigma_{\alpha\beta} r_{\beta}$ becomes of the order of the first term. Subsequently, the growth of $\langle r_{\perp}^2(t)\rangle$ switches from diffusive growth as in Eq.~(\ref{grd}) to exponential growth at the rate $\sim \lambda_1$. We find by equating the two contributions to the rate of change of $r_{\perp}^2$ that, in accord with the considerations in terms of $\theta$, the crossover time $t_c$ obeys the estimates
\begin{eqnarray}&&\!\!\!\!\!\!\!
\frac{dr_{\perp}^2}{dt}\sim z^2 \kappa_{\alpha z \alpha z}+\lambda_1 r_{\perp}^2,\ \
z^2 \kappa_{\alpha z \alpha z}\sim \lambda_1 r_{\perp}^2(t_c),\ \ \ t_c \sim  \frac{1}{\lambda_1},\ \ \ r_{\perp}(t_c)\sim z \sqrt{\frac{\kappa_{\alpha z \alpha z}}{\lambda_1}}\sim \theta^* z.
\end{eqnarray}
The above results tell that at $t\gg \lambda_1^{-1}$ the evolution of $\bm r$ occurs horizontally starting from effective initial separation $\theta^* z$. It will be useful later to have a description of this observation via
the Jacobi matrix $W$ defined by
\begin{eqnarray}&&\!\!\!\!\!\!\!
\partial_t W=\sigma W,\ \ W_{ik}(0)=\delta_{ik},\ \ \bm r(t)=W(t)\bm r(0),\ \ \dot{\bm r}=\sigma\bm r. \label{defined}
\end{eqnarray}
The previous results for the evolution $[W(t){\hat z}]_i=W_{iz}(t)$ of initially vertical separation ${\hat z}$ can be written as
\begin{eqnarray}&&\!\!\!\!\!\!\!
W_{\alpha z}(t)\approx \theta^*\exp(\lambda_1 t)(\cos\theta(t), \sin \theta(t)),\ \ W_{zz}(t)\ll W_{\alpha z}(t);\ \ \ \ t\gg \lambda_1^{-1}, \label{properties}
\end{eqnarray}
where by isotropy of small-scale turbulence $\theta(t)$ is uniformly distributed over $[0, 2\pi]$.

We conclude that at small horizontal separations, the change in $\bm r$ consists of rotational diffusion. This is given by the diffusion of the two-dimensional vector $n_i$ obtained by dropping the last term in Eq.~(\ref{alp}). The steady-state distribution corresponding to diffusion is uniform so that $g(r, \theta)$ is constant at $\theta\ll \theta^*$. In contrast, $g(r, \theta)$ monotonously decreases with $\theta$ at $ \theta^*\lesssim \theta \leq \pi/2$.  Put differently, for the time $t^*(\bm r)$ in
Eq.~(\ref{time}), we have, for $\theta\ll \theta^*$,
\begin{eqnarray}&&\!\!\!\!\!\!\!\!\!\!\!\!\!
t^*(\bm r)\sim \frac{c'}{\lambda_1}+t^*(r, \theta^*)\approx t^*(r, \theta^*),
\end{eqnarray}
where the first term gives the time of reaching $\theta^*$ with $c'\sim 1$. The actual value of $c'$ is irrelevant due to $t^*(r, \theta^*)\gg c'/\lambda_1$. We find that $t^*(\bm r)$, and thus also $g(r, \theta)$, are independent of $\theta$ at $\theta\ll \theta^*$. This corresponds to asymptotic relation $g(r, \theta)\sim g(r, \theta^*)$ at $\theta\lesssim \theta^*$ or flattenning of $g(r, \theta)$ below $\theta^*$.
The matching of the asymptotic regions $\theta\ll \theta^*$ and $\theta\gg \theta^*$ is beyond the scope of this paper, since the region of $\theta\sim \theta^*$ is small and gives negligible contribution in the collision kernel which is our main interest. The region $\theta\sim \theta^*$ becomes relevant at larger $Fr$ though, where $\theta^*$ is not so small.


{\bf DNS and separability of the RDF}---The angle dependence of the RDF was shown to be separable, i.e. given by a product of functions of $r$ and $\theta$, at $\theta>\theta^*$, see Eq.~(\ref{eq:alpha}). We have also seen that $g(r, \theta)$ becomes constant at $\theta<\theta^*$. Thus it is plausible that the RDF is separable at all $\theta$. This is indeed confirmed by the DNS so that e.g. $g(r, \theta)/g(r, \pi/2)$ is $r-$independent. The angular dependence of $g(r, \theta)/g(r, \pi/2)$ at a fixed distance $r$ in the smoothness range is shown in the left panel of Fig. \ref{fig:angular}. The Figure also confirms that the dependence of the RDF on $\theta$ is $\sim \sin^{-\alpha} \theta$ (solid lines) for $\theta > \theta^*$, separable from $r$.

\begin{figure}
\includegraphics[width=17cm,trim={1mm 2mm 1mm 2mm},clip]{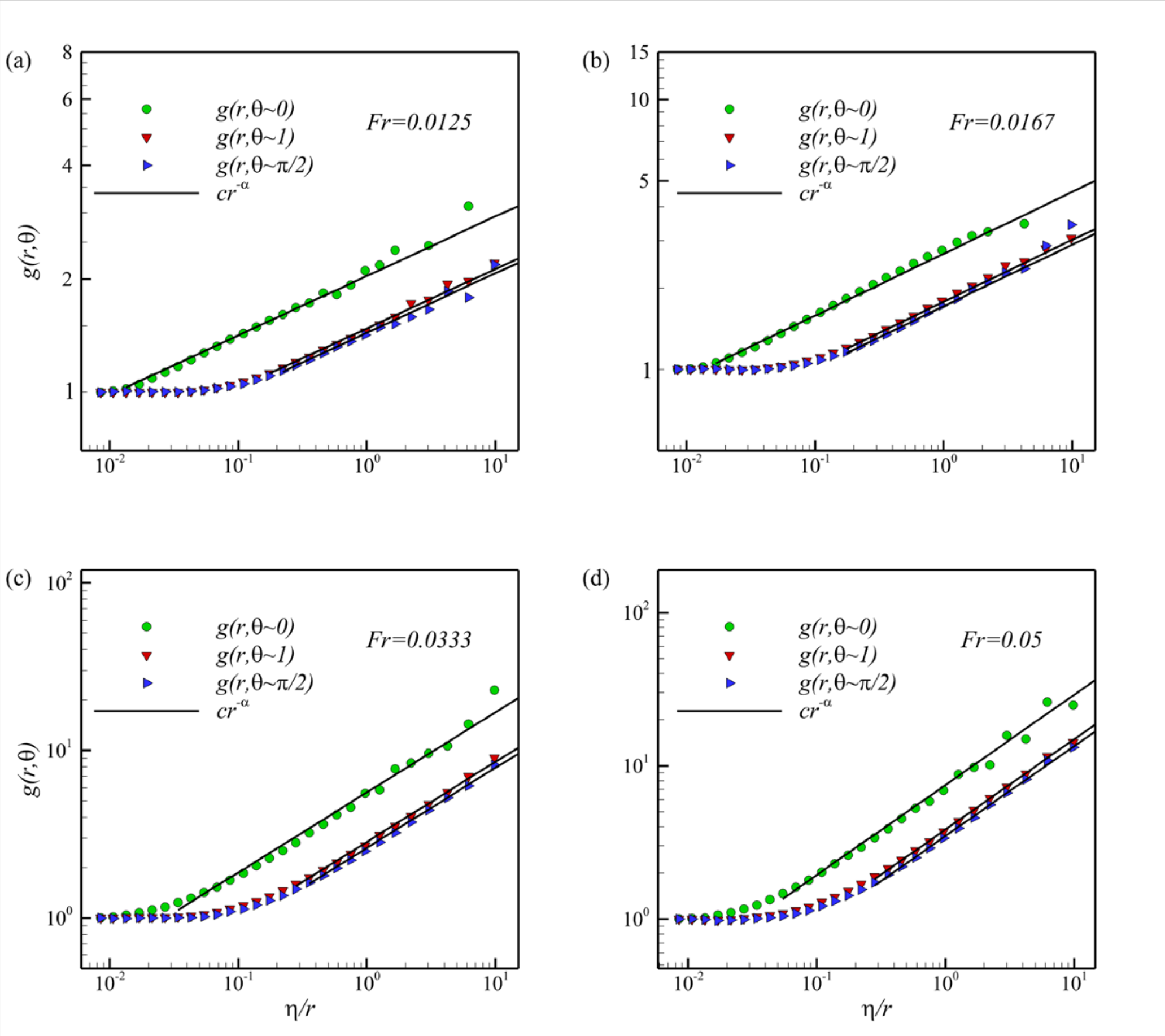}
\caption{Plots of $ g(r, \theta)$ versus $ r$ at representative angles $\theta=0,\ 1,\ \pi/2$ and (a) $Fr=0.0125$, (b) $Fr=0.0167$ , (c) $Fr=0.03$, and (d) $Fr=0.05$. For all the three angles at each given $Fr$, the plots contain linear pieces whose slopes coincide with the numerical accuracy. Similar properties hold if other values of $\theta$ are used. This allows us to conclude that $g(r, \theta)$ is $r^{-\alpha}$
times a function of $\theta$, which is confirmed directly in Fig. \ref{theta}. In contrast, the cutoff defined scale at which the power-law dependence on $r$ breaks down depends on $\theta$. The presented data and data for other angles confirm that the cutoff scale for $\theta>\theta^*$ is $l_c$, and for $\theta<\theta^*$, the scale is of order $g\tau^2$.}
\label{distance}
\end{figure}

\begin{figure}
\includegraphics[width=8.5cm,trim={2mm 2mm 2mm 2mm},clip]{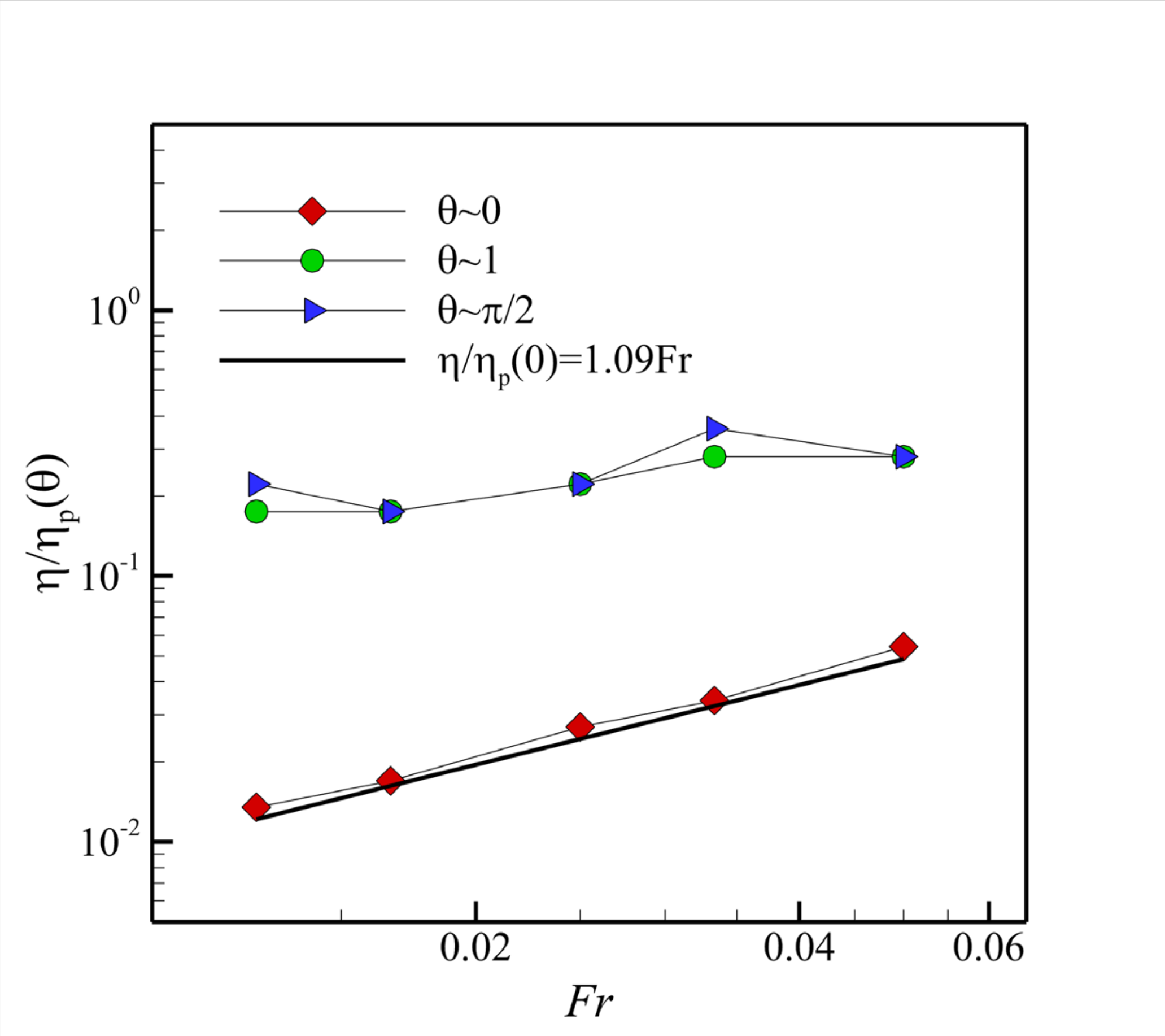}
\caption{The cutoff scale $\eta_p(\theta)$ in radial variable for $g(r, \theta)\propto r^{-\alpha}$ law. The cutoff is considered for $\theta=0$, $1$, and $\pi/2$. The behavior of the cutoff
scale agrees with the angular dependence of the correlation length of velocity gradients. For vertical separation, the cutoff scale is very close to $g\tau^2=\eta/Fr$ (we use $St=1$). For separations that are not close to the vertical, the correlation length $l_c$ does not depend much on $Fr$ and is about ten times larger than $\eta$. The asymptotic matching condition provided after Eq.~(\ref{product}) gives $l_c/(g\tau^2)\propto\theta^*\propto \sqrt{Fr}$ which agrees with the observations.  }
\label{cutoff}
\end{figure}

Further insight into the separability is attained by considering the dependence of the RDF on $r$ for three representative angles $\theta =0, 1$ and $\pi/2$.  The DNS results, shown in Fig. \ref{distance}, demonstrate clearly that the dependence of $g(r, \theta)$ on $r$ is $\sim r^{-\alpha}$ for all Froude numbers. The cutoff scale at which the power-law dependence on $r$ breaks down depends on $\theta$. We estimated the cutoff scale as the scale where the RDF starts to deviate from the power-law dependence by 10 percent of the RDF value. For $\theta\ll \theta^*$, the cutoff scale is found to be fit well by $g\tau^2$, see Fig. \ref{cutoff}. This scale can be understood by observing that it is the length that a particle sedimenting at speed $g\tau$ passes in its velocity relaxation time $\tau$. Thus the particle reacts to flow averaged in $z-$direction over the length $g\tau^2$ which must then provide the smoothness scale in the vertical direction. The cutoff scale $g\tau^2$ scales inversely proportional with $Fr$. In contrast for $\theta\gtrsim \theta^*$ the cutoff scale, as discussed previously, does not depend much on $Fr$. The scale is $\theta-$independent and is about ten times larger than $\eta$, i.e. of order $L_c$, as shown in Fig. \ref{cutoff}. Thus correlations exists over the same region as the correlations of the gradients of the particles' flow do, see subsection \ref{asgr} and cf. the general theory in \cite{itzhak1}.

\begin{figure}
\includegraphics[width=8.5cm,trim={2mm 2mm 2mm 2mm},clip]{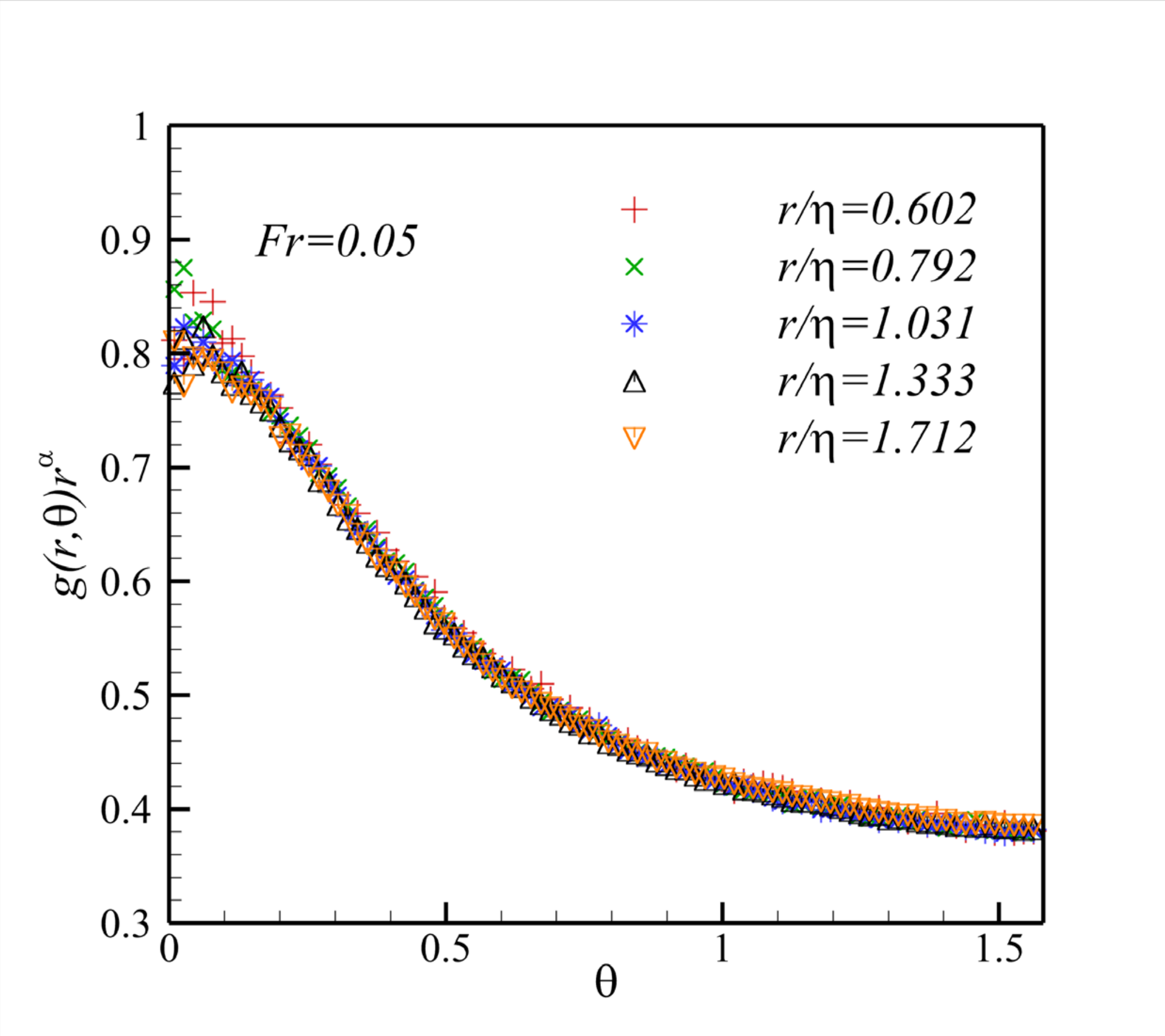}
\caption{Direct demonstration that $g(r, \theta)$ is the product of $r^{-\alpha}$ and a function of $\theta$ at $Fr = 0.05$. The existence of this function is demonstrated as a collapse of the curves $g(r, \theta)r^{-\alpha}$ for different $r$ onto a single curve that depends on $\theta$.}
\label{theta}
\end{figure}

Finally, the direct dependence of $g(r,\theta)r^\alpha$ on $\theta$ is demonstrated for $Fr=0.05$ in Fig. \ref{theta}, which clearly shows the collapse of $g(r,\theta) r^\alpha$ for five different $r$. Similar collapse was observed for the other Froude numbers as well. A juxtaposition of the above data produces the final result fitting all the data
\begin{eqnarray}&&\!\!\!\!\!\!\!\!\!\!\!\!\!
g(r, \theta)=r^{-\alpha} h(\theta);\ \ h(\theta)= \left(\frac{l_c}{\sin\theta}\right)^{\alpha},\ \ r<l_c=10 \eta, \ \ \theta\gg \theta^*;\ \ h(\theta)= \left(g\tau^2\right)^{\alpha},\ \ r<g\tau^2, \ \ \theta\ll \theta^*.
\label{product}
\end{eqnarray}
Here $l_c$ is of order of $L_c$ and $l_c=10 \eta$ is an empirical observation that holds at the considered range of $Fr$ with $\alpha\ll 1$. The refinements at $\alpha\sim 1$ are considered later. 

{\bf Estimate of $\theta^*$}---The asymptotic matching at $\theta\simeq \theta^*$ gives a possible definition of $\theta^*$ as
\begin{eqnarray}&&\!\!\!\!\!\!\!\!\!\!\!\!\!
\frac{1}{\sin^{\alpha}\theta^*}\equiv \frac{g(r, \theta=0)}{g(r, \theta=\pi/2)}=\left(\frac{g\tau^2}{l_c}\right)^{\alpha};\ \ \ \ \sin\theta^*=\frac{l_c}{g\tau^2}=\frac{Fr}{St^2} \frac{l_c}{\eta}, \label{thet}
\end{eqnarray}
see Eq.~(\ref{product}). Since $l_c=10 \eta$ then for $St=1$ the above formula implies that $\theta^*\ll 1$ demands quite small $10Fr\ll 1$. Our theory assumes $\theta^*\ll 1$ which is empirically observed to hold at $Fr\leq 0.033$ where $\theta^*=10 Fr/St^2$ can be considered as a small number. However, already $Fr=0.05$ gives a non-small $\theta^*$.

{\bf Sum rule and $\theta^*$}---As an example of the use of the above results we demonstrate that contribution of angles $\theta\lesssim \theta^*$ into the angle averaged RDF is the negligible as we demonstrated implicitly in Eq.~(\ref{RDFa}). The reason is that despite that the RDF is maximal at zero angle, the weight of these angles, proportional to $\sin\theta$ makes their contribution negligible. We have $\int_0^{\theta^*}h(\theta)\sin\theta d\theta\sim  \left(g\tau^2\right)^{\alpha}(\theta^*)^2$ which is much smaller than the contribution of $\theta>\theta^*$ given by $l_c^{\alpha}$ due to $\theta^*\sim l_c/(g\tau^2)\ll 1$ and $\alpha\ll 1$. We have from Eqs.~(\ref{sdb}) and (\ref{product}) that,
\begin{eqnarray}&&\!\!\!\!\!\!\!\!\!\!\!\!\!\!
r^{\alpha}{\tilde g}(r)=\int_0^{\theta^*}h(\theta)\sin\theta d\theta+\int_{\theta^*}^{\pi/2} h(\theta)\sin\theta d\theta\sim 
l_c^{\alpha} \int_{\theta^*}^{\pi/2}\sin^{1-\alpha}\theta d\theta\approx l_c^{\alpha}.
\end{eqnarray}
However for not too small $Fr$ and $\alpha$ the angles $\theta\lesssim \theta^*$ become relevant as measured by the deviation of $d\equiv \int_0^{\pi/2}\sin^{1-\alpha}\theta d\theta$ from $1$ shown in Fig.\ref{angles}.

\begin{figure}
	\includegraphics[width=17cm,trim={1mm 2mm 1mm 2mm},clip]{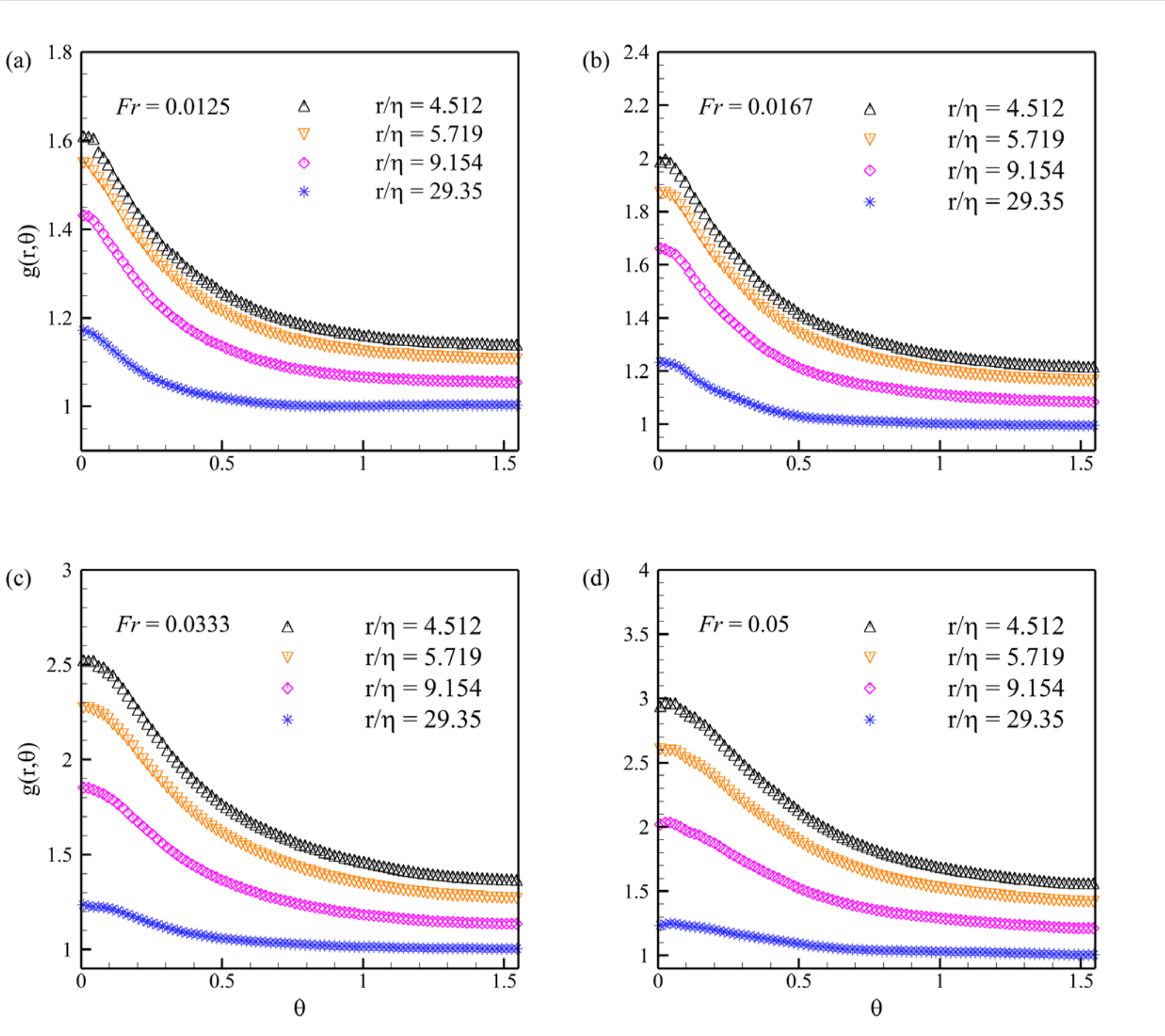}
	\caption{(Color) Angle dependence of the RDF in the inertial range at fixed $St=1$ and increasing $Fr$ of (a) $Fr=0.0125$, (b) $Fr=0.0167$, (c) $Fr=0.0333$, (d) $Fr=0.05$. }
	\label{pl}
\end{figure}

{\bf Correlations in the inertial range}---The correlations extend significantly into the inertial range for reasons explained previously, see Fig. \ref{pl}.



\section{Bidisperse RDF} \label{ask}

In this section, we consider the RDF $g_{12}(\bm r)$ of droplets of different radii $a_1$ and $a_2$ where $\Delta \tau=\tau_2-\tau_1$ is finite. Here, in contrast with the $\tau_1=\tau_2$ case, $g_{12}(0)$ is finite.
The reason is that particles at the same point have finite velocity difference $\bm g\Delta \tau$. Thus the trajectories $\bm x_i(t, 0)$ in Eq.~(\ref{reps}) diverge, making the integral finite.
The consideration in this section differs from \cite{nature} due to the anisotropy of the flow.

We demonstrate that position correlations are appreciable for particles with close sizes only. A straightforward generalization of Eq.~(\ref{time}) is (it is shown in a moment that $\sum_{i=1}^3\lambda_i$ of the flow with any $\tau_i$ can be used here)
\begin{eqnarray}&&\!\!\!\!\!\!\!\!\!\!\!\!\!
\ln g_{12}(\bm r)=2 \left|\sum_{i=1}^3\lambda_i\right| t_{12}^*(\bm r), \label{dps}
\end{eqnarray}
where we introduce the average time $t_{12}^*(\bm r)$ that the distance $\bm r(t)\equiv \bm x_2(t, \bm r)-\bm x_1(t, 0)$ spends inside the range of correlations of $\nabla\cdot \bm v_i$. Since $g_{12}(\bm r)$ is maximal at zero then we can use $g_{12}(0)-1$ as the measure of preferential concentration: if $g_{12}(0)-1\ll 1$ then there is no preferential concentration and vice versa. We observe that due to the difference in sedimentation velocities $g|\Delta \tau|$, the correlation length $g\tau^2$ of $\nabla\cdot\bm v$ in the vertical direction is passed in time $\tau^2/|\Delta \tau|$. Therefore, $t_{12}^*(0)$ cannot be larger than $\tau^2/|\Delta \tau|$. We find from Eq.~(\ref{dps}) that deviations of $g_{12}(r)$ from one can be appreciable only for $|\Delta \tau|$ obeying $\left|\sum\lambda_i\right| \tau^2/|\Delta \tau|\gtrsim 1$. However, it was demonstrated in \cite{fi2015} that $\left|\sum\lambda_i\right| \tau\propto Fr^2$. We conclude that the deviations from $g_{12}(\bm r)\approx 1$ can be appreciable only for $|\Delta \tau|/\tau\lesssim Fr^2\ll 1$. For $|\Delta \tau|\gtrsim \tau_1$ the attractors of the particles with relaxation times $\tau_1$ and $\tau_2$ are significantly displaced with respect to each other so that there are no positive correlations in the particles positions.

The above implies that we can limit the consideration to the case of $|\Delta \tau|\ll \tau_1$. Thus we can use either of $\tau_1$ or $\tau_2$ for both $\sum_{i=1}^3\lambda_i$ and the correlations' range above, or use their symmetrized versions.  We have from Eq.~(\ref{veldif}) where the half sum of the flow gradients can be taken as $\sigma\equiv \nabla\bm v_1$
\begin{eqnarray}&&\!\!\!\!\!\!\!\!\!\!\!\!\!
\dot{\bm r}=\sigma \bm r+\bm g\Delta \tau.
\end{eqnarray}
The solution can be written with the help of the Jacobi matrix defined in Eq.~(\ref{defined}) as
\begin{eqnarray}&&\!\!\!\!\!\!\!\!\!\!\!\!\!
 \bm r(t)=W(t)\bm r(0)-g\Delta \tau\int_0^t W(t)W^{-1}(t'){\hat z} dt'. \label{si}
\end{eqnarray}
We observe that $W(t)W^{-1}(t')$ is the matrix that describes the evolution of distances from time $t'$ to $t$: for any vector $\bm f$, the vector $\bm f(t)\equiv W(t)W^{-1}(t')\bm f$ obeys $\dot{\bm f}=\sigma\bm f$ with $\bm f(t')=\bm f$. Thus, considering times larger than $\lambda_1^{-1}$, we can use Eq.~(\ref{properties}) for $W(t)W^{-1}(t'){\hat z}$ with $t-t'$ instead of $t$. For the horizontal component, using that $\lambda_1\tau=Fr l_c/(10\eta)$, see Eq.~(\ref{lau}), we find that
\begin{eqnarray}&&\!\!\!\!\!\!\!
g\Delta \tau\left(\int_0^t W(t)W^{-1}(t'){\hat z} dt'\right)_{\perp}\sim r_c\exp\left(\lambda_1 t\right);\ \ \ r_c\equiv \frac{g|\Delta \tau|\theta^*}{\lambda_1}\sim \frac{ St^2}{Fr^{3/2}}\frac{|\Delta \tau|}{\tau}\frac{10\eta^2}{l_c}, \label{rc}
\end{eqnarray}
where the vertical component is much smaller and we introduced the crossover scale $r_c$. The horizontal component of this contribution to $\bm r(t)$ becomes of the order $l_c$ in time $t_{12}^*(0)\approx \lambda_1^{-1} \ln(l_c/r_c)$ as found by setting $\bm r(0)=0$ in Eq.~(\ref{si}). This gives
\begin{eqnarray}&&\!\!\!\!\!\!\!\!\!\!\!\!\!
g_{12}(0)=\left(\frac{l_c}{r_c}\right)^{\alpha},  \label{sd}
\end{eqnarray}
see Eqs.~(\ref{eq:alpha}) and (\ref{dps}). The order one indeterminacy in the definitions of the scales $l_c$ and $r_c$ is irrelevant by $\alpha\ll 1$, cf. previous considerations.
We consider the condition $g_{12}(0)>\kappa$  for having appreciable preferential concentration of different size particles. Here we introduced the threshold $\kappa>1$ defined by the condition that the preferential concentration in the case of interest is considered negligible for $g_{12}(0)<\kappa$. The condition $g_{12}(0)>\kappa$ gives
\begin{eqnarray}&&\!\!\!\!\!\!\!
\frac{|\Delta \tau|}{\tau}\lesssim  \frac{l_c^2 Fr^{3/2}}{10\eta^2 St^2}\kappa^{-1/\alpha}, \label{thres}
\end{eqnarray}
where $l_c/\eta\sim 10$, cf. with the less detailed condition in the beginning of the Section. We see that since the condition of applicability of both flow description and white noise approximation, given by Eq.~(\ref{lv}), implies that $(l_c/\eta)Fr/St^2\ll 1$ then the above condition implies $|\Delta \tau|/\tau\ll 1$.

We find from Eq.~(\ref{si}) that $\bm r(t)$ grows as $r(t)\sim \max[r(0), r_c]\exp(\lambda_1 t)$. In other words finite difference of relaxation times creates the minimum separation time $\lambda_1^{-1} \ln(l_c/r_c)$ so that
$t_{12}^*(\bm r)$ is the minimum of $\lambda_1^{-1} \ln(\l_c/r_c)$ and the separation time for $\tau_1=\tau_2$. We find from Eq.~(\ref{dps}) that  
\begin{eqnarray}&&\!\!\!\!\!\!\!\!\!\!\!\!\!
g_{12}(\bm r)=\left(\frac{l_c}{r \sin\theta+r_c}\right)^{\alpha},\ \ \theta\gg \theta^*;\ \ \ g_{12}(\bm r)=\left(\frac{l_c}{r \sin\theta^*+r_c}\right)^{\alpha},\ \ \theta\lesssim \theta^*, \label{ds}
\end{eqnarray}
where the precise way of matching at $r\sin\theta\sim r_c$ (or $r\sin\theta^*\sim r_c$ at smaller $\theta$) is irrelevant. This inaccuracy or the inherent inaccuracy in the definition of $r_c$ (which is defined up to a factor of order one) does not change the answer appreciably since $\alpha\ll 1$, cf. above and similar formula in \cite{nature}. However, the extension of this theory to higher $Fr$ and $\alpha$ might demand refinement of the matching. We have the interpolation formula
\begin{eqnarray}&&\!\!\!\!\!\!\!\!\!\!\!\!\!
g_{12}(\bm r)=\left(\frac{l_c}{r \sin(\theta+\theta^*)+r_c}\right)^{\alpha}.
\end{eqnarray}
We see that the difference in $\tau_i$ produces effective smoothening of the multifractal at the angle-dependent scale $r_c/\sin(\theta+\theta^*)$. At this scale, for each $\theta$, the power-law in $r$ stops at decreasing $r$. This smoothening scale appears also in the theory at small $St$ and no gravity \cite{nature} however here it depends on the angle due to the anisotropy. This scale characterizes the spatial displacement of the random attractors of droplets with radii $a_1$ and $a_2$ due to the difference in their inertia, cf. \cite{chun,mei,cals}. For the observability of the multifractal at all, it is necessary that the smoothness scale is much
smaller than the correlation length of the steady state fluctuations of the particles concentration. It is readily seen that for $\theta\gg \theta^*$ this gives the condition $r_c\ll l_c$ and for vertical direction $\theta\lesssim \theta^*$ it gives the same condition. The condition $r_c\ll l_c$ is weaker that the condition of significant preferential concentration given by Eq.~(\ref{thres}). We remark that for having $g_{12}(\bm r)$ symmetric in $\tau_i$, as it must, we can symmetrize the exponent $\alpha$ in Eq.~(\ref{ds}) with respect to $\tau_i$ using $(\alpha_1+\alpha_2)/2$. This does not change the answer appreciably in the considered limit.

We sum up the picture of separation of two particles of different sizes implied by the above study. For initial separations larger than $r_c$, the separation coincides with that of particles with identical size. For initial separations much smaller than $r_c$, the separation occurs in two stages. In the first stage occurring at times smaller than $\lambda_1^{-1}$, we have linear growth of separation with time according to $\bm r(t)=\bm r(0)+\Delta \tau\bm g t$; see Eq.~(\ref{si}). This growth continues until $t\sim\lambda_1^{-1}$ when $r(t)$ becomes of order $r_c$. Then, after a transient whose duration is of order $\lambda_1^{-1}$, the growth becomes exponential with an exponent of $\lambda_1$. Finally, we stress that $g_{12}(0)$ is finite. It diverges at coinciding particle sizes according to $g_{12}(0)\sim |\Delta\tau|^{-\alpha}$, see Eq.~(\ref{sd}).

\section{Geometric collision kernel at $Fr\to 0$: reduction to RDF} \label{RDf}

In this section, we derive the formula for the geometric collision kernel $\gamma_{12}$ that describes geometrical cross section of the droplets' collisions. This kernel is defined as the rate per unit volume at which the droplets with radii $a_1$ and $a_2$ approach each other by the distance $a_1+a_2$, if the hydrodynamic interactions occurring at small separations are neglected. The passage to the physical rate of collisions is realized with the help of the collision efficiency, see e.g. \cite{nature}. The formula for a poly-disperse solution is then obtained by integrating over $a_1$ and $a_2$, see \cite{st}.

Here we provide the formula for $\gamma_{12}$ in terms of statistical properties of turbulence and the RDF. The reduction to the RDF holds thanks to the existence of the particles' flow. There are simplifications in comparison with the case of incompressible flow of particles, see \cite{st,Wang20081,Wang20082}. Thanks to Gaussianity of the gradients of the droplets' flow the reduction does not involve unknown factors.

We demonstrate in Appendix \ref{kernel} that concentration fields $n_i$ in the well-known formula \cite{st},
\begin{eqnarray}&&\!\!\!\!\!\!\!\!\!\!\!\!\!
\gamma_{12}
\!=\!-\int_{w_r<0, r=a_1+a_2}  \langle n_2(0) n_1(\bm r)w_r(\bm r)\rangle dS;\ \ w_r(\bm r)\equiv \left(\bm v_1(\bm r)-\bm v_2(0)\right)\cdot\bm r/r, \label{rate}
\end{eqnarray}
can be considered as exact microscopic concentrations, given by sums of $\delta-$functions, without employing coarse-graining. Here $w_r$ is the radial component of the velocity difference of colliding droplets (hereafter "collision" refers to approach by the distance $a_1+a_2$). The integral in Eq.~(\ref{rate}) is taken over the surface of the ball of radius $a_1+a_2$ centered at the origin, and the angular brackets stand for the spatial averaging, which we assume to be equal to the average over the statistics of velocity by ergodicity (strictly speaking space average includes about $(L/l_c)^3$ independent realizations of the averaged random variable where $L$ is the linear size of the system. This might leave some inaccuracies which can be removed by further averaging over time, cf. \cite{lee} and references therein). Changing the integration variable in Eq.~(\ref{rate}) from $\bm r$ to $-\bm r$ the symmetry $\gamma_{12}=\gamma_{21}$ is verified. In the case where the fluctuations of concentration are negligible, Eq.~(\ref{rate}) reduces to the formula of \cite{st}, that describes the total inward flux of particles with radius $a_1$ through the sphere of collision with radius $a_1+a_2$ in unit volume.

We simplify Eq.~(\ref{rate}) by using the fact that $n_i$ obeys the continuity equation
\begin{eqnarray}&&\!\!\!\!\!\!\!\!\!\!\!\!\!
\partial_tn_i+\nabla\cdot(n_i\bm v_i)=0,
\end{eqnarray}
where $i=1, 2$. In the steady state (we make standard Smoluchowski-type assumption that all collisions occur in the quasi-steady state, cf. however \cite{mo}), this equation implies the stationarity condition
\begin{eqnarray}&&\!\!\!\!\!\!\!\!\!\!\!\!\!
0\!=\!-\partial_t\langle n_1(t, \bm x)n_2(t, \bm x')\rangle\!=\!\nabla\! \cdot\! \langle n_1n_2 \bm v_1\rangle\!+\!\nabla'\! \cdot\! \langle n_1n_2 \bm v_2\rangle=\nabla \cdot \langle n_1(\bm x)n_2(\bm x')\left(\bm v_1(\bm x)-\bm v_2(\bm x')\right)\rangle,
\end{eqnarray}
where we used spatial homogeneity of the statistics. In this equation, the nabla operators $\nabla$ and $\nabla'$ stand for derivatives over $\bm x$ and $\bm x'$, respectively. For identical size particles, this identity is the counterpart of the Monin-Yaglom relation of the passive scalar turbulence \cite{reviewt}. By taking $\bm x=\bm r$ and $\bm x'=0$, we find that
\begin{eqnarray}&&\!\!\!\!\!\!\!\!\!\!\!\!\!
\nabla \cdot \langle n_2(0) n_1(\bm r) \left(\bm v_1(\bm r)-\bm v_2(0)\right)\rangle=0.
\end{eqnarray}
Finally, by integrating over the volume of the ball with radius $a_1+a_2$ and using the divergence theorem, we find that
\begin{eqnarray}&&\!\!\!\!\!\!\!\!\!\!\!\!\!
\int  \langle n_2(0) n_1(\bm r)w_r\rangle dS=0.
\end{eqnarray}
Thus, we can write Eq.~(\ref{rate}) in the form
\begin{eqnarray}&&\!\!\!\!\!\!\!\!\!\!\!\!\!
\gamma_{12}=\frac{1}{2}\int_{r=a_1+a_2}  dS \langle n_2(0) n_1(\bm r)\left| w_r\right|\rangle .\label{rat}
\end{eqnarray}
The derivation of a similar identity in the incompressible isotropic case was done in \cite{st}. This identity is useful because averaging conditioned on the sign of $w_r$ is more difficult.
When there is no flow description of the motion, Eqs.~(\ref{rate}) and (\ref{rat}) differ.

\subsection{Monodisperse case}

First, we consider the mono-disperse case, $a_1=a_2=a$, where we designate $n_1=n_2=n$, $\bm v_1=\bm v_2=\bm v$. Using for $w_r$ in Eq.~(\ref{rat}), the approximation $w_r\approx 2a{\hat r}\sigma{\hat r}$ holding because of $r\ll \eta$ with $\sigma_{ik}=\nabla_k v_i(0)$, we have
\begin{eqnarray}&&\!\!\!\!\!\!\!\!\!\!\!\!\!
\gamma_{11}\!=\!a\int_{r=2a}  dS \langle n(0) n(\bm r)\left|{\hat r}\sigma{\hat r}\right|\rangle.\label{gammaf}
\end{eqnarray}
We observe that there is separation of the time-scales of variations of the concentration and velocity gradients due to which we can perform independent averaging
over $n(0) n(\bm r)$ and $\left|{\hat r}\sigma{\hat r}\right|$ in Eq.~(\ref{gammaf}). Indeed, the characteristic time of the variations of the concentration is the inverse of the modulus of the sum of Lyapunov exponents
$|\sum\lambda_i|$. The correlation time of $\sigma$ is $\tau$, see Eq.~(\ref{lin}), which is smaller than
$|\sum\lambda_i|^{-1}$ by a factor of $Fr$; see \cite{fi2015}. Thus, conditioning on an instantaneous value of $\sigma$ does not change much the distribution of $n(0)n(\bm r)$ and these quantities can be considered as independent giving,
\begin{eqnarray}&&\!\!\!\!\!\!\!\!\!\!\!\!\!
\gamma_{11}\!=\!a\int_{r=2a}  dS \langle n(0) n(\bm r)\rangle\langle \left|{\hat r}\sigma{\hat r}\right|\rangle, \label{go}
\end{eqnarray}
cf. \cite{nature}. We can readily study $\langle \left|{\hat r}\sigma{\hat r}\right|\rangle$ using the Gaussianity of $\sigma$ considered in the previous section. Using the average of the modulus of the Gaussian random variable with zero mean, we have from Eq.~(\ref{d01}) with $\sigma=\nabla \bm v$,
\begin{eqnarray}&&\!\!\!\!\!\!\!\!\!\!\!\!\!
\langle \left({\hat r}\sigma{\hat r}\right)^2\rangle\!=\!\frac{c_0Fr \sin^4\theta}{2\tau^2},\ \ \langle |{\hat r}\sigma{\hat r}|\rangle\!=\!\sqrt{\frac{c_0Fr}{\pi\tau^2}}\sin^2\theta,\label{velt}
\end{eqnarray}
where $\theta$ is the polar angle. This formula is only valid for separation vectors that are not too close to the vertical. Indeed, for vertical separations, the leading order approximation of the flow gradients matrix
by the horizontal components gives zero (see Eq.~(\ref{velt}) at $\theta=0$), necessitating higher order corrections. It is demonstrated in the following section on the angular structure of the radial distribution function that Eq.~(\ref{velt}) holds for $\theta$ much larger than the cutoff angle $\theta^*\ll 1$ (of order $\sqrt{Fr}$) and breaks down at $\theta\lesssim \theta^*$, cf. Eq.~(\ref{eq:alpha}) and \cite{fi2015}.  However, the contribution of angles smaller than $\theta^*$ in the integral in Eq.~(\ref{go}) is negligible. The reason is that the increase of $\langle n(0) n(\bm r)\rangle$ at small $\theta$, derived below, does not compensate for the smallness of the considered region of integration, see the sum rule in the previous section, and velocity difference at $\theta\lesssim \theta^*$ is depleted. Thus $\gamma_{11}$ can be approximately calculated with the help of Eq.~(\ref{velt}) which effectively gives a complete description of the dependence of $\gamma$ on the velocity of the collision.

\subsection{Bidisperse case} \label{fdji}

To obtain the velocity difference of colliding droplets with different sizes, we consider the difference in velocity fields at nearby points separated by $\bm r$, where $r\ll \eta_p$,
\begin{eqnarray}&&\!\!\!\!\!\!\!\!\!\!\!\!\!
\bm v_1(\bm x\!+\!\bm r)\!-\!\bm v_2(\bm x)\!\approx\! (\bm r\!\cdot\! \nabla)[\bm v_1(\bm x)\!+\!\bm v_2(\bm x)]/2
\!+\!\bm g\Delta \tau,\label{veldif}
\end{eqnarray}
where $\Delta \tau=\tau_1-\tau_2$ and we used symmetrized version of Eq.~(\ref{smoothnes}). The decomposition shows that the velocity difference of colliding particles is caused by both the difference in spatial positions at the time of the collision and the difference in flows of different-size particles $\bm g\Delta \tau$. We have using independence of concentration and velocity difference in Eq.~(\ref{rat}),
\begin{eqnarray}&&\!\!\!\!\!\!\!\!\!\!\!\!\!
\gamma_{12}\!=\!\frac{1}{2}\int_{r=a_1+a_2}\delta v_{12}(\theta)\langle n_2(0) n_1(\bm r)\rangle dS,\ \ \delta v_{12}\!=\!\langle\left|y\right|\rangle,\ \
y\!=\!(a_1\!+\!a_2){\hat r}\sigma^s {\hat r}\!-\!g\Delta \tau\cos\theta, \label{cold1}
\end{eqnarray}
where $\sigma^s$ is a symmetrized matrix $(\sigma_1+\sigma_2)/2$ and $\sigma_i$ is the matrix of the derivatives of $\bm v_i$. Since the sedimentation velocity of the particle is assumed to be much larger than the Kolmogorov scale velocity then it is readily seen that gravity dominates the radial components of the velocity difference: $y\!\approx\bm -g\Delta \tau\cos\theta$ at not too small $|\Delta \tau|/\tau$. Indeed, we observe from Eq.~(\ref{velt}) that the first term in $y$ is of order $(a_1\!+\!a_2)\sqrt{Fr}/\tau$ (we use Kolmogorov theory in the estimates otherwise there is a power of $Re_{\lambda}$). Thus, we have
\begin{eqnarray}&&\!\!\!\!\!\!\!\!\!\!\!\!\!
y\approx -g\Delta \tau\cos\theta;\ \ \ \frac{|\Delta \tau \cos\theta|}{\tau}\gg \epsilon_0,\ \ \epsilon_0\equiv \frac{(a_1\!+\!a_2)\sqrt{Fr}}{g\tau^2}=\left(\frac{a_1\!+\!a_2}{\eta}\right)\frac{Fr^{3/2}}{St^2}\ll 1. \label{lrg}
\end{eqnarray}
Thus, in the first term in $y$, we can assume that $|\Delta \tau|/\tau \lesssim \epsilon_0\ll 1$ and set $\tau_1=\tau_2$ (the angles with very small $\cos\theta$ give negligible contribution into the integral in Eq.~(\ref{cold1}) and can be neglected).
Using the previously proved Gaussianity of $\nabla\bm v$, we find that $y$ can be considered as a Gaussian random variable with non-zero mean. The averaging of its modulus gives
\begin{eqnarray}&&\!\!\!\!\!\!\!\!\!\!\!\!\!
\delta v_{12}(\theta)=\exp\left(-\frac{(g\Delta \tau \cos\theta)^2 \tau_1 \tau_2 }{c_0Fr(a_1+a_2)^2\sin^4\theta}\right)\sqrt{\frac{c_0 Fr}{\pi\tau_1\tau_2}}(a_1+a_2)\sin^2\theta
+er\!f\left(\frac{g\Delta \tau \cos\theta\sqrt{\tau_1\tau_2}}{\sqrt{c_0Fr}(a_1+a_2)\sin^2\theta}\right)g\Delta \tau\cos\theta , \label{modl}
\end{eqnarray}
where $er\!f(x)$ is the error function, see the cumbersome calculation in Appendix \ref{ash}. The symmetry $\delta v_{12}(\theta)=\delta v_{21}(\pi-\theta)$ guarantees $\gamma_{12}=\gamma_{21}$. We have for equal size particles and for velocity difference $\delta v_{12}^0(\theta)$ at $g=0$, respectively,
\begin{eqnarray}&&\!\!\!\!\!\!\!\!\!\!\!\!\!
\delta v_{11}(\theta)=2a_1\sqrt{\frac{c_0 Fr}{\pi\tau_1^2}}\sin^2\theta, \ \ \delta v_{12}^0(\theta)=\sqrt{\frac{c_0 Fr}{\pi\tau_1\tau_2}}(a_1+a_2)\sin^2\theta, \label{modl1}
\end{eqnarray}
where $\delta v_{11}(\theta)$ reduces to Eq.~(\ref{velt}) on observing that $\delta v(\theta)=2a\langle |{\hat r}\sigma{\hat r}|\rangle$. We can rewrite Eq.~(\ref{modl}) more compactly
\begin{eqnarray}&&\!\!\!\!\!\!\!\!\!\!\!\!\!
\delta v_{12}(\theta)=\exp\left(-\frac{(g\Delta \tau \cos\theta)^2 }{\pi(\delta v_{12}^0(\theta))^2}\right)\delta v_{12}^0(\theta)
+ er\!f\left(\frac{g\Delta \tau \cos\theta}{\sqrt{\pi}\delta v_{12}^0(\theta)}\right)g\Delta \tau\cos\theta. \label{modl11}
\end{eqnarray}
The exponent in the first term above decays very rapidly with the growth of $|\Delta\tau|$, effectively becoming non-zero only for a very narrow vicinity of $\theta=\pi/2$. Similarly, the error function becomes equal to the sign of $\Delta \tau\cos\theta$ for not too small $\Delta \tau\cos\theta$. We find refinement of Eq.~(\ref{lrg}),
\begin{eqnarray}&&\!\!\!\!\!\!\!\!\!\!\!\!\!
\delta v_{12}(\theta)\approx \delta v_{12}^0(\theta)+\frac{\left(g\Delta \tau\cos\theta\right)^2}{\pi \delta v_{12}^0(\theta)} ,\  \ \frac{\left|g\Delta \tau\cos\theta\right|}{\delta v_{12}^0}\ll 1;\ \
\delta v_{12}(\theta)\approx \left|g\Delta \tau\cos\theta\right|,\ \ \frac{\left|g\Delta \tau\cos\theta\right|}{\delta v_{12}^0}\gg 1,
\end{eqnarray}
where we performed in Eq.~(\ref{modl1}) expansion of the exponent in the first term and of the error function in the second term using that $er\!f(x)\approx 2x/\sqrt{\pi}$ at small $|x|$. We observe from the first term that the leading order correction due to the difference of settling velocities of droplets of different sizes is quadratic in $g$. The asymptotic forms match at
$\left|g\Delta \tau\cos\theta\right|\sim \delta v_{12}^0$ and roughly we have $\delta v_{12}(\theta)=\max[\delta v_{12}^0(\theta), \left|g\Delta \tau\cos\theta\right|]$, cf. Eq.~(\ref{cold1}).

The correlation function of concentration $\langle n_1(0) n_2(\bm r)\rangle$ is equivalent to the RDF $g_{12}(\bm r)$ by $\langle n_1(0) n_2(\bm r)\rangle=\langle  n_1\rangle \langle n_2\rangle g_{12}(\bm r)$, see Appendix \ref{kernel}. Thus we can rewrite Eq.~(\ref{cold1}) as ($\gamma_{12}=\gamma_{21}$)
\begin{eqnarray}&&\!\!\!\!\!\!\!\!\!\!\!\!\!
\gamma_{12}\!=\!\frac{1}{2}\int_{r=a_1+a_2}\delta v_{21}(\theta)\langle n_1(0) n_2(\bm r)\rangle dS\!=\!\pi(a_1+a_2)^2\langle  n_1\rangle \langle n_2\rangle
\int_{0}^{\pi}\delta v_{21}(\theta)g_{12}(a_1+a_2, \theta) \sin\theta d\theta. \label{cold}
\end{eqnarray}
The increase of the collision rate due to the multifractality of the distribution is described by positivity of $g(\bm r)-1$ at small $r$. If the droplets were distributed in space uniformly, then $g(\bm r)$ would be one. In fact, this value holds for separations in the inertial range, $g_{12}(\bm r)\approx 1$  at $r\gtrsim \eta_p(\theta)$, because the droplets are effectively independent at these scales, cf. Fig. \ref{angles}. However, for collisions, scales $a_1+a_2$ that are much smaller than $\eta_p$ are relevant, increasing the rate of collisions by weighted integral of $g_{12}(a_1+a_2, \theta)$ over $\theta$. For the same size particles the $g$ factor describes preferential concentration on the same random multifractal attractor. For different size particles $g$ describes decay of the preferential concentration since particles are on different attractors in space the distance between which increases with size difference, cf. \cite{cals}.


\section{Collision kernel formula}

In the previous sections, we have derived the limiting $Fr\to 0$ form of all quantities that enter the collision kernel under the condition of validity of the flow description. This description applies at $Fr\leq 0.033$, however it does not apply at $Fr=0.05$ (so that the upper limit of applicability is between $0.033$ and $0.05$). The obtained formulae can be readily used to write down the collision kernel.  Here, as an illustration, we calculate the collision kernel for equal size particles and demonstrate that collision rate is formed by purely horizontal collisions.

We have from Eq.~(\ref{cold}) for $a_1=a_2=a$ and $\delta v(\theta)=2a\langle |{\hat r}\sigma{\hat r}|\rangle$ (see remark after Eq.~(\ref{modl1})) that collision kernel of equal size droplets with radius $a$ obeys
\begin{eqnarray}&&\!\!\!\!\!\!\!\!\!\!\!\!\!
\gamma\!=\!16\pi a^3\langle  n\rangle^2  (2a)^{-\alpha}
\int_{0}^{\pi/2}\langle |{\hat r}\sigma{\hat r}|\rangle h(\theta) \sin\theta d\theta, \label{cols}
\end{eqnarray}
where we used Eq.~(\ref{product}). We saw in subsection \ref{sumr} that $\int_{0}^{\pi/2} h(\theta) \sin\theta d\theta$ is determined by angles larger than $\theta^*$. The velocity difference $2a\langle |{\hat r}\sigma{\hat r}|\rangle$ further decreases the role of $\theta\lesssim \theta^*$, since at these angles the velocity is predominantly tangential, see Sec. \ref{angl}, and more generally the velocity difference is an increasing function of $\theta$, cf. Eq.~(\ref{velt}). Thus in the integral in Eq.~(\ref{cols}) we can neglect angles $\theta\lesssim \theta^*$. We find using Eqs.~(\ref{velt}), (\ref{d01}) and (\ref{product}) that
\begin{eqnarray}&&\!\!\!\!\!\!\!\!\!\!\!\!\!
\gamma_{11}=16 \pi a^3 \langle n\rangle^2\left(\frac{l_c}{2a}\right)^{\alpha}\sqrt{\frac{c_0Fr}{\pi\tau^2}}\int_0^{\pi/2}\sin^{3-\alpha}d\theta \approx \frac{16\pi a^3 \langle n\rangle^2}{3\tau}\left(\frac{l_c}{2a}\right)^{\alpha }\sqrt{\frac{\int E(k)kdk}{2g}}=\frac{16\pi a^3 \langle n\rangle^2}{3\tau}\left(\frac{l_c}{2a}\right)^{\alpha }\sqrt{\frac{\alpha}{3\pi}}, \label{csosas}
\end{eqnarray}
where we used the definition of the Froude number and $\alpha\ll 1$. Thus $2^{-\alpha}\approx 1$ (see discussion in subsection \ref{sumr}); factor of two in the denominator is kept for transparency and asymptotic continuation to non-small $\alpha$. Since the contribution of $\theta\lesssim \theta^*$ in the collision kernel is small then we conclude that collision angles are predominantly much larger than $\theta^*$, which by the results of Sec. \ref{angl} implies that the collisions occur horizontally. This result is intuitive since the particles' separation changes mainly horizontally.

The collision kernel provided by Eq.~(\ref{csosas}), for particles of fixed size (and thus fixed $\tau$), depends on $Fr$ via the preferential concentration $\propto(l_c/2a)^{\alpha}$ and change of collision velocity $\sqrt{\alpha}$. Remarkably both dependencies reduce to one parameter $\alpha$, which is a non-trivial prediction that could hardly be obtained otherwise, cf. \cite{bec2014} (remind that the dependence of $l_c$ on $Fr$ is weak in the considered range of $Fr$, see Fig. \ref{cutoff}). For considering the numerical values of $\gamma$ we observe that the Kolmogorov scale \cite{frisch} for air is a single-valued function of $Fr$,
\begin{eqnarray}&&\!\!\!\!\!\!\!\!\!\!\!\!\!
\frac{1}{\eta^3}=\frac{\epsilon^{3/4}}{\nu^{9/4}}=\frac{Fr g}{\nu^2};\ \ \eta= Fr^{-1/3} 280 \mu m,
\end{eqnarray}
where we used the air viscosity $\nu=0.15\  cm^2/s$, e. g. $\eta=760  \mu m$ for $Fr=0.05$ and $\eta=873  \mu m$ for $Fr=0.033$. We find using that $l_c$ is order of magnitude larger than $\eta$ that $l_c/2a\sim 100$. Correspondingly the values of the product $(l_c/2a)^{\alpha}\sqrt{\alpha}$ can be quite large. Using as a reference the collision kernel of \cite{st} for tracers $\gamma_t$ we have for the parameter dependence of the ratio,
\begin{eqnarray}&&\!\!\!\!\!\!\!\!\!\!\!\!\!
\frac{\gamma}{\gamma_t}\!\sim \!\frac{1}{St} \left(\frac{l_c}{2a}\right)^{\alpha }\sqrt{\alpha}.
\end{eqnarray}
We find at $St=1$ (where the tracers' kernel must be understood as no more as a reasonable unit for measuring the collisions rate) and $a=50 \mu m$ that for $Fr=0.033$ the ratio is larger than three, see \cite{bec2014} for other information on the collision kernel. The collision kernel formula, given by Eq.~(\ref{csosas}), can be used at $Fr\leq 0.033$. Our previous considerations indicate that this formula applies at $Fr=0.05$ provided $St\geq 1$. Finally the formula can be interpolated to higher $Fr$, possibly demanding higher $St$ for its validity, which is left for future work.
%
%
%
%

\section{Facing the high Reynolds number limit} \label{face}

We demonstrated that increase of Reynolds number at fixed $St$ and $Fr$ destroys the applicability of the flow description of particles' motion. This happens because the characteristic size of turbulent vortices affecting the particles increases with $Re_{\lambda}$. At some critical Reynolds number $Re_{cr}$, it reaches $g\tau^2$ which is the length traversed by the sedimenting particle within its reaction time $\tau$. At $Re_{\lambda}\gtrsim Re_{cr}$  a single vortex can affect the particle's motion significantly. This creates situations where geometrical centers of the particles would intersect at the same point when having different velocities, the sling effect not describable by the flow. The value of $Re_{cr}$ depends on $Fr$ and $St$.  It seems quite certain that for $Fr$ and $St$ relevant for the rain formation problem, $Re_{cr}(Fr, St)$ is much smaller than the Reynolds numbers that actually hold in clouds. Thus theoretical description of behavior of droplets in clouds demands that we face the problem of describing the collision kernel in $Re_{\lambda}\to \infty$ limit.

We consider the RDF's scaling exponent $\alpha$ as a function of $Re_{\lambda}$ at fixed $Fr$ and $St$. It seems inevitable that this is an increasing function however we will only assume that it has monotonous behavior. Then since $0\leq \alpha\leq 3$ there is a finite limit (we assume that the RDF obeys a power-law at any $Re_{\lambda}$)
\begin{eqnarray}&&\!\!\!\!\!\!\!\!\!\!\!\!\!
\alpha^*(Fr, St)\equiv \lim_{Re_{\lambda}\to\infty}\alpha(Fr, St, Re_{\lambda}).
\end{eqnarray}
It is plausible that $\alpha$ that holds in clouds is $\alpha^*$ i.e. $Re_{\lambda}^{cr}(Fr, St)$ at which the limit saturates, defined by
\begin{eqnarray}&&\!\!\!\!\!\!\!\!\!\!\!\!\!
\alpha(Fr, St, Re_{\lambda})\approx \alpha(Fr, St, Re_{\lambda}^{cr})\approx \alpha^*(Fr, St),\ \ Re_{\lambda}\geq Re_{\lambda}^{cr}(Fr, St),
\end{eqnarray}
is much smaller than $Re_{\lambda}\sim 10^4$ holding in the clouds. In fact, at $Fr=0.05$, $St=1$, the value of $\alpha=0.588$ that we observed  and $Re_{\lambda}=70$ is indistinguishable from that observed by \cite{bec2014} at
$Re_{\lambda}=460$ (these authors report the correlation dimension which equals $3-\alpha$). Similarly the data of \cite{Ireland} for $Fr=0.052$ and $St=1$ demonstrate that $\alpha$ does not depend significantly on $Re_{\lambda}$ for a set of Reynolds numbers ranging from $90$ to $398$. Moreover, in the same range,  \cite{Ireland} observe essentially no dependence on $Re_{\lambda}$ for $0\leq St\leq 3$. Some data are given also for  $Re_{\lambda}=597$ where the same conclusion holds.

The data described above indicate that it is plausible that the dependence of $\alpha(Fr, St, Re_{\lambda})$ on the Reynolds number saturates at quite low $Re_{\lambda}$. If true, measurements at low $Re_{\lambda}$ would allow to predict the RDF for the high-Reynolds number turbulence in the clouds. Moreover, our low-$Re_{\lambda}$ could, at least in some cases, provide then an accurate prediction for $\alpha$ in clouds. This would remind of the observations of  \cite{ishihara,sreeni}. These works studied the dependence of moments of turbulent velocity gradients on $Re_{\lambda}$. They found that the power-law dependence, which is predicted to become valid asymptotically at large Reynolds numbers \cite{frisch}, sets in already at $Re_{\lambda}\sim 100$.

However a word of caution is in order. Numerical measurements of  \cite{ishihara,gotoh} of the dissipation range spectrum, see Eq.~(\ref{spf}), show the possibility of very slow dependencies on $Re_{\lambda}$. Since the dissipation range spectrum determines $\alpha$, at least under conditions of moderate intermittency, see Eq.~(\ref{c0}), then this makes it quite plausible also that there is a slow dependence of $\alpha$ on $Re_{\lambda}$ that simply cannot be detected over the range of $Re_{\lambda}$ studied thus far. For instance, this would imply that at $Fr=0.05$ and $St=1$, the value of $\alpha$ in clouds could differ appreciably from $\alpha=0.588$, cf. above. Present level of knowledge does not allow to decide which of the possibilities described holds.

In both cases, whether the dependence of $\alpha$ on $Re_{\lambda}$ saturates at low $Re_{\lambda}$ or is slow, it seems that the low-$Fr$ theory's prediction $\alpha=3\pi\int_0^{\infty} E(k)kdk/(2g)$ is a good estimate for the actual value in clouds. For instance, using the fit for $\alpha$ provided in Fig. \ref{ouralpha}, would propose that $\alpha=13 Fr$ is a good estimate for clouds at least for $Fr\leq 0.05$.


\section{Conclusions and outlook} \label{outlook}

In this work, we propose to use as a starting point for the study of collision kernel of droplets in warm clouds the moderate Reynolds number theory at small $Fr$. Indeed, it seems to us most natural to employ smallness of $Fr$ in order to get insight into what happens in clouds, since this is the only small parameter of this problem. The Froude number of cloud turbulence is smaller than one even for strong turbulence with $\epsilon=2000 cm^2/s^3$ where $Fr=0.5$.

The work \cite{fi2015} showed that a detailed study of $Fr\to 0$ limit is realizable. Smallness of $Fr$ implies that the droplet sweeps fast through many turbulent vortices before their impact on its motion is appreciable. In other words, action of one vortex on the particle is negligible and only many vortices are able to have a finite effect on the particle motion. This reminds of the ordinary Brownian motion where the Brownian particle does not change appreciably its momentum in one collision with particle of the fluid, and many collisions are necessary to produce a finite change of the momentum. Thus, similarly, an effective white noise description is possible.  The present work completes \cite{fi2015}, both by providing the remaining pieces of the theory and by confirming the theory by direct numerical simulations of the Navier-Stokes equations.

By itself, the above is remarkable, since quantitative predictions for problems involving the Navier-Stokes turbulence are feasible only very rarely. Here these are made possible thanks to the effective white noise features described above. The white noise is determined by one constant only - the amplitude, which is why universality ensues. However this progress must be taken with a grain of salt. The theory applies only when intermittency of turbulence is negligible, i.e. when the Reynolds number is moderate. Increasing intermittency of turbulence destroys the theory not via stronger bursts, but rather via increase of characteristic sizes of regions of calm and quiescent flow. We remind that these two aspects of intermittency go together: increase of regions of calm flow and at the same time increased probability of strong bursts \cite{frisch}.

\subsection{Current status of small $Fr$ theory of weakly intermittent turbulence}

We summarize the theory at small Froude number. In the leading order, the particle velocity is a sum of the local velocity of the turbulent flow and the sedimentation velocity in still air $\bm g\tau$. This gives velocity of the particle as a function of its position and thus defines a flow. However it is necessary to consider corrections to this approximation since the flow provided by it, is incompressible and leads to no clustering of particles in space. The theory of $Fr\to 0$ limit stands on the observation that these corrections are due to accumulated action of many vortices. More precisely these are vortices encountered by the particle along vertical path with length $g\tau^2$ made during relaxation time $\tau$ due to sedimentation at the speed $g\tau$. Within Kolmogorov theory the number of the encountered vortices is estimated as $(g\tau^2/\eta)\sim St^2/Fr$. Velocity of all vortices on this path is averaged which has the effect of creating velocity that smoothly depends on coordinates. Moreover, as in the case of Brownian particle whose velocity changes due to many independent collisions with the particles of the fluid, the motion allows a Langevin-type description. This description is quantitatively accurate allowing to consistently study the impact that intermittent features of turbulence have on the particles. This produces a significant result: the motion is influenced most by large quiescent regions of turbulence whose size $L_c$ is much larger than the Kolmogorov scale. These larger vortices are able to interact longer with the sedimenting particles and are still not too rare so that the probability of encountering them can be neglected. The increase of $Re_{\lambda}$ at fixed $Fr$ increases the probability of larger regions and the size $L_c$ due to intermittency. Thus the larger $Re_{\lambda}$ is, the larger the size of the vortices that interact with the particles is. When the size becomes of order $g\tau^2$ the theory breaks down because the number of vortices encountered by the sedimenting droplet along the path of length $g\tau^2$ is of order one.

The quantitative manifestation of the above picture is that $Fr$, for which the asymptotic theory holds, obeys the phenomenological condition $Fr\leq c (Re_{\lambda}/70)^{-\Delta_{\lambda}}$ with $0.033<c<0.05$. This condition derives from observation that validity condition behaves as a power of $Re_{\lambda}$ and that at $R_{\lambda}=70$ the theory is accurate at $Fr=0.033$ and less accurate at $Fr=0.05$. The exponent $\Delta_{\lambda}$ here is obtained from the Reynolds number scaling of a correlation length of velocity gradients $L_c$ which is defined in Eq.~(\ref{corl}), where $L_c$ is the effective size of regions of quiescent turbulence described above.

The small $Fr$ theory of moderately intermittent turbulence is defined as the theory that holds under the assumption that the condition above is obeyed. This theory can be considered proved by the numerical simulations. These are the simulations of this work, of the previous work \cite{fi2015} and of \cite{bec2014,Ireland} at $Fr=0.01$.  Despite that this theory does not apply in clouds, it might find uses in other cases.

The theory needs to be considered only for droplet sizes with $St\gtrsim 1$, since $St\ll 1$ (at $Fr<1$) was studied in detail, see all cases in \cite{fi2015}. The condition of the theory validity at $Re_{\lambda}=70$, given by $Fr\leq 0.033$ confines the mean energy dissipation rate to values smaller than $50 cm^2/s^3$. Stratiform clouds with so weak turbulence are usually non-precipitating so the consideration is mainly a theoretical limit. The theory predicts and the simulations confirm that the correlation codimension $\alpha$ of the multifractal, formed by droplests in space, is twice the Kaplan-Yorke codimension $\left|\sum_{i=1}^3\lambda_i/\lambda_3\right|$. Moreover a spectral formula holds that provides $\alpha$ as an integral of the energy spectrum.

The spectral formula does not hold at $Fr=0.05$ where the white noise description of the interaction of the droplet with the turbulent vortices is inaccurate. However we found that the correlation codimension is still twice the Kaplan-Yorke codimension, as characteristic of weakly compressible flows \cite{nature,itzhak1}. Checking if $\alpha=2 \left|\sum_{i=1}^3\lambda_i/\lambda_3\right|$ extends to $Fr>0.05$ is significant since the Lyapunov exponents are simpler for studies than $\alpha$. Thus we performed preliminary simulations at $Fr=0.5$ and $St=1$. We found $\alpha=0.664$ that within five per cent accuracy agrees with the observed value of $2\left|\sum_{i=1}^3\lambda_i/\lambda_3\right|$. This hints that $\alpha=2\left|\sum_{i=1}^3\lambda_i/\lambda_3\right|$ might hold for all combinations of $St$ and $Fr$ with $Fr\leq 0.5$. The testing of this hypothesis will be published elsewhere.

We have derived here the angular dependence of the RDF which is strong in the considered range of $Fr\leq 0.05$, see Fig. \ref{fig:angular}. Considered as a function of $\theta$ the RDF has a peak at $\theta=0$ with width $\theta^*$ of order $Fr$. The vertical alignment ratio $g(r, \theta=0)/g(r, \theta=\pi/2)$, that characterizes "preferential vertical alignment" \cite{Gustavsson} of particle pairs, is given by $(g\tau^2/L_c)^{\alpha}$, as can be seen from Eq.~(\ref{product}) (thus if $Re$ grows so that $L_c$ becomes of order $g\tau^2$, the distribution becomes isotropic, cf. above). It is greater than one and reaches maximum of about two at all considered $Fr$, see Figs. \ref{fig:angular} and \ref{theta}. This anisotropy is however of not much influence on the rate of collisions. These, of course, happen at all angles however the contribution of angles smaller than $\theta^*$ is negligible.


\subsubsection{Outlook on the droplets behavior at small $Fr$}

We describe the outlook on the droplets behavior at small $Fr$ (defined as $Fr$ for which the flow description works) and the results of the present work in this context.
The flow of droplets $\bm v$, as the driving homogeneous turbulence flow $\bm u$, is homogeneous. The two flows coincide on large scales, however are very different on small scales. The small-scale flow of droplets is strongly anisotropic. We demonstrate here that $\bm v$ has an angle-dependent smoothness scale, the scale below which velocity differences are obtained from derivatives and $\nabla \bm v$ are roughly constant. Smoothness holds over a stripe whose vertical dimension is $g\tau^2$ and horizontal dimension is $L_c$. Both dimensions are much larger than the smoothness scale of turbulence \cite{frisch}, the Kolmogorov scale $\eta$. These properties hold due to gravity and differ much from the case of $St\ll 1$ and negligible gravity.

The other difference of $\bm v$ and $\bm u$ is that $\bm v$ is much calmer at small scales \cite{fi2015}. The gradients of $\bm v$ are smaller than those of $\bm u$ by $\sqrt{Fr}/St$, at $St\gtrsim 1$. Moreover fluctuations of $\nabla\bm v$ are much less vigorous than those of $\nabla\bm u$: the particle-vortex interaction is short so the droplets' flow gradients are sums of independent effects of large number of turbulent vortices. As a result, $\nabla\bm v$ are Gaussian. This is much different from the intermittent statistics of $\nabla\bm u$. 

Here we use the Gaussianity for finding the velocity difference of colliding droplets, overcoming the old calculational problem \cite{st}, and reduce the geometric collision kernel to the RDF. This kernel describes geometric cross section and is defined as the rate at which the droplets with radii $a_1$ and $a_2$ approach by the distance $a_1+a_2$ if the hydrodynamic interactions, occurring at close distances, are neglected. The passage to the actual rate of collisions, modified by the interactions, is done by multiplying with the collision efficiency factor, see \cite{kh} for numerical tables and the modern discussion and also \cite{pk}. Theoretical calculation of the collision efficiency is hardly realizable: the flow perturbation due to the droplets does not necessarily have a small Reynolds number and also \cite{ujhd} hydrodynamic interactions of droplets at very close distances are very different from the well-studied interactions of rigid particles \cite{ujhd}, cf. \cite{fjd}. In fact, the recent observation of \cite{yavuz} of increased preferential concentration of droplets in turbulence at scales comparable with the particle size might be holding due to this difference, which study is beyond the scope of this work.

It is readily seen \cite{s,Ireland} and briefly rederived here that the geometric collision kernel (below simply collision kernel) involves the radial component of the relative velocity of droplets at the collision, conditioned that the velocity is negative. This conditional average is not convenient in theoretical calculations however. For incompressible flow it was demonstrated that the average absolute value of the velocity divided by two can be used instead of the conditional average \cite{st}. Here, we demonstrate similar reduction for the compressible droplets' flow using Yaglom-type relation \cite{reviewt}. Generally, this useful passage from conditional to unconditional averaging does not hold and it is desirable to compare these quantities numerically, cf. \cite{Wang20081,sc}.

The angle-dependent RDF $g(r, \theta)$ gives probability to find a pair of droplets at a given distance $r$ and polar angle $\theta$, where the $z-$axis is directed upwards. Thus $g(r, \theta)r^2\sin\theta dr d\theta$ is proportional to the fraction of pairs whose separation vector $\bm r$ has spherical coordinates belonging to the intervals $(r, r+dr)$ and $(\theta, \theta+d\theta)$. We use the usual normalization where $g(\bm r)=1$ at large $r$, beyond the correlation length of the particle density. The axial symmetry implies independence of the RDF of the azimuthal angle. It was demonstrated in \cite{fi2015} that since the particle separation is in the leading order horizontal, then the RDF depends on $r$ and $\theta$ only via the product $r\sin\theta$ that provides the horizontal component of the separation. Moreover the dependence on $r$ must obey a power-law with a negative exponent $\alpha$, introduced above, because of multifractality. This gives \cite{fi2015} that the RDF must be proportional to $(r\sin\theta)^{-\alpha}$. However this formula cannot be true at any $\theta$: it gives unphysical divergence of the RDF at $\theta \to 0$ and fixed finite $r$. We derive here how the divergence is regularized by the next order corrections. These destroy the conservation of the vertical distance causing $g(r, \theta)$ to flatten at angles below the critical angle $\theta^*$ of order $Fr$, cf. \cite{Ireland}. We demonstrate theoretically and numerically that at $Fr\leq 0.033$ the asymptotic behavior of the RDF at small $r$ is separable as described by Eq.~(\ref{product}). The scale $l_c$ in that equation is of order of the correlation length $L_c$ describing spatial extent of correlations of turbulent velocity gradients, see Eq.~(\ref{corl}). This scale is observed to be about ten times larger than the Kolmogorov scale $\eta$ and ten times smaller than $g\tau^2$ in the considered range of $Fr$. The last inequality guarantees matching of the asymptotic forms of $h(\theta)$ at $\theta\sim\theta^*$, that demands $\theta^*\sim l_c/(g\tau^2) \sim Fr St^{-2}l_c/\eta$, see Eq.~(\ref{thet}). For parameters of our simulations this gives $\theta^*\sim 0.1$ in complete accord with the observations.

The power law dependence of the RDF on $r$ is the manifestation of multifractality and, on general grounds, it must hold in the range of correlations of gradients of the particles' flow \cite{itzhak1}. Indeed, we observed that the cutoff scale, where the power-law dependence on $r$ breaks down, depends on $\theta$. It is $l_c$ for all $\theta\gg \theta^*$ and the much larger scale $g\tau^2$ at $\theta\ll \theta^*$ which coincides with the range of correlations of the flow gradients.

Separability of $g(r, \theta)$ is predicted by the theory. However we observed numerically that $g(r, \theta)$ is separable also at $Fr=0.05$ where the theory does not apply. Here $h(\theta)$ has behavior similar to that in Eq.~(\ref{product}), with neither $\alpha$ (about $0.588$) nor $\theta^*$ (about 0.266) much smaller than one. The scales $l_c$ and $L_c$ are of the same order and can be used interchangeably in Eq.~(\ref{product}) when $\alpha\ll 1$. At $Fr=0.05$ the difference is appreciable and $l_c=cL_c$ must be used in Eq.~(\ref{product}) with $c\sim 1$.

The behavior described by Eq.~(\ref{product}) is very different from $St\ll 1$, no gravity, case where $g(r)\sim (\eta/r)^{\alpha'}$ holds isotropically at $r\lesssim \eta$, see the derivation of $\alpha'$ in \cite{nature}. The power-law scaling holds in the range of smoothness of the droplets' flow, whose anisotropy can be seen by observing that the sedimenting droplet integrates information on the flow on time-scale $\tau$ during which it passes through the flow the distance $g\tau^2$. Thus, the gradients of the droplets' flow are roughly instantaneous gradients of turbulence averaged in space over an interval $g\tau^2$ in the $z-$direction. Therefore, in the vertical direction, the gradients of the droplets' flow change on the scale $g\tau^2$. For horizontal scale of variations, we observe that
another particle at horizontal distance of order $\eta$ interacts with different turbulent vortices. However, this does not tell that the smoothness scale in the horizontal direction is $\eta$. Indeed, the averaging smoothens the difference between the vortices observed by the particles since these obey the same statistics (if the average is over an infinite interval, the two particles would experience effectively the same average action of the flow on them). Thus, we find from the DNS that $g\tau^2\sim 10 l_c$ and $l_c\sim 10\eta$ in  Eq.~(\ref{product}). At $Fr=0.05$ the increase of the smoothness scale of the flow from $\eta$ to $l_c$ increases the angle-averaged RDF and the collision kernel by the factor of $(l_c/\eta)^{\alpha}\simeq 2$. The described smoothness properties hold beyond the flow description framework and must be included in the future studies.

The RDF $g(\bm r)$ is proportional to the pair correlation function $\langle n(0) n(\bm r)\rangle$ and the above properties imply strong anisotropy of the steady state density $n$. The transition from the power-law behavior in $r$ in Eq.~(\ref{product}) to the large-scale behavior $g(\bm r)\approx 1$ is sharp and occurs in less than a decade. Therefore the cutoff scales in Eq.~(\ref{product}) are also the correlation scales of the fluctuations on $n$. We find that the correlation length of $n$ is $l_c$ for $\theta\gg \theta^*$ and $g\tau^2$ for $\theta\ll \theta^*$. Independences of $\langle n(0) n(\bm r)\rangle$ of $z$ at $\theta\gg  \theta^*$ and of $\theta$ at $\theta\ll\theta^*$ along with its increase in vertical direction provide different aspects of columns that are formed by particles. Our study here implies that characteristic height of columns is $g\tau^2$ and their horizontal dimension is of order $l_c$. This multifractal structure is very different from $St\ll 1$, no gravity case, where the multifractal is isotropic and there are no simple preferred structures. The weak compressibility of $\bm v$ allows for full description \cite{itzhak1} of the statistics of $n$ via the considered $\langle n(0) n(\bm r)\rangle$.

Theory of $Fr\ll 1$ case provides a useful reference point to compare the results with. For instance this theory, that holds at $Fr\leq 0.033$ , predicts that $\alpha$ is independent of properties of the particles as measured by $St$. However, in \cite{bec2014}, it was observed that $\alpha$ is independent of $St$ also at $Fr=0.05$ provided that $St\geq 1$ in the considered range of Stokes numbers (for $St=0.5$ deviations were observed in \cite{fi2015}). This agrees with \cite{fi2015} who observed that for $St\geq 1$ and $Fr=0.05$ the Lyapunov exponents are well described by the $St$ dependence of the small $Fr$ theory via $E(k)$. 

We make a conjecture on the behavior of $\alpha$ as a function of $Fr$ at a fixed $St$. We make an analogy with mulitfractality's dependence on $St$ at negligible gravity ($Fr=\infty$). At $Fr=\infty$ the correlation codimension $\alpha$ increases quadratically with $St$ at $St\ll 1$ where the flow exists \cite{arxiv,nature}. It reaches maximum at $St\sim 1$ where the flow description breaks down. Then $\alpha$ decreases to zero, vanishing identically starting from a finite $St$ where the multifractality breaks down \cite{bec} (the results of \cite{bec2014} demonstrate that $\alpha$ is still positive at $St>6$). Similarly, for a fixed $St\geq 1$ we can propose that $\alpha$ grows linearly with $Fr$ at $Fr\ll 1$, reaches a maximum at $Fr\sim 0.1$ where the flow description breaks down and then decreases with $Fr$ until it becomes zero where the multifractality of droplets distribution disappears.
This picture is consistent with data of \cite{bec2014} at $St>2$ (for smaller $St$ inertial effects complicate the picture) who find that $\alpha$ at $Fr=0.3$, that is weakly dependent on $St$, is smaller than its value at $Fr=0.05$. Thus at $Fr=0.3$ the correlation codimension already passed the maximum. The low $Fr$ theory also provides good qualitative and reasonable quantitative description of the DNS's results for the RDF provided in \cite{Wang20082}. Qualitatively, \cite{Wang20082} give that the RDF does not change with numerical accuracy between $50$ and $60$ microns which agrees with the size independence of $\alpha$ (we use that variations of the size, which also determine the RDF,  do not change the RDF appreciably by $\alpha\ll 1$). The theory also explains the change of the qualitative dependence of the RDF on size at about $30-40$ microns: in this range the ratio of $\tau$ and the time of passage through one turbulent vortex due to sedimentation passes one. Thus the limit of flow description explains many observations.

\subsection{Interpolation to clouds}

Stronger intermittency associated with higher $Re_{\lambda}$ creates more of  large, spatially extended vortices which the sedimenting droplet cannot pass during the velocity relaxation time $\tau$. Individual interaction with such vortices is strong and it destroys the flow description at increasing  $Re_{\lambda}$, thereby making smallness of $Fr$ unusable. This seemingly signifies that the low-$Re_{\lambda}$ theory described here would be useless at high $Re_{\lambda}$, including those of the warm clouds. However the validity of this conclusion depends on the quantity considered.

We recall that some properties characterizing high Reynolds number turbulence can be measured at rather moderate $Re_{\lambda}$. Besides the scaling exponents of the structure functions of velocity in the inertial range, which are less relevant to the problem at hand, these are scaling properties of the dissipation statistics. Thus \cite{ishihara,sreeni} observed that intermittency exponents of moments of the dissipation field can be observed in the flow that by itself is only weakly intermittent having $Re_{\lambda}\sim 100$. It was demonstrated that the moments obey power-law dependence on $Re_{\lambda}$ with constant exponents.

Similarly, here we observe that the existing data indicate that the scaling exponent of the RDF is independent of the Reynolds number in the $100-500$ range, at least. This independence can be understood by assuming the validity of our conjecture that the exponent is well-described by $\alpha=2\left|\sum_{i=1}^3\lambda_i/\lambda_3\right|$, see above. The Lyapunov exponents $\lambda_i$ probably depend on $Re_{\lambda}$ appreciably however their ratio might be constant, cf. \cite{mj}. Thus if $\lambda_i$ obey a power-law dependence on $Re_{\lambda}$, then this dependence is identical, see discussion in \cite{recent}, and the resulting $\alpha$ is constant. Thus the exponent $\alpha$, derived at moderate $Re_{\lambda}$ theoretically in \cite{fi2015} and proved numerically here, might be the same as $\alpha$ in clouds where the high $Re_{\lambda}$ invalidates the theory (e. g. $\alpha(Fr=0.05, St=1)=0.588$). Of course, a slow dependence of $\alpha$ on $Re_{\lambda}$ can be also envisioned and extensive numerical simulations are required to study the conjecture.

The application of the theory to droplets' behavior in warm clouds demands extension of results to higher $Re_{\lambda}$ and higher $Fr$, the former discussed above. The extension to higher $Fr$ involves the inclusion of the sling effect - existence of rare regions of space where several streams of particles meet and the flow description breaks down, see the Introduction. The observations of \cite{bec2014} demonstrate that the multifractality in space still holds at $Fr\leq 0.3$, and our preliminary data indicate that the multifractality holds also at $Fr=0.5$. Thus, seemingly, the spatial distribution of droplets is multifractal for all $Fr$ of interest. In this situation, the decomposition of space into regions of sling effect and smooth flow (see the Introduction) can be effective. Another issue, that needs to be dealt with in the extension to higher $Fr$, is the decomposition of the collision kernel into the product of the average magnitude of velocity of the colliding particles and the RDF. That relies on the flow description and needs a correction, which can be studied numerically. Resolution of the described questions might eventually produce a well-working formula for the collision kernel of droplets in clouds, capable of helping us in rain prediction.

\subsection{Resume}

The small Froude number theory of motion of particles in turbulence is exact. It gives quantitative, proved results for the Navier-Stokes turbulence with moderate Reynolds numbers. Therefore we hope that, as it is usual with such results, this theory, developed in \cite{fi2015} and here, will become a significant reference point in the future studies. The application to the problem of droplet collisions in warm clouds demands interpolation from the moderate $Re_{\lambda}$ theory to the high $Re_{\lambda}$ in clouds. We demonstrated here that, at least for the radial distribution function, the interpolation seems feasible and conjectured the numerical value of the RDF's scaling exponent. 

\vspace{0.2in}
\begin{center}
\bf{ACKNOWLEDGMENTS}
\end{center}

I. F. cordially thanks Jong-Jin Baik and Hyunho Lee for fruitful and helpful discussions, Michael Wilczek for discussions and organizing a course where some ideas provided here were sharpened, and Tobias Baetge for discussions. C. L. acknowledges the support by Samsung Science \& Technology Foundation under Project Number SSTF-BA1702-03.

{}

\newpage
\begin{appendices}

\section{Separation of close droplets} \label{spso}

Here we consider the separation of droplets of the same size when the separation vector $\bm r(t)$ is in the viscous range, $r\ll \eta$. In fact, it is demonstrated in the main text that the scale of smoothness of the flow $L_c$ is larger than $\eta$ so that the described evolution has a larger region of validity. We demonstrate that the separation can be described quantitatively with the help of white noise. We formulate the Fokker-Planck equation on the evolution of the probability density function of the distance. This equation provides the Lyapunov exponent of the droplets. The exposition here is a somewhat more detailed formulation of \cite{fi2015}, which allows us to provide more details for some quantities needed in the main text.

The separation below the smoothness scale of $\bm v$ is exponential. We designate the positions of the particles by $\bm x(t)$ and $\bm x(t)+\bm r(t)$. Then, $r\ll \eta$ implies that in the distance equation $\dot {\bm r}=\bm v(t, \bm x+\bm r)-\bm v(t, \bm x)$, we can use the Taylor expansion that gives
\begin{eqnarray}&&
\dot{\bm r}=\sigma \bm r, \ \ \sigma_{ik}(t, \bm x_0)=\nabla_k v_i[t, \bm x(t, \bm x_0)]. \label{sep}
\end{eqnarray}
We defined $\bm x(t, \bm x_0)$ as the trajectory that passes through $\bm x_0$ at $t=0$. Below, we often omit $\bm x_0$ in writing. It is seen from Eq.~(\ref{lin}) that $\sigma$ are turbulent flow gradients averaged over the interval of order $g\tau^2$ in the vertical direction. Thus the gradients $\sigma$ vary over scales $g\tau^2$  in the vertical direction and, as demonstrated in the main text, over $L_c$ in the horizontal direction. The uniform applicability of Eq.~(\ref{sep}) for all directions of $\bm r$ demands $r\ll L_c$, however, for vertical separations, much larger $r$ can be considered.

The rate of change in the distance $r=|\bm r|$ is given by the radial component of the velocity $\sigma \bm r$; therefore,
\begin{eqnarray}
&&\frac{d\ln r}{dt}={\hat n}\sigma{\hat n},\ \ {\hat n}\equiv\frac{\bm r}{r},\ \ \frac{1}{t}\ln\left(\frac{r(t)}{r(0)}\right)=\frac{1}{t}\int_0^t {\hat n} \sigma{\hat n} dt'. \label{radi}
\end{eqnarray}
The equation for orientation ${\hat n}$ is obtained by using the above equation in $\dot {\bm r}=\dot r{\hat n}+rd{\hat n}/dt=\sigma\bm r$, which gives
\begin{eqnarray}&&
\frac{d{\hat n}}{dt}=\sigma{\hat n}-{\hat n}[{\hat n} \sigma{\hat n}].\label{orientation}
\end{eqnarray}
The law of large numbers (or rather ergodicity; see below) implies that with probability one, the last term in Eq.~(\ref{radi}) tends to a constant, given by statistical average, in the limit $t\to\infty$
\begin{eqnarray}
&&
\lambda_1=\lim_{t\to\infty}\frac{1}{t}\ln\left(\frac{r(t)}{r(0)}\right)=\lim_{t\to\infty} \langle{\hat n}(t)\sigma(t){\hat n}(t)\rangle. \label{firstl}
\end{eqnarray}
The limit $t\to\infty$ in the RHS is needed because it takes finite time for the statistics of ${\hat n}(t)\sigma(t){\hat n}(t)$ to become stationary. To clarify this point, consider the average at $t=0$:
\begin{eqnarray}
&&
\langle {\hat n}\sigma{\hat n}\rangle(t=0)=\langle \sigma_{ik}(t=0)\rangle \langle {\hat n}_i(t=0){\hat n}_k(t=0)\rangle=0,\nonumber
\end{eqnarray}
where we considered ${\hat n}(t=0)$ independent of $\sigma$ and used $\langle \sigma_{ik}(t=0)\rangle=\langle\nabla_kv_i(\bm x, 0)\rangle=0$ by spatial uniformity. In contrast, $\langle {\hat n}\sigma{\hat n}(t)\rangle\neq 0$ when $t>0$, both because $\sigma$ and ${\hat n}$ become correlated and because $\langle \sigma_{ik}(t)\rangle$ becomes non-vanishing. To see the latter point in the simplest context, consider the trace of the average $\langle tr \sigma(t)\rangle$, which vanishes at $t=0$ by $\langle\nabla\cdot \bm v(\bm x, 0)\rangle=0$. When $t>0$, since the particles move to regions with negative $\nabla\cdot \bm v$ and out of regions with positive $\nabla\cdot \bm v$ (preferential concentration), $\langle tr\sigma\rangle$ becomes negative in the particle's frame \cite{arxiv}.

The average in Eq.~(\ref{firstl}) is the time-average of the process ${\hat n}(t) \sigma(t){\hat n}(t)$, which becomes stationary at large times. Since ${\hat n}(t) \sigma(t){\hat n}(t)$ is uniquely determined by $\bm x_0$ and ${\hat n}(t=0)$, the average could depend on the latter. However, it is usually the case that the limit is independent of both $\bm x_0$ and ${\hat n}(t=0)$. This independence holds up to the set of $\bm x_0$ and ${\hat n}(t=0)$ with zero volume; that is, the spatial volume of $\bm x_0$ for which the independence does not hold is zero, and the area of the unit ball covered by ${\hat n}(t=0)$ for which the independence does not hold is zero. Disregarding the latter, which does not contribute to the statistical averages of non-degenerate observables, we find that $\lambda_1$ provides global characteristics of the separation of infinitesimally close trajectories independently of where the trajectories start and what their initial orientation was. 
These results are implied by the generalization of the Oseledets theorem \cite{oseledets}. The theorem establishes constancy of the limit for almost all points with respect to the natural measure (steady-state density) and not the spatial volume. However, the statistics with respect to the natural measure and the volume coincide \cite{gawedzki}; therefore, the constancy holds for almost all $\bm x(0)$ in space \cite{gawedzki}.

With regard to finding $\lambda_1$, we observe that the distance $\bm r$ does not change much over the correlation time $\tau$ of $\sigma$ in Eq.~(\ref{sep}). This is the consequence of the condition of definability of the flow $\tau^2\langle\sigma^2\rangle\ll 1$. We integrate the equation of $\bm r$ from $t$ to $t+\delta t$
\begin{eqnarray}&&\!\!\!\!\!\!\!\!
\bm r(t+\delta t)=\bm r(t)+\int_t^{t+\delta t} \sigma(t')\bm r(t')dt'.\nonumber
 \end{eqnarray}
If $\delta t$ is much smaller than the inverse of the typical value $\sigma_c$ of $\sigma$, then the last term is small. The asymptotic series in $\sigma_c \delta t\ll 1$ is obtained by solving the equation by iterations.  Neglecting the higher order terms, we have
\begin{eqnarray}&&\!\!\!\!\!\!\!\!
\bm r(t+\delta t)=\bm r(t)+\int_t^{t+\delta t} \sigma(t')dt'\bm r(t).\label{it}
\end{eqnarray}
We impose the condition $\delta t\gg \tau$, which is possible due to $\sigma_c\tau\ll 1$.  Since by causality $\bm r(t)$ is determined by $\sigma$ at times earlier than $t$, then it can be considered independent of $\sigma (t')$ in the integral in (\ref{it}), except for $t$ in the vanishingly small vicinity of $t=-\delta t$ of the order of the correlation time $\tau$ of $\sigma$. Neglecting this vicinity, we can consider the integral $\int_t^{t+\delta t} \sigma(t')dt'$ and $\bm r(t)$ in Eq.~(\ref{it}) as statistically independent. Since the integral is taken over a time interval much larger than the correlation time $\tau$, it is roughly the sum of a large number of $\delta t/\tau$ independent identically distributed random variables. We conclude that the statistics of $\int_t^{t+\delta t} \sigma(t')dt'$ is Gaussian (the proof can be obtained using the cumulant expansion theorem \cite{ma}). The latter is determined uniquely by the mean and dispersion; therefore, we can introduce the effective description of the evolution with the white-noise random matrix $\xi'$,
\begin{eqnarray}&&
\frac{d\bm r}{dt}=\xi'\bm r,\label{white}
\end{eqnarray}
where the integrals $\int_t^{t+\delta t} \sigma(t')dt'$ and $\int_t^{t+\delta t} dt'\xi'(t')$ have the same statistics, that is, the same mean and dispersion.

\subsection{The mean} \label{average}

We demand $\langle \xi'_{ik}\rangle=\langle \sigma_{ik}\rangle$, where
\begin{eqnarray}&&\!\!\!\!\!\!\!\!\!
\langle \sigma_{ik}\rangle=\lim_{t\to\infty}\langle \sigma_{ik}[t, \bm x(t, \bm x_0)]\rangle
=-\int_0^{\infty} \langle tr\sigma(0)\sigma_{ik}(t)\rangle_c dt .\nonumber
\end{eqnarray}
Here, we used the formula for Lagrangian averages via the correlation function \cite{ff,dtl}. We introduced the  unnecessary cumulant or dispersion, designated by $c$, in order to avoid the divergent integrals shown below:
we have $\langle tr\sigma(0)\rangle=0$, so $\langle tr\sigma(0)\sigma_{ik}(t)\rangle_c=\langle tr\sigma(0)\sigma_{ik}(t)\rangle$. Using Eq.~(\ref{trace}), we find a leading order in weak compressibility:
\begin{eqnarray}&&\!\!\!\!\!\!\!\!\!\!\!\!\!\!\!
\langle \sigma_{ik}\rangle=\int_0^{\infty} dt\int_{-\infty}^0 dt' \exp \left( \frac{t'}{\tau} \right) \langle tr \sigma_l^2(t')(\sigma_l)_{ik}(t)\rangle_c. \nonumber
\end{eqnarray}
Since the correlation time of $\sigma$ is $\tau$, we find $\tau\left|\langle \sigma_{ik} \rangle\right|\sim [\langle \sigma^2\rangle\tau^2]^{3/2}\sim Fr^{3/2}\ll 1$.
There can be a further factor of smallness when $i$ or $k$ equal $z$; see the next subsection. Here, we are only interested in the estimate from above. Parity and axial symmetry confine the possible form of the average to
\begin{eqnarray}
&&\!\!\!\!\!\!\!\!\!\langle \sigma_{ik}\rangle=a\delta_{ik}+b{\hat z}_i{\hat z}_k,\label{avr}
\end{eqnarray}
where $a$ and $b$ are constants. The term proportional to $\epsilon_{ikl}{\hat z}_l$, which is consistent with the axial symmetry, is forbidden by parity. By performing similar transformations for
$\langle tr\sigma\rangle$ and using Eq.~(\ref{trace}), we have
\begin{eqnarray}&&\!\!\!\!\!\!\!\!\!
\tau\langle tr \sigma \rangle=-\tau\int_0^{\infty} dt\int_{-\infty}^0 dt' \int_{-\infty}^t dt''\exp \left( \frac{t'+t''-t}{\tau} \right)  
\langle tr \sigma_l^2(t')tr \sigma_l^2(t'')\rangle_c\propto Fr^2\ll \tau \langle \sigma_{ik}\rangle.
\end{eqnarray}
This is of a higher order in $Fr\ll 1$ than a general component of $\langle \sigma_{ik}\rangle$. Thus, in the leading order, $\langle \sigma_{ik}\rangle$ is traceless:
\begin{eqnarray}
&&\!\!\!\!\!\!\!\!\!\langle \sigma_{ik}\rangle=a\left[\delta_{ik}-3{\hat z}_i{\hat z}_k\right],\ \ a\sim Fr^{3/2}/\tau.\label{avr0}
\end{eqnarray}
We conclude that the average multiplied by $\tau$ is at most of order $Fr^{3/2}$. This makes the average negligible in comparison with the dispersion by a factor of $Fr^{1/2}$. However, for possible future extensions
of the theory to the region where $Fr$ is small (however not too small), $Fr^{1/2}\sim 1$, the mean must be considered.

\subsection{Dispersion}

The dispersion is given by
\begin{eqnarray}
&&\left\langle \left[\int_t^{t+\delta t}dt_1 \sigma_{ik}(t_1)\right]\left[\int_t^{t+\delta t}dt_2 \sigma_{pr}(t_2)\right]\right\rangle_c\approx \delta t \int_{-\infty}^{\infty} dt \left[\langle \sigma_{ik}(0)\sigma_{pr}(t)\rangle_{st}-\langle\sigma_{ik}\rangle\langle \sigma_{pr}\rangle\right], \label{com}
\end{eqnarray}
where the steady-state correlation function of $\sigma$ in the particle's frame is given by
\begin{eqnarray}
&&\langle\sigma_{ik}(0)\sigma_{pr}(t)\rangle_{st}
=\lim_{t'\to \infty} \langle \sigma_{ik}[t', \bm x(t', \bm x_0)]\sigma_{pr}[t'+t, \bm x(t'+t, \bm x_0)]\rangle.
\end{eqnarray}
The limit $t\to\infty$ is necessary for $\sigma(t)$ (which is degenerate at $t=0$ where its average is zero) to become stationary. However, at weak compressibility, we can use the above Lagrangian trajectories of the solenoidal component of the flow instead of $\bm x(t', \bm x_0)$. Then, by performing spatial averaging over $\bm x_0$ passages to the variable $\bm x(t', \bm x_0)$, we find
\begin{eqnarray}&&
\langle\sigma_{ik}(0)\sigma_{pr}(t)\rangle_{st}\approx \langle\sigma_{ik}(0)\sigma_{pr}(t)\rangle,\label{simpl}
\end{eqnarray}
where the RHS is the usual spatial average. The average in the RHS tends to zero at large times since
$\langle\sigma_{ik}(0)\sigma_{pr}(t)\rangle\approx \langle\sigma_{ik}(0)\rangle\langle\sigma_{pr}(t)\rangle$ and $\langle\sigma_{ik}(0)\rangle=0$. However, if we put Eq.~(\ref{simpl}) into Eq.~(\ref{com}), then the integral becomes divergent: at large $t$, non-zero $\langle\sigma_{ik}\rangle\langle \sigma_{pr}\rangle$ is compensated by the non-zero infinite time limit of $\langle\sigma_{ik}(0)\sigma_{pr}(t)\rangle_{st}$, which we neglected due to small compressibility. The consistent approximation to the considered order gives
\begin{eqnarray}
&&\left\langle \left[\int_t^{t+\delta t}dt_1 \sigma_{ik}(t_1)\right]\left[\int_t^{t+\delta t}dt_2 \sigma_{pr}(t_2)\right]\right\rangle_c
\approx \delta t \int_{-\infty}^{\infty} dt \langle \sigma_{ik}(0)\sigma_{pr}(t)\rangle.
\end{eqnarray}
This is consistent with $\langle \sigma\rangle^2\tau^2\sim Fr^3$ derived in the previous subsection and $\langle \sigma^2\rangle\tau^2\sim Fr$ derived in Eq.~(\ref{sg}). The mean is much smaller than dispersion:
$\langle \sigma\rangle^2/\langle \sigma^2\rangle\sim Fr^2\ll 1$. We find that the statistics of $\bm r$ can be described quantitatively by Eq.~(\ref{white}), with $\xi'$ written as the sum of average (\ref{avr}) and dispersion $\xi$,
\begin{eqnarray}
&&\!\!\!\!\!\!\!
\dot{\bm r}=a\left[\bm r-3r_z{\hat z}\right]+\xi \bm r,\ \ \langle \xi_{ik}(t)\xi_{pr}(t')\rangle=\kappa_{ikpr}\delta(t-t'), \ \  \kappa_{ikpr}=
 \int_{-\infty}^{\infty} dt \langle \sigma_{ik}(0)\sigma_{pr}(t)\rangle, \label{whitedescr}
\end{eqnarray}
where $\kappa_{ikpr}$  is defined in Eq.~(\ref{kappa}). The mean is smaller than dispersion by $Fr^{1/2}$, and it can be neglected (for instance, the mean would change $\lambda_1\tau$ below by $Fr^{3/2}$, which is negligible by $\lambda_1\tau\sim Fr$). Thus, the growth of separation predominantly occurs in the horizontal plane, with $r_z$ staying constant (or rather growing parametrically slower) and $\bm r_{\perp}=(x, y)$ obeying the closed equation
$\dot{\bm r}_{\perp}=\xi_{\perp} \bm r_{\perp}$,
where $\xi_{\perp}$ is the confinement of $\xi$ to the plane. We found the evolution of $\bm r_{\perp}$ in the well-studied incompressible Kraichnan model in two dimensions \cite{review}. The probability density function $P(y, t)$ of $y(t)=t^{-1}\ln (r_{\perp}(t)/r_{\perp}(0))$ obeys
\begin{eqnarray}&&\!\!\!\!\!\!\!\!\!\!\!\!
P(y, t)=\sqrt{\frac{t}{2\pi D }}\exp\left[-\frac{t\left(y-D\right)^2}{2D}\right],\ \ \lim_{t\to\infty}P(y, t)=\delta(y-\lambda_1),\ \ \lambda_1\equiv D,
\end{eqnarray}
where $\lambda_1$ is the Lyapunov exponent and $D$ is defined in Eq.~(\ref{d0}). Due to the approximate incompressibility, we have, for the third Lyapunov exponent, $\lambda_3\approx -\lambda_1$. The second Lyapunov exponent in this approximation is determined by the vertical direction and is vanishing. The non-zero $\sum_{i=1}^3\lambda_i$ appears in the higher order approximation \cite{fi2015}.

Therefore, we wrote the Lyapunov exponent in terms of the spectrum $E(k)$ (which defines $D$), which characterizes the instantaneous statistics of turbulence instead of different time statistics, which determines the Lyapunov exponent of passive tracers.
This fits the physics of the description: the particles' drift through the flow makes them pass many correlation lengths $\eta$ of the turbulent gradients during their relaxation time $\tau$. As a result, the particles react to the accumulated action of a large number of independent turbulent vortices, considering turbulence as a frozen Gaussian field,
which is completely characterized by the power. The power is the integral of the pair-correlation function of gradients as ``seen" by the particle falling at speed $\bm g\tau$, which is what determines $\kappa_{ikpr}$.

\section{Gaussian statistics of gradients}\label{gradients}

Here we demonstrate that in contrast to the complex statistics of the gradients of the transporting turbulent flow of the fluid, which is intermitent and very non-Gaussian \cite{frisch}, the statistics of the gradients of the particle flow is Gaussian.  This observation was made in \cite{fi2015}; here, we provide a somewhat different formulation. From Eq.~(\ref{lin}), we have
\begin{eqnarray}&&\!\!\!\!\!\!\!\!\!\!\!\!\!
\nabla_k v_i(0, \bm x)=\int_{-\infty}^0 s_{ik}(t)\exp\left[\frac{t}{\tau}\right]\frac{dt}{\tau},\label{integral}
\end{eqnarray}
where $s_{ik}(t)=\nabla_k u_i[t, \bm x(t)]$ is the gradient of the turbulent flow taken in the frame of the particle that passes through $\bm x$ at $t=0$. It was observed in \cite{fi2015} that the integral has Gaussian statistics. \
Indeed, we have
\begin{eqnarray}&&\!\!\!\!\!\!\!\!\!\!\!\!\!
\int_{-\infty}^0 s_{ik}(t)\exp\left[\frac{t}{\tau}\right]\frac{dt}{\tau}\sim \int_{-\tau}^0 \frac{s_{ik}(t) dt}{\tau}.
\end{eqnarray}
The RHS is the sum of a large number $\tau/\tau_g$ of independent random variables, implying Gaussianity by the central limit theorem (The statistics of $s_{ik}(t)$ at $|t|\lesssim \tau_g$ is somewhat different due to the final
condition that the trajectory passes through $\bm x$ at $t=0$. This does not change the distribution of $\nabla\bm v$ similarly to how the distribution of the sum of a large number of independent random varaibles does not depend
on several terms in the sum.) A more formal proof can be obtained using the cumulant expansion theorem \cite{ma} for the characteristic function of the RHS of Eq.~(\ref{integral}).

Thus, at $St^2\gg Fr$, the statistics of $\nabla_k v_i$ is fixed uniquely by the average and dispersion that determine the Gaussian distribution. The average of $\nabla_k v_i(0, \bm x)$ is zero by spatial homogeneity. The dispersion is
\begin{eqnarray}&&\!\!\!\!\!\!\!\!\!\!\!\!\!
\langle \nabla_k v_i \nabla_r v_p\rangle \!\approx \!\int_{-\infty}^0\!\! \frac{dt_1dt_2}{\tau^2} \exp\left[\frac{t_1\!+\!t_2}{\tau}\right]\langle s_{ik}(t_1)s_{pr}(t_2)\rangle.
\end{eqnarray}
Since $\tau_g$ is much smaller than $\tau$, we can put $t_2\approx t_1$ in the exponent and change the limits of integration:
\begin{eqnarray}&&\!\!\!\!\!\!\!\!\!\!\!\!\!
\langle \nabla_k v_i \nabla_r v_p\rangle \!\approx \!\int_{-\infty}^0\!\! \frac{\kappa_{ikpr} dt_1}{\tau^2} \exp\left[\frac{2t_1}{\tau}\right]=\frac{\kappa_{ikpr}}{2\tau}, \label{corv}
\end{eqnarray}
where the tensor $\kappa_{ikpr}$ is defined by
\begin{eqnarray}&&\!\!\!\!\!\!\!\!\!\!\!\!\!
\kappa_{ikpr}=\int_{-\infty}^{\infty}\langle s_{ik}(0)s_{pr}(t)\rangle dt
= \int_{-\infty}^{\infty}\langle \nabla_ku_i(0)\nabla_r u_p(\bm g\tau t)\rangle dt.\label{kappa}
\end{eqnarray}
This tensor coincides with a similar tensor introduced in \cite{fh} to describe the separation of heavy inertial particles when
sedimentation is negligible but they move fast through the flow due to large inertia. The correlation function in the last line of Eq.~(\ref{kappa}) is the spatial correlation function taken at the separation $\bm g\tau t$, not the different time correlation function in the first line.
The tensor $\kappa_{ikpr}$ was found in \cite{fi2015}.
We describe the properties derived there.
We have $\kappa_{ikpr}=0$ if one of its indices is the index of the vertical coordinate $z$. Thus in the leading order in the Froude number, the flow is horizontal: it does not have vertical components or derivatives of
horizontal components in the vertical direction. This implies that the distance $\bm r$ between droplets inside
the viscous range, which obeys $\dot{\bm r}=(\bm r\cdot\nabla)\bm v$, changes only horizontally in the leading order \cite{fi2015}. The $z-$component of the separation is preserved in time in this order.
When all the indices differ from $z$, we have
\begin{eqnarray}&&\!\!\!\!\!\!\!\!\!\!\!\!\!
\kappa_{ikpr}\!=\!D\left(3\delta_{ip}\delta_{kr}\!-\!\delta_{ik}\delta_{pr}\!-\!\delta_{ir}\delta_{kp}\right),\label{kap}
\end{eqnarray}
where $D$ can be written via the energy spectrum of turbulence $E(k)$ as
\begin{eqnarray}&&\!\!\!\!\!\!\!\!\!\!\!\!\!
D\!=\!\frac{\pi\int_0^{\infty}\!\!E(k)kdk}{8g\tau}=\frac{c_0Fr}{\tau}; \ \ \ c_0\equiv \frac{\pi\nu^{1/4}\int_0^{\infty}\!\!E(k)kdk}{8\epsilon^{3/4}}, \label{d0}
\end{eqnarray}
where $c_0$ is introduced in Eq.~(\ref{d01}). Thus, we find that for the horizontal components of the flow gradients (that is, $x$ or $y$ derivatives of the $x$ or $y$ components of the particles' flow), the covariance matrix $\Gamma$ defined by $\langle \nabla_k v_i \nabla_r v_p\rangle=\Gamma_{ik, pr}$ is
\begin{eqnarray}&&\!\!\!\!\!\!\!\!\!\!\!\!\!
\Gamma_{ik, pr}=\frac{\pi\int_0^{\infty}E(k)kdk\left(3\delta_{ip}\delta_{kr}-\delta_{ik}\delta_{pr}-\delta_{ir}\delta_{kp}\right)}{16 g\tau^2}.
\end{eqnarray}
This matrix is degenerate $\Gamma_{ii, pr}=\Gamma_{ik, pp}=0$ because in this order, $\nabla_xv_x+\nabla_yv_y=0$. Thus, there are only three random variables that describe the gradients given by the components of the zero
trace $2\times 2$ matrix. The above results can be written as Eq.~(\ref{d01}).

\section{Geometric collision kernel} \label{kernel}

Here we consider the rate of collisions in a bidisperse solution of droplets with radii $a_1$ and $a_2$. The solution is assumed to be dilute so the collisions are pair-wise. In accord with the disregard of hydrodynamic
interactions in the geometric collision kernel, the droplets are considered to move independently obeying Eq.~(\ref{flo}) until their centers are at a distance of $a_1+a_2$ when the "collision" occurs. We have
\begin{eqnarray}&&\!\!\!\!\!\!\!\!\!\!\!\!\!
\frac{d\bm x_i}{dt}=\bm v_1(t, \bm x_i(t)),\ \ \frac{d\bm x_k}{dt}=\bm v_2(t, \bm x_k(t)),\label{def0}
\end{eqnarray}
where $\bm x_i$ are coordinates of droplets with radius $a_1$, $\bm x_k$ are coordinates of droplets with radius $a_2$, and $v_i(\bm x, t)$ are the corresponding radius-dependent flows.
Droplet $i$ with radius $a_1$ collides with droplet $k$ with radius $a_2$ in the time interval $(t, t+\Delta t)$, provided that at time $t$, we have $(\bm v_i-\bm v_k)\cdot (\bm x_i-\bm x_k)<0$ and
$(\bm x_i-\bm x_k)^2-(a_1+a_2)^2<-2(\bm v_i-\bm v_k)\cdot (\bm x_i-\bm x_k)\Delta t$ (this is readily seen from $d(\bm x_i-\bm x_k)^2/dt=2(\bm v_i-\bm v_k)\cdot (\bm x_i-\bm x_k)$). Thus, the total number $dN$ of collisions that occur in time interval $\Delta t$ is
\begin{eqnarray}&&\!\!\!\!\!\!\!\!\!\!\!\!\!
dN=\sum_{ik} \theta\left((\bm v_k\!-\!\bm v_i)\!\cdot\! (\bm x_i\!-\!\bm x_k)\right)\theta\left((a_i+a_k)^2
-2(\bm v_i-\bm v_k)\cdot (\bm x_i-\bm x_k)\Delta t-(\bm x_i-\bm x_k)^2\right),
\end{eqnarray}
where $\theta(x)$ is the step function, and the sum is taken over all droplets with radius $a_1$ indexed by $i$ and with radius $a_2$ indexed by $k$.
Taking the derivative over $\Delta t$, we find that the total rate $\Gamma_{12}$ of collisions of droplets with sizes $a_1$ and $a_2$ is
\begin{eqnarray}&&\!\!\!\!\!\!\!\!\!\!\!\!\!
\Gamma_{12}\!=\!2\sum_{ik} (\bm v_k\!-\!\bm v_i)\!\cdot\! (\bm x_i\!-\!\bm x_k)\theta\left((\bm v_k\!-\!\bm v_i)\!\cdot\! (\bm x_i\!-\!\bm x_k)\right)\delta\left((a_i+a_k)^2-(\bm x_i-\bm x_k)^2\right).\label{rate1}
\end{eqnarray}
This gives the rate of collisions in the whole volume of the flow. We can write it differently by introducing the concentration fields:
\begin{eqnarray}&&\!\!\!\!\!\!\!\!\!\!\!\!\!\!\!\!
n_1(t, \bm x)\!=\!\sum_i\! \delta(\bm x_i(t)\!-\!\bm x),\ \ n_2(t, \bm x)\!=\!\sum_k\! \delta(\bm x_k(t)\!-\!\bm x).\label{def}
\end{eqnarray}
We observe that by introducing integration over the unit vector ${\hat r}$,
\begin{eqnarray}&&\!\!\!\!\!\!\!\!\!\!\!\!\!
\int d{\hat r} \int d\bm x n_2(\bm x, t) n_1(\bm x+(a_1+a_2){\hat r}, t)=\sum_{ik} \int d{\hat n} \delta(\bm x_k(t)+(a_1+a_2){\hat r}-\bm x_i(t))
\nonumber\\&&\!\!\!\!\!\!\!\!\!\!\!\!\!=\sum_{ik} \frac{2\delta\left((a_1+a_2)^2-(\bm x_i(t)-\bm x_k(t))^2\right)}{a_1+a_2},\label{iden}
\end{eqnarray}
where we used the identity
\begin{eqnarray}&&\!\!\!\!\!\!\!\!\!\!\!\!\!
\int  \delta(\bm x-r{\hat r})d{\hat r}=\frac{\delta(x-r)}{x^2}=\frac{2\delta(x^2-r^2)}{x}.
\end{eqnarray}
This equation is readily proved by multiplying both sides with $r^2$ and integrating over $r$ using $\int \delta(\bm x-r{\hat r})r^2 dr d{\hat r}=1$, where $r^2 dr d{\hat r}=d\bm r$ with $\bm r=r{\hat r}$. Using Eq.~(\ref{iden}) we can rewrite Eq.~(\ref{rate1}) as,
\begin{eqnarray}&&\!\!\!\!\!\!\!\!\!\!\!\!\!\!\!\!
\Gamma_{12}\!=\!\!\int_{w_r<0}\!\!\! (a_1\!+\!a_2)^2|\left(\bm v_1(\bm x\!+\!(a_1\!+\!a_2){\hat r})-\bm v_2(\bm x)\right)\cdot{\hat r}| n_2(\bm x) n_1(\bm x\!+\!(a_1\!+\!a_2){\hat r}) d{\hat r}d\bm x.
\label{rate2}
\end{eqnarray}
Division by the total volume $\Omega$ gives Eq.~(\ref{rate}) of the main text. The formalism presented here is useful for writing the probability density function $P_{12}(\bm r)$ of finding a droplet of radius $a_2$ at a distance $\bm r$ from a droplet with radius $a_1$. We have
\begin{eqnarray}&&\!\!\!\!\!\!\!\!\!\!\!\!\!
P_{12}(\bm r)=\sum_{k}\langle\delta(\bm x_k-\bm x_i-\bm r)\rangle=\langle n_2\rangle g_{12}(\bm r).\label{rdf}
\end{eqnarray}
By using definition of concentration in Eq.~(\ref{def}), we have
\begin{eqnarray}&&\!\!\!\!\!\!\!\!\!\!\!\!\!
\langle n_1(0) n_2(\bm r)\rangle\!=\!\!\int \!\!\frac{d\bm x}{\Omega} n_1(\bm x)n_2(\bm x\!+\!\bm r)\!=\!\!\sum_{ik}\!\frac{\langle\delta(\bm x_k\!-\!\bm x_i\!-\!\bm r)\rangle}{\Omega}
=\langle n_1\rangle \sum_{k}\langle\delta(\bm x_k-\bm x_i-\bm r)\rangle=\langle  n_1\rangle \langle n_2\rangle g_{12}(\bm r),
\end{eqnarray}

which we provided in Sec. \ref{properties1}.

\section{Velocity difference in bidisperse case} \label{ash}

We have from Eq.~(\ref{integral}) for the symmetrized velocity gradient introduced after Eq.~(\ref{cold1}),
\begin{eqnarray}&&\!\!\!\!\!\!\!\!\!\!\!\!\!
\sigma^s_{ik}(\bm x)\!=\!\int_{-\infty}^0 \frac{dt}{2} \left( \frac{s_{ik}^1(t)\exp\left[t/\tau_1\right]}{\tau_1}
\!+\!\frac{s_{ik}^2(t)\exp\left[t/\tau_2\right]}{\tau_2}\right),\nonumber
\end{eqnarray}
where $s^l_{ik}$ is $\nabla_k u_i$ taken on the trajectory of the droplet with relaxation time $\tau_l$, passing  through $\bm x$ at $t=0$ with $l=1, 2$. The statistics of $\sigma_s$, similar to that of $\sigma_i$, is Gaussian with zero mean.
The dispersion is found using Eq.~(\ref{corv}), and the $\tau-$independent dimensionless tensor ${\tilde \kappa}_{ikpr}\equiv \tau \kappa_{ikpr}$ is introduced as
\begin{eqnarray}&&\!\!\!\!\!\!\!\!\!\!\!\!\!
\langle \sigma^s_{ik} \sigma^s_{pr}\rangle=\frac{{\tilde \kappa}_{ikpr}}{8\tau_1^2}+\frac{{\tilde \kappa}_{ikpr}}{8\tau_2^2}+\frac{I}{2},\ \
I=\int_{-\infty}^0\frac{dt_1dt_2}{\tau_1\tau_2} \exp\left[\frac{t_1}{\tau_1}+\frac{t_2}{\tau_2}\right]\langle s^1_{ik}(t_1)s^2_{pr}(t_2)\rangle.\label{dspr}
\end{eqnarray}
We consider I. Since the correlation time $\tau_g$ of $s$ is much smaller than $\tau$, we can write
\begin{eqnarray}&&\!\!\!\!\!\!\!\!\!\!\!\!\!
I\!=\!\int_{-\infty}^0\!\!\frac{dt_1}{\tau_1\tau_2} \exp\left[t_1\left(\frac{1}{\tau_1}\!+\!\frac{1}{\tau_2}\right)\right]\int_{-\infty}^{t_1}\!\!\langle s^1_{ik}(t_1)s^2_{pr}(t_2)\rangle dt_2
+\int_{-\infty}^0\!\!\frac{dt_2}{\tau_1\tau_2} \exp\left[t_2\left(\frac{1}{\tau_1}\!+\!\frac{1}{\tau_2}\right)\right]\int_{-\infty}^{t_2}\!\!\langle s^1_{ik}(t_1)s^2_{pr}(t_2)\rangle dt_1,\nonumber
\end{eqnarray}
where the contributions are symmetrized in $I$. In contrast to the monodisperse case, the correlation function $\langle s^1_{ik}(t_1)s^2_{pr}(t_2)\rangle$ depends on both $t_1$ and $t_2-t_1$, and not only on the difference $t_2-t_1$. This is because the trajectories of the droplets starting at the same point diverge in time because of the size difference. Thus, we consider the divergence of the trajectories.
The distance $\bm r$ between two droplets with size $a_1$ and $a_2$ obeys
\begin{eqnarray}&&\!\!\!\!\!\!\!\!\!\!\!\!\!
\dot{\bm r}=\bm v_1(\bm x+\bm r)-\bm v_2(\bm x)\approx\sigma^s \bm r+\bm g\Delta \tau,\label{disat}
\end{eqnarray}
where we assumed $r\ll \eta_p$. At times of order $\tau_i$ the $\sigma^s$ term cannot produce significant changes in $\bm r$ because the typical value of $\sigma^s\tau$ of order $\sqrt{Fr}$ is assumed to be much smaller than one; see Eq.~(\ref{velt}). Therefore, we can assume that the distance between the trajectories in the integrand of $I$ obeys $\bm r(t)=\bm g\Delta \tau t$. Correspondingly, we have $r(t\sim\tau)\sim g\tau\Delta \tau\lesssim (a_1+a_2)\sqrt{Fr}$, where we used the condition on $\Delta \tau$ for which we perform the calculation. We have $r(t\sim\tau)\ll \eta$, so the displacement of the trajectories at times relevant in $I$ is negligible. Using $\kappa_{prik}=\kappa_{ikpr}$, we find that
\begin{eqnarray}&&\!\!\!\!\!\!\!\!\!\!\!\!\!
I\!=\!\int_{-\infty}^0\!\!\frac{dt_1}{2\tau_1\tau_2^2} \exp\left[t_1\left(\frac{1}{\tau_1}\!+\!\frac{1}{\tau_2}\right)\right]{\tilde \kappa}_{ikpr}
+\int_{-\infty}^0\!\!\frac{dt_2}{2\tau_1^2\tau_2} \exp\left[t_2\left(\frac{1}{\tau_1}\!+\!\frac{1}{\tau_2}\right)\right]{\tilde \kappa}_{prik}\!=\!\frac{{\tilde \kappa}_{ikpr} }{2\tau_1\tau_2}.\nonumber
\end{eqnarray}
By using Eq.~(\ref{dspr}), we conclude that
\begin{eqnarray}&&\!\!\!\!\!\!\!\!\!\!\!\!\!
\langle \sigma^s_{ik} \sigma^s_{pr}\rangle=\frac{{\tilde \kappa}_{ikpr}}{8}\left(\frac{1}{\tau_1}+\frac{1}{\tau_2}\right)^2.
\end{eqnarray}
For $\tau_1=\tau_2=\tau$, this reduces to the previously derived Eq.~(\ref{corv}). We conclude using Eq.~(\ref{kap}) that the dispersion of $y$ is
\begin{eqnarray}&&\!\!\!\!\!\!\!\!\!\!\!\!\!
\langle y^2\rangle_c=\frac{c_0Fr(a_1+a_2)^2\sin^4\theta}{8}\left(\frac{1}{\tau_1}+\frac{1}{\tau_2}\right)^2.
\end{eqnarray}
where $c_0$ is defined in Eq.~(\ref{d0}). By using the average of the modulus of the Gaussian variable with non-zero mean,
\begin{eqnarray}&&\!\!\!\!\!\!\!\!\!\!\!\!\!
\delta v_{12}=\sqrt{\frac{c_0Fr}{\pi}} \frac{(a_1+a_2)(\tau_1+\tau_2)\sin^2\theta}{2\tau_1\tau_2}
\exp\left(-\frac{4g^2\tau_1^2\tau_2^2(\Delta \tau)^2\cos^2\theta}{c_0Fr(a_1+a_2)^2(\tau_1+\tau_2)^2\sin^4\theta}\right)\nonumber \\&&\!\!\!\!\!\!\!\!\!\!\!\!\!
+g\Delta \tau\cos\theta \times er\!f\left(\frac{2g\tau_1\tau_2\Delta \tau\cos\theta}{\sqrt{c_0Fr}(a_1+a_2)(\tau_1+\tau_2)\sin^2\theta}\right),
\end{eqnarray}
where $er\!f(x)$ is the error function. This formula is derived assuming that $|\Delta \tau|/\tau \lesssim \epsilon_0\ll 1$. However, if this formula is used at $|\Delta \tau|/\tau \gg \epsilon_0$, it reproduces Eq.~(\ref{lrg}). Thus, in fact, the formula can be used at any $\Delta\tau$. This completes the calculation of the average modulus of velocity of the colliding droplets. Since the first term is
relevant only at $|\Delta \tau|/\tau \ll 1$, we can simplify and symmetrize the formula as given by Eq.~(\ref{modl}).

\end{appendices}

\end{document}